\documentclass[iop,apj,tighten,numberedappendix,twocolappendix]{emulateapj}
\usepackage{natbib}
\newcommand{\figref}[1]{Fig.~\ref{fig:#1}}
\newcommand{\figtextref}[1]{Figure~\ref{fig:#1}}
\newcommand{\tabref}[1]{Tab.~\ref{tab:#1}}
\newcommand{\tabtextref}[1]{Table~\ref{tab:#1}}
\newcommand{\secref}[1]{\S\ref{sec:#1}}
\newcommand{\subsecref}[1]{\S\ref{subsec:#1}}
\newcommand{\subsubsecref}[1]{\S\ref{subsubsec:#1}}
\newcommand{\appref}[1]{App.~\ref{app:#1}}

\newcommand{\kms}[0]{$\mathrm{km s^{-1}}$}

\begin{document}

\title{Mergers in Galaxy Groups - I. Structure and Properties of Elliptical Remnants}
\author{Dan S. Taranu, John Dubinski, and H.K.C. Yee}
\affil{Department of Astronomy and Astrophysics, University of Toronto, 50 St. George Street, Toronto, Ontario, Canada, M5S 3H4}

\begin{abstract}
We present collisionless simulations of dry mergers in groups of three to twenty-five galaxies to test the hypothesis that elliptical galaxies form at the centers of such groups. Mock observations of the central remnants confirm their similarity to ellipticals, despite having no dissipational component. We vary the profile of the original spiral's bulge and find that ellipticals formed from spirals with exponential bulges have too low Sersic indices. Mergers of spirals with de Vaucouleurs (classical) bulges produce remnants with larger Sersic indices correlated with luminosity, as with SDSS ellipticals. Exponential bulge mergers are better fits to faint ellipticals, whereas classical bulge mergers better match luminous ellipticals. Similarly, luminous ellipticals are better reproduced by remnants undergoing many ($>$5) mergers, and fainter ellipticals by those with fewer mergers. 
	The remnants follow tight size-luminosity and velocity dispersion-luminosity (Faber-Jackson) relations ($<$0.12 dex scatter), demonstrating that stochastic merging can produce tight scaling relations if the merging galaxies also follow tight scaling relations. The slopes of the size-luminosity and Faber-Jackson relations are close to observations but slightly shallower in the former case. Both relations' intercepts are offset - remnants are too large but have too low dispersions at fixed luminosity. Some remnants show substantial (v/$\sigma >$ 0.1) rotational support, although most are slow rotators and few are very fast rotators (v/$\sigma >$ 0.5).
	These findings contrast with previous studies concluding that dissipation necessary to produce ellipticals from binary mergers of spirals. Multiple, mostly minor and dry mergers can produce bright ellipticals, whereas significant dissipation could be required to produce faint, rapidly-rotating ellipticals.
\end{abstract}


\section{Introduction}
\label{sec:introduction}
Merging of spiral galaxies is a promising mechanism for producing elliptical galaxies. Although it was perhaps not until \citet{Too77} that interacting spirals became widely accepted as elliptical progenitors, simulations of interacting spirals date back at least to \citet{TooToo72} and arguably as far as \citet{Hol41}. Much of this work has focused on major mergers (mass ratio $<$3:1) of pairs of spiral galaxies on parabolic orbits. While such binary major mergers certainly are observed in the local universe - and will likely be the ultimate fate of the Milky Way and M31 - they may not be as common as minor hierarchical mergers. 

Observational evidence and numerical simulations suggest that most L* galaxies are found in groups \citep{McGBalBow09}, where the central galaxy is likely to have experienced multiple mergers and have several surviving satellites. Furthermore, late-type galaxies in groups follow a Schechter luminosity function \citep{Sch76} similar to those in other environments \citep{CroFarNor05,RobPhideP10}. Thus, if high-redshift groups are composed primarily of spiral galaxies, they are likely dominated by several bright spirals with a larger number of fainter satellites, as in our Local Group. Our hypothesis is that groups of three or more spiral galaxies with luminosity distributions following a Schechter function will naturally merge to produce a central elliptical galaxy, and possibly fainter satellites.

We aim to test this hypothesis with numerical experiments. More specifically, we test whether the properties of the central galaxies formed through collisionless mergers in groups of spirals are consistent with observations of local ellipticals. This first paper in a series outlines our methodology and demonstrates that the results are both qualitatively and quantitatively different from both the more prevalent studies on binary mergers and also the less abundant literature on galaxy group mergers. We present results on morphological and kinematical measures as well as two dimensional scaling relations. Paper II \citep{TarDubYee13b} will focus on the three dimensional fundamental plane scaling relation. To further motivate this endeavour, we will outline some of the key results of the past several decades of work in this field.

\citet{CarCavSan81} and \citet{IshMatTaj83} were amongst the first to produce simulations of mergers in groups of galaxies (10-20 each), using 20 and 100 particles per galaxy, respectively. \citet{Bar85} introduced separate stellar and dark matter profiles, with 30 and 270 particles for each component, respectively. \citet{Bar89} added stellar disks and bulges to the galaxy models, taking advantage of the newly invented N-body tree code \citep{BarHut86} to increase resolution to 4096 luminous and dark particles each. While the arrangement of orbits was somewhat artificial (a pair of triple systems each consisting of a binary orbiting a more massive single galaxy), the study showed that mergers of realistic galaxies in compact groups can be rapid and produce central remnants with de Vaucouleurs profiles and shell structures, similar to local ellipticals.

A problem with the collisionless merger scenario was identified by \citet{Car86}. Collisionless mergers cannot increase central phase space density, which is necessary if disks are to merge and produce ellipticals with higher central densities than their progenitors, as is often the case. \citet{HerSpeHey93} directly addressed this issue with simulations of binary mergers of spirals with bulges \citep{Her93}, which, unlike mergers of bulgeless spirals \citep{Her92}, are capable of producing sufficiently centrally dense remnants.

\citet{WeiHer96} extended this methodology to mergers of groups of six equal-mass spirals. The resolution of these simulations was a factor of 18 higher than \citet{Bar89}, with almost 150,000 particles per galaxy, and included a comparison sample of pair mergers. Group mergers were shown to produce remnants with some rotation, in contrast to the non-rotating remnants typical of dry binary mergers \citep{CoxDutDiM06}. Both varieties of mergers produced early-type galaxies well fit by de Vaucouleurs profiles. However, group mergers with bulges did not maintain centrally concentrated profiles, exhibiting the same low central phase space density as found in bulgeless pair mergers. This may be seen as once again disfavoring the group merging scenario; however, it should be noted these mergers were of equal-mass galaxies and hence of very large mass ratios, which would tend to maximize this problem. Furthermore, \citet{KorFisCor09} argues that ellipticals have lower central densities ('cores') than expected from inward extrapolation of their outer surface brightness profiles, which is consistent with the group merging scenario and essentially the opposite of the central phase space density problem, although \citet{KorFisCor09} attribute these cores to scouring of inner stars by inspiraling supermassive black holes rather than purely stellar dynamical processes.

Since \citet{WeiHer96}, the relatively rapid pace of advancements in the field of simulations of merging in galaxy groups has slowed, with the focus shifting to studies of hydrodynamical processes in group environments. This can be partially attributed to the findings of \citet{RobCoxHer06} that collisionless binary mergers are unable to produce ellipticals following a tilted fundamental plane relation, although \citet{AceVel05} found an appreciable tilt by merging spirals sampled from an appropriate Schechter luminosity function. \citet{RobCoxHer06} also found that collisionless binary mergers could only produce very slow-rotating remnants and that dissipation was required to produce significantly rotationally-supported ellipticals. However, between \citet{WeiHer96} and \citet{RobCoxHer06}, very few studies have tested whether these results apply to multiple collisionless mergers as well. Galaxy clusters \citep{Dub98} and starbursts in groups \citep{Bek01} have been considered. \citet{CioLanVol07} simulated consecutive mergers of spheroidal galaxies, roughly approximating hierarchical group merging. While this approach has provided useful estimates of the growth of stellar mass and size, the use of purely spheroidal progenitors is questionable given the prevalence of disks at high redshift. \citet{HopHerCox09} combined the results of binary merger simulations with cosmological merger trees, using empirical halo occupation models and spiral scaling relations to predict the evolution of early-type scaling relations. However, the models only incorporated multiple mergers by allowing binary merger remnants sufficient time to relax dynamically, whereas \citet{MosMacSom12} find that halos undergoing multiple mergers are likely to have two mergers in quick succession. Galaxies in groups are thuse likely to undergo multiple mergers within a relatively short periods rather than a steady stream of isolated mergers.

More recently, fully cosmological simulations of mergers in a group or several groups of galaxies have been performed by, for example, \citet{KhaCatSch08} and \citet{FelCarMay11}. Such simulations naturally incorporate hierarchical merging, typically by using the 'zoom-in' method of re-simulating a small sub-volume of a large dark matter-only cosmological simulation at higher resolution. \citet{NaaJohOst09} and \citet{OseNaaOst12} demonstrated that ellipticals can form in groups, with minor mergers being an important mechanism in controlling the evolution of the central galaxy size and mass. \citet{FelCarMay11} showed success in producing not only a central elliptical but early-type satellites by simulating a single group using this method. However, such ab initio simulations encounter difficulties achieving sufficient spatial resolution to produce realistic spiral galaxies. Typical softening lengths in such simulations are between 500 to 1000 pc, which significantly alters the inner profile of ellipticals, especially at low masses. Increasing the resolution can mitigate this problem but also greatly increases computational cost and limits the possible sample size to a few groups.

It is clear that there is a gap in the literature on multiple mergers in groups, even though multiple mergers likely create brightest cluster galaxies \citep{Dub98}. By contrast, observations of elliptical galaxies have advanced tremendously in recent years, providing public catalogs of morphologies of thousands of nearby ellipticals and spirals alike \citep[e.g.][]{BlaSchStr05,HydBer09a,NaiAbr10,SimMenPat11}, mainly based on Sloan Digital Sky Survey (SDSS) images. The SAURON project \citep{deZBurEms02} and its volume-limited successor survey Atlas3D \citep{CapEmsKra11} have provided integral-field kinematics of hundreds of early type galaxies at a comparable resolution to SDSS. It is increasingly necessary to match the large samples of observations with simulations and explore the vast parameter space of conditions in galaxy groups.

To meet the requirement of a large simulation sample, it is currently necessary to focus on dry merging and gravitational dynamics alone. Hydrodynamical simulations are more computationally expensive and add numerous parameters to initial conditions: disk gas fractions, gas disk scale heights and lengths relative to the stellar disk, the presence of a gaseous halo, etc. More importantly, existing literature has yet to establish the effects of collisionless gravitational dynamics in group mergers on central remnant structure. There is observational evidence suggesting that dry merging contributes significantly to the growth of massive galaxies, particularly ellipticals \citep[e.g.][]{vDoWhiBra10}. Even if exclusively dry merging is not the most common mechanism for forming ellipticals, many ellipticals will have experienced at least one dry merger in their lifetimes and it is instructive to ask what collisionless dynamics alone would predict before moving on to hydrodynamical processes.

The remainder of the paper is structured as follows: \secref{simulations} motivates and details the methods used in creating the simulations, while \secref{analysis} details the analysis methodology and pipeline, with additional tests presented in \appref{analysis_testing}. A more detailed examination of numerical convergence can be found in \appref{numerics}. Key results on morphology, scaling relations and kinematics of central remnants are presented in \secref{results}. The implications of these results on theories of elliptical galaxy formation are detailed in \secref{discussion}, with reference to prior studies on the subject. The conclusions are summarized in \secref{conclusions}.

\section{Simulations}
\label{sec:simulations}
The simulations are designed to extend the methodology of binary galaxy merger simulations to groups of galaxies. This section details the parameters of the group sample (\subsecref{ic_groupsample}), as well as the two key ingredients in the initial conditions: group configuration (\subsecref{ic_groupconfig}) and galaxy models (\subsecref{ic_galaxymodels}). Finally, the code and parameters used for the simulations are described in \subsecref{simcode}.

Our choice of initial condition parameterization is designed to evenly sample the parameter space of groups which are likely to produce a central elliptical remnants, rather than be a unbiased sampling of real, nearby galaxy groups. This approach is similar to that used in binary mergers simulations, in which the orbits are typically nearly parabolic, with some cosmologically-motivated distribution of pericentric distance and disk alignment. In our case, we model group-sized halos at the turnaround radius at z=1-2, such that the subhalos are likely to contain spiral galaxies which will eventually merge to form one central elliptical.

We use two galaxy models designed to reproduce the surface brightness profile and rotation curve of M31 and scale these models according to the Tully-Fisher relation \citep{TulFis77}. The only parameter we vary between the models is the profile of the bulge, which has a substantial impact on the structure of the merger remnant (~\subsecref{morphology}). While this approach does not reproduce the variety of spiral galaxies found in the local universe, let alone at high redshift, it maintains the simplicity of the initial conditions. We do not vary the bulge fraction in the progenitors, as pre-formed bulges are required to produce sufficient central densities in the merger remnant. We discuss these choices further in \secref{discussion}.

Although the simulations are nominally scale-free, as with any system of units having G=1, our simulations assume units of length in kpc, velocity in 100 \kms, time in 9.73 million years and mass in $2.325 \times 10^9 M_{\odot}$. The luminosity function sampling and initial group radius impose a unique, preferred scaling to each simulation, such that mergers of groups with the same number of galaxies but different luminosities are not simply re-scaled versions of each other.

\subsection{Group Sample}
\label{subsec:ic_groupsample}

We create groups with total luminosities from 0.1-10L* and masses between $2 \times 10^{11}-2 \times 10^{13} M_{\odot}$. We incorporate several basic assumptions consistent with observations and cosmological simulation predictions. More massive groups contain more galaxies on average, with galaxies preferentially located closer to the center of the group. The group as a whole is initially collapsing, with galaxies located within $R_{max}=2 \times R_{200,z=2}$ but having insufficient orbital energy to prevent collapse (i.e. the groups are sub-virial). We simulate each group configuration twice, with each simulation containing either spiral galaxies with exponential bulges or classical bulges (but not both), referring to the former sample as B.n$_{s}$=1 and the latter as B.n$_{s}$=4 for short.

There are 3 sets of simulations, each with different random seeds for the initial conditions. Each set has 7 target luminosity (or mass) bins, ranging from 1/8 to 8 L* and increasing by factors of 2. Each bin contains 8 groups, for a total of $3 \times 7 \times 8=168$ simulations, of which $3 \times 7 \times 2=42$ are mergers of spirals with equal masses, while the remaining 126 are sampled from a realistic luminosity function. Since each simulation is run twice (with different spiral bulge profiles), there are nominally 336 simulations, but only 168 different sets of galaxy masses and orbits.

Each group has a number of galaxies between $N_{min}=3$ and $N_{max}=2 + (5/6) \times 10 \cdot (L/L*)^{1/2}$. Within each group luminosity bin, the number of galaxies in each simulation varies linearly from the minimum (3) to the maximum, so that L* groups have between 3 and 10 galaxies and the largest groups have 25 galaxies. This range of galaxy numbers roughly covers the number of bright galaxies one would expect in poor groups. The mass range covered by the groups is $2.0 \times 10^{11} M_{\odot}$ to $3.0 \times 10^{13} M_{\odot}$. 

\begin{table}
\caption{Range of Numbers of Galaxies Initially in Each Group}
\begin{tabular}{cccccc}
\hline 
Group Mass (M*) & N$_{min}$ & N$_{max}$,F & N$_{min}$,M & N$_{max}$ & \\
\hline
1/8 & 3 & 3 & 4 & 4 \\ 
1/4 & 3 & 4 & 5 & 6 \\ 
1/2 & 3 & 5 & 6 & 8 \\ 
1 & 3 & 6 & 7 & 10 \\
2 & 3 & 7 & 9 & 13 \\
4 & 3 & 9 & 12 & 18 \\
8 & 3 & 12 & 16 & 25 \\
\hline
\end{tabular}
\tablecomments{Each simulation sample is divided into those groups with relatively Few (F) or Many (M) galaxies for their mass, with three groups in either category per mass bin. The minimum and maximum number of galaxies in a group is listed, as well as the maximum for the F and the minimum for the M subsamples.}
\label{tab:ngals}
\end{table}

We further subdivide the sample into groups with relatively many mergers (the Many-merger or 'M' subsample) or relatively few (Few-merger or 'F'). The groups in each mass bin with the three lowest initial galaxy counts are part of the F subsample, while the groups with the three largest initial galaxy counts qualify for the M subsample. Because the maximum number of galaxies changes in each mass bin, the dividing line between the Many-merger and Few-merger subsamples depends on mass and is not a fixed number of galaxies or mergers. Each mass bin also contains two groups with equal-mass galaxies ('Eq'), one with three galaxies ('F-Eq') and the other with the same number of galaxies - N$_{min}$,M - as the fourth group in the LF-sampled simulations (i.e., the group in the 'M' subsample with the fewest galaxies). The number of galaxies in a representative number of groups is listed in \tabref{ngals}. 

\subsection{Group Configuration}
\label{subsec:ic_groupconfig}

Once the target luminosity and number of galaxies are selected, each group is initialized through the following steps:

\begin{enumerate}
\item Randomly select luminosities for all of the galaxies from a restricted range of the spiral galaxy luminosity function.
\item Set the maximum radius within which to spawn galaxies, $R_{max}=2R_{200,z=2}$.
\item Place the most luminous galaxy in the center.
\item Place all other galaxies in order of decreasing luminosity.
\item Compute the group's gravitational potential energy.
\item Assign random velocities to the satellite galaxies, applying an inward and radial bias and re-selecting any velocities with $v > v_{esc}$.
\end{enumerate}

Galaxy luminosities are randomly sampled from the inclination- and extinction-corrected spiral luminosity (Schechter) function of \citet{Sha07}. For the r-band, the faint-end slope $\alpha=-1.26$ and $M*_{r}=-20.99+5\log(h=0.71)$, or $M*_{r}=-21.73$, which is nearly identical to our standard M31 model's absolute magnitude of $M_{r}=-21.69$. We set a minimum luminosity of $0.01L*$ for the spirals, as we do not expect the luminosity function to continue to arbitrarily faint magnitudes. Luminosities are drawn from a restricted range of the luminosity function with a width equal to $(N_{gals}+2)/10$ dex, such that the integral under the curve is equal to the target group luminosity. This limited range produces groups with smaller magnitude differences between the brightest central galaxy and the next brightest satellite, making major mergers more likely - especially in groups with few galaxies. We avoid simulating groups with a single luminous spiral and several much fainter satellites, since these groups would only produce relatively minor mergers and would be unlikely to produce ellipticals. The 42 groups with equal-mass spirals ('Eq') are not sampled from an LF and instead have exactly the target luminosity, split evenly between three or a larger number of galaxies.

Once a luminosity is determined for each group galaxy, the galaxies are randomly assigned locations within the group in order of decreasing luminosity. The most luminous galaxy is placed at the center of the halo, while subsequent galaxies are given a random radius with a likelihood inversely proportional to radius, i.e., $\rho \propto r^{-1}$, and with $r<R_{max}$, where $R_{max}=2R_{200,z=2}$. $R_{200,z=2}$ is the radius at which a volume enclosing the total group mass has a mean density of 200 times the critical density at z=2. Next, a random polar and azimuthal angle is given. The minimum distance between galaxies is set by $0.5 \times R_{max}*(M_{galaxy}/M_{group})^{1/3}$, which allows galaxy halos to be in contact but not overlap significantly.

All of the galaxies are given preferentially inward and radial orbits. The group itself is sub-virial to ensure collapse and no satellite is given a speed $|v| > v_{escape}$. This is accomplished by giving each group a target virial ratio $Q_{target}=-2T/W=0.5$, where W is the gravitational potential energy of the group. This is equivalent to a zero-energy parabolic orbit in a galaxy pair, where T=-W. The group velocity dispersion is determined by $Q_{target}$: $\sigma=(-Q \times (W/a)/M)^{0.5}$, where $a=3s-2\beta$ and $\beta$ is an orbital anisotropy parameter. Each galaxy's radial velocity is then $\sigma*s$, where s is a number randomly selected from a unit Gaussian centered on s=0.5. On average more than 70\% of galaxies will have inward radial velocities. The azimuthal and polar velocities are given by $\sigma(s \times (1-\beta)^{0.5})$, where $\beta=0.5$ directs most of the velocity of the galaxy radially. 

In two of the three sets of simulations, the initial conditions in groups of similar mass (1/8 to 1/2, 1 to 4, and 8) are correlated, in the sense that galaxy positions are seeded in the same order (but not individual galaxy masses or orbits). This is intended to test the effect of adding additional galaxies to otherwise similar initial conditions. In the third set, all of the initial conditions are completely randomized. We note no statistically significant differences between the partially and completely random initial conditions for the relations presented in this paper; however, we will note some differences in the fundamental plane parameters in Paper II.

Finally, each galaxy has its own massive, extended dark halo. In practice, the these individual halos overlap in their outer regions, leaving little or no 'empty' space between galaxies. We have also experimented with including a separate dark matter halo for the group, not associated with any particular galaxy, but find that group galaxies then (unrealistically) merge with this invisible dark halo rather than with each other.

\subsection{Galaxy Models}
\label{subsec:ic_galaxymodels}
Initial spiral galaxy models are created using the GalactICS galaxy initial condition code \citep{WidDub05}. This code generates equilibrium models of galaxies with a bulge, disk and halo through spherical harmonic expansions of analytic potentials. Although the models begin in equilibrium and do not require additional time to settle, they have been tested in isolation. All models remain (statistically) unchanged for at least a Hubble time, even at the lowest resolution (55,000 particles per galaxy), which the vast majority of galaxies exceed.

The models are similar to the 'M31c' model of \citet{WidDub05}, with some parameters adjusted following the approach of \citet{WidPymDub08} to better reproduce the surface brightness profile and rotation curve of M31. M31 was chosen as a well-studied, nearby spiral having a sufficiently massive bulge to produce concentrated merger remnants. The models are bar-stable and contain a massive, non-rotating bulge, as well as a dark matter halo. The first variant uses a nearly exponential bulge with $\mathrm{n_{s}}=0.93$, which will be referred to as exponential for convenience. A second variant uses an $\mathrm{n_{s}}=4$ de Vaucouleurs or classical bulge but otherwise identical parameters.

The halo density profile is designed to match an NFW \citep{NavFreWhi97} profile at large radii and smoothly drop to zero at large radii. The halo has a 6.07 kpc scale radius, $\rho \propto r^{-1}$ inner cusp and $r^{-2.3}$ outer slope, an outer radius of 300 kpc and a total mass of 1185 units, or $2.75 \times 10^{12} M_{\odot}$. This profile produces a 30:1 ratio in dark:baryonic (stellar) mass, which is a factor of two larger than estimates for M31 and the Milky Way \citep[e.g.][]{Wat10}, but smaller than global estimate for the universal dark:stellar mass ratio.

The disk has a 5.8 kpc scale radius and a 750 pc $\mathrm{sech^2}$ scale height, equivalent to a 375 pc exponential scale height. The disk is cut off past 6 scale radii, or 35 kpc, for a total mass of 25 simulation units, or $5.8 \times 10^{10} M_{\odot}$. We adopt a disk stellar mass-to-light ratio of $(M/L_{r})_{D}=3.4$. 

Each bulge has a 1.5 kpc effective radius. The exponential bulge and de Vaucouleurs bulge have masses of 14.75 units ($3.4 \times 10^{10} M_{\odot}$) and 15 units ($3.5 \times 10^{10} M_{\odot}$), respectively. The bulge-to-total mass ratio $B/T_{M}$ is about 33\% in both models, larger than the 20-30\% estimate for the R-band B/T ratio $B/T_{R}$ in M31 \citep{CouWidMcD11}. The models could compensate by using a lower $(M/L_{r})_{B}$; however, the bulge kinematics favor a lower value of 1.9 \citep{WidDub05}, which we adopt here. While $(M/L_{r})_{B}$ does not affect the simulations, the resulting bulge-to-total light ratio $B/T_{r}$ is 50\%, and so mock images are more strongly weighted to bulge stars than disk stars.

\begin{figure}
\includegraphics[width=0.49\textwidth]{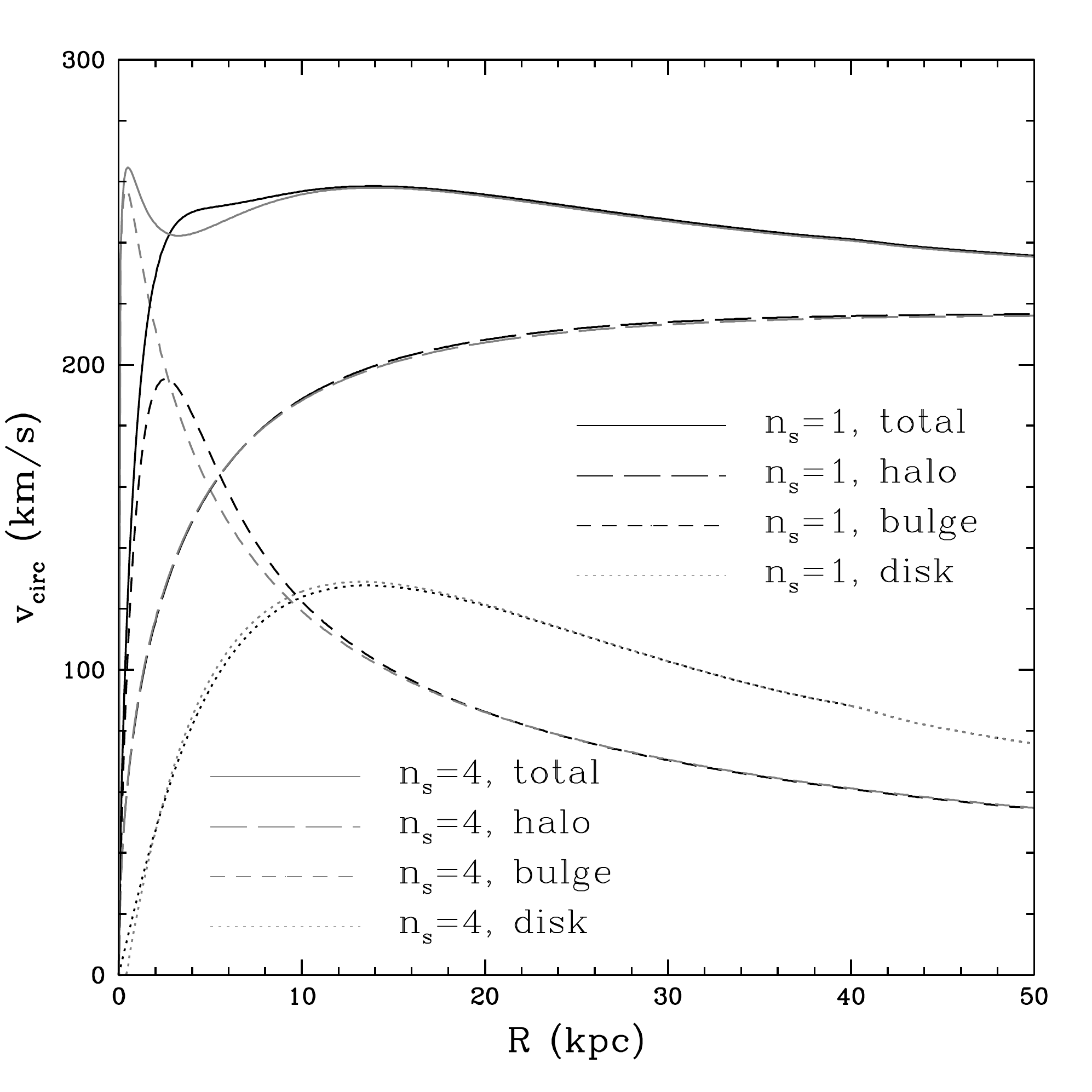}
\caption{The rotation curve of the M31 models used in these simulations. The rotation curve is dominated by the bulge inside in the inner 5 kpc and by the halo thereafter. The more concentrated $\mathrm{n_{s}=4}$ bulge also produces a sharper, inner peak in the rotation curve at 1 kpc.
\label{fig:m31rotationcurve}}
\end{figure}

The rotation curve for both models is shown in \figref{m31rotationcurve}. The bulge dominates within the inner 4-5 kpc and the halo thereafter, with the disk contribution typically half that of the halo. The non-maximal disk is both consistent with recent observations of spiral galaxies (see \citet{vdKFre11} for a review) and also promotes bar stability. Although the exponential bulge can be torqued into a bar, the intrinsic bar stability means that any remnant properties such as rotation are a result of the merging process and not secular instabilities.

To scale our model to different masses, we multiply all masses by a factor $m$ while retaining the same $M/L_{R}$. Velocities are scaled by $m^{0.29}$, assuring that the galaxies follow a Tully-Fisher relation $V \propto L^{0.29}$ \citep{CouDutvdB07}. To maintain virial equilibrium ($R \propto M/\sigma^{2}$, or, $\mathrm{log}(R) \propto \mathrm{log}(M) - \mathrm{2log}(\sigma)$), particle distances from the center of the galaxy are scaled by a factor of $R \propto M/\sigma^{2} \propto m^{1 - 2 \times 0.29}$, or, $R \propto m^{0.42}$. As a result, surface brightness scales weakly with mass - $L/R^{2} \propto m^{0.16}$ - consistent with the Tully-Fisher relation's assumption of nearly constant effective and/or central surface brightnesses. We do not incorporate scatter into the input galaxy scaling relations, so that scatter in the merger remnant scaling relations is both a lower limit and dependent on the formation process (merging) and bulge profile, rather than an additional input parameter like the Tully-Fisher relation's scatter. Similarly, we use the same bulge fraction for all galaxies. We deliberately avoid using bulgeless disks, as existing literature (e.g., \citet{Her93}) shows that bulgeless disk mergers do not produce sufficiently high central densities. We will further discuss the implications of these choices in \secref{discussion}.

The lowest resolution model has 5,000 bulge, 10,000 disk and 40,000 halo particles, for a 1:2:8 bulge:disk:halo ratio, and 15,000 stellar particles. More massive galaxies have larger particle counts by factors of two, up to a maximum of 480,000 disk particles. Most groups have at least three galaxies with 60,000 stellar particles and only a few tens of galaxies have fewer than 30,000 stellar particles. By scaling resolution this way, stellar particles all have the same mass within a factor of three, while dark particles are not more than 10 times more massive than star particles, limiting spurious numerical artifacts. \appref{numerics} discusses the effects of numerical resolution in greater detail; in summary, this resolution is more than sufficient for the more massive galaxies and adequate for the least massive satellites.

\subsection{Simulation Code and Parameters}
\label{subsec:simcode}

\begin{figure}
\includegraphics[width=0.225\textwidth]{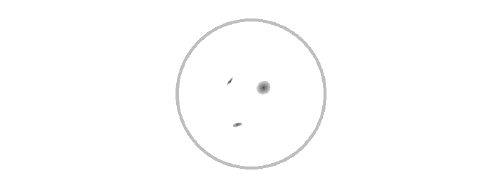}
\includegraphics[width=0.225\textwidth]{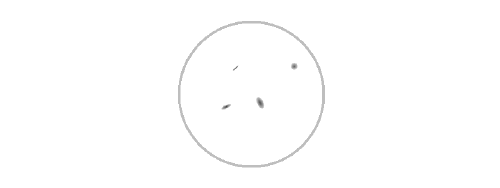}
\includegraphics[width=0.225\textwidth]{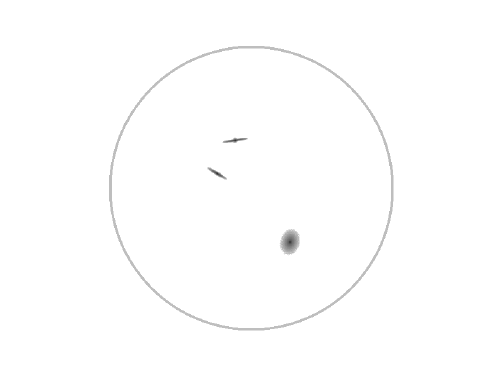}
\includegraphics[width=0.225\textwidth]{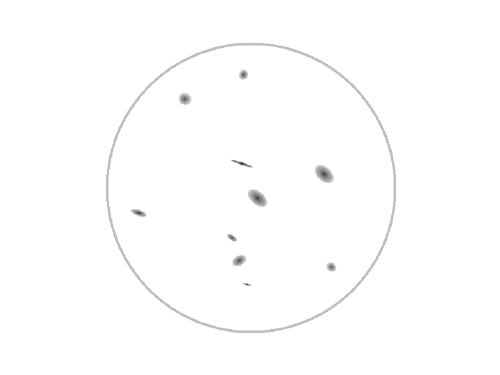}
\includegraphics[width=0.225\textwidth]{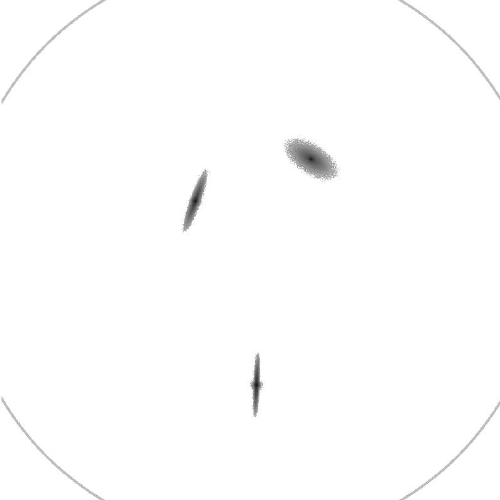}
\includegraphics[width=0.225\textwidth]{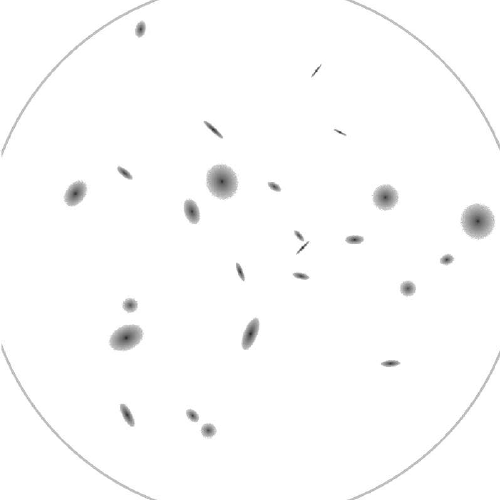}
\caption{Logarithmic surface density maps of the initial conditions for four of the simulated groups. The groups have a nominal total luminosity of 0.125, 1, and 8 L* in each row respectively. The number of galaxies in each group varies from three (leftmost column) to a mass-dependent maximum $N_{max}$ (rightmost column). The maximum radius within which galaxies are spawned (equivalent to 2$R_{200,z=2}$) is shown in gray. Images are 1 Mpc across.
\label{fig:initcondimages}}
\end{figure}

\begin{figure}
\includegraphics[width=0.225\textwidth]{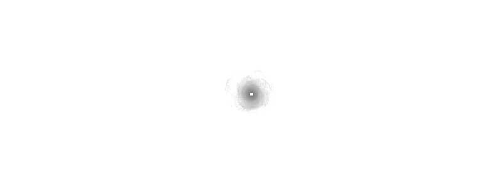}
\includegraphics[width=0.225\textwidth]{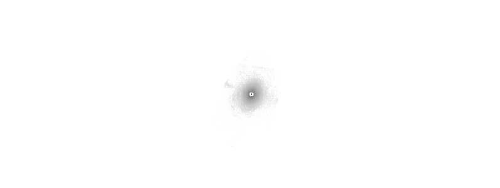}
\includegraphics[width=0.225\textwidth]{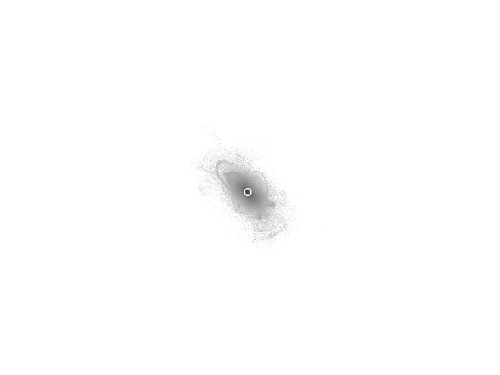}
\includegraphics[width=0.225\textwidth]{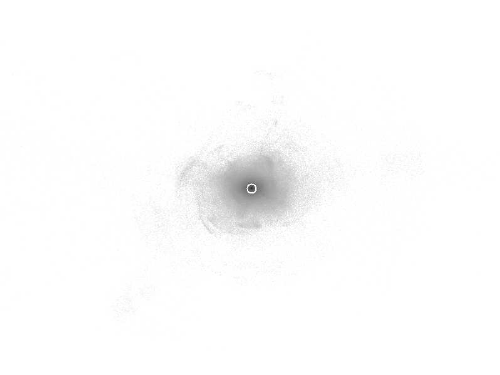}
\includegraphics[width=0.225\textwidth]{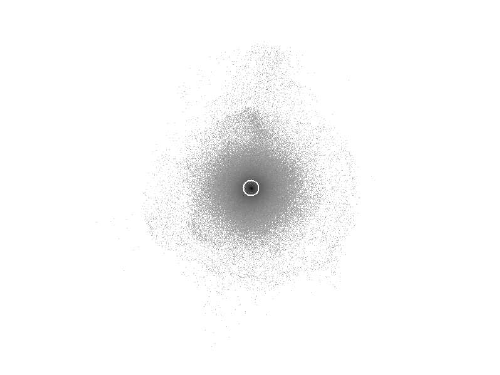}
\includegraphics[width=0.225\textwidth]{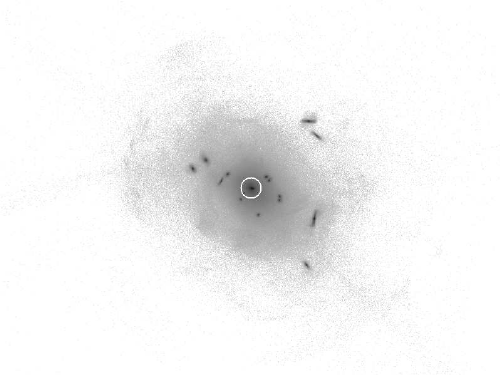}
\caption{Logarithmic surface density maps of nine of the simulated groups after the full simulation time of 52,000 steps (about 10 Gyr). The groups shown are the same as in \figref{initcondimages}. The effective radius of the central galaxy in each group is shown in white. Images are 1 Mpc across, as in \figref{initcondimages}.
\label{fig:finalimages}}
\end{figure}

Each group is simulated for 10 Gyr with the parallel N-body tree code PARTREE \citep{Dub96}. Figures \ref{fig:initcondimages} and \ref{fig:finalimages} show a typical evolution for one such group. The simulations use 52,000 fixed timesteps of 0.02 units - about 195,000 years - and a softening length (spatial resolution) of 100 pc. We use an opening angle of $\theta=0.9$ with forces computed to quadrupole order. While this opening angle is somewhat larger than typical values of 0.7 to 0.8, PARTREE calculates forces between nearby particles in different trees directly, eliminating the source of the largest force errors. For $\theta=1.0$, PARTREE has been shown to produce median force errors under $0.2\%$, with $90\%$ of force errors under $0.5\%$ \citep{Dub96}; force errors with $\theta=0.9$ are considerably smaller.

\section{Analysis}
\label{sec:analysis}
The simulations are analyzed at three different epochs after 5.0, 7.7 and 10.3 Gyr, which correspond to formation redshifts of about 0.5, 1, and 2, respectively, if one assumes that the group formed at t=0 Gyr. Since the only redshift-dependent parameter in the initial conditions is the maximum radius of the group, analysis of the same group at different epochs is equivalent to assuming a different age for the group. Also, since galaxies are given an initial separation sufficient to prevent their halos from overlapping significantly, it typically takes 1-2 Gyr for the first mergers to occur. Groups with fewer galaxies complete the merger process after another 2-3 Gyr and so are not sensitive to the choice of formation time, while groups with more galaxies continue slowly accreting lower-mass satellites and growing even after 10 Gyr. Although we do not introduce additional galaxies into the simulation to mimic cosmological accretion, we note that the long merging time for less massive galaxies still allows for late-time mergers in richer groups.

\subsection{Analysis Pipeline}
\label{subsec:analysis_methods}
Once the simulations are complete, we create mock r-band photometry and kinematics of each group at the three different epochs, placing the group at a mock redshift of 0.025 (about 100 Mpc away). In brief, we create SDSS-like photometry of the central galaxy out to 8 effective radii, including a sky background and appropriate signal-to-noise ratio. We use GALFIT ~\citep{PenHoImp02,PenHoImp10} to fit a single Sersic profile to each galaxy. We also use GALMORPH \citep{HydBer09a} to fit a de Vaucouleurs profile to the sky- and satellite-subtracted image, both for comparison to general Sersic fits and to the de Vaucouleurs fits of \citet{HydBer09b}. Finally, we create spatially resolved kinematics at the same scale, and use these maps to measure kinematical quantities within the central region and the effective radius of the central galaxy. Although our simulations do not resolve faint satellites particular well, our pipeline is able to recover the properties of the central ellipticals with precision comparable to SDSS observations.

Simulations are processed with our own imaging pipeline, which is intended to create images of the central galaxy in each group equivalent to those produced by the SDSS. We convert mass to luminosity to create nominal r-band images, using fixed stellar mass-to-light ratios for the bulge and disk components. We then extract a one-dimensional profile of the central galaxy in circular bins, masking out the central regions of satellite galaxies. A single Sersic profile is then fit to produce a rough estimate of the effective radius of the central galaxy ($R_{eff,est}$). 

\begin{figure}
\includegraphics[width=0.48\textwidth]{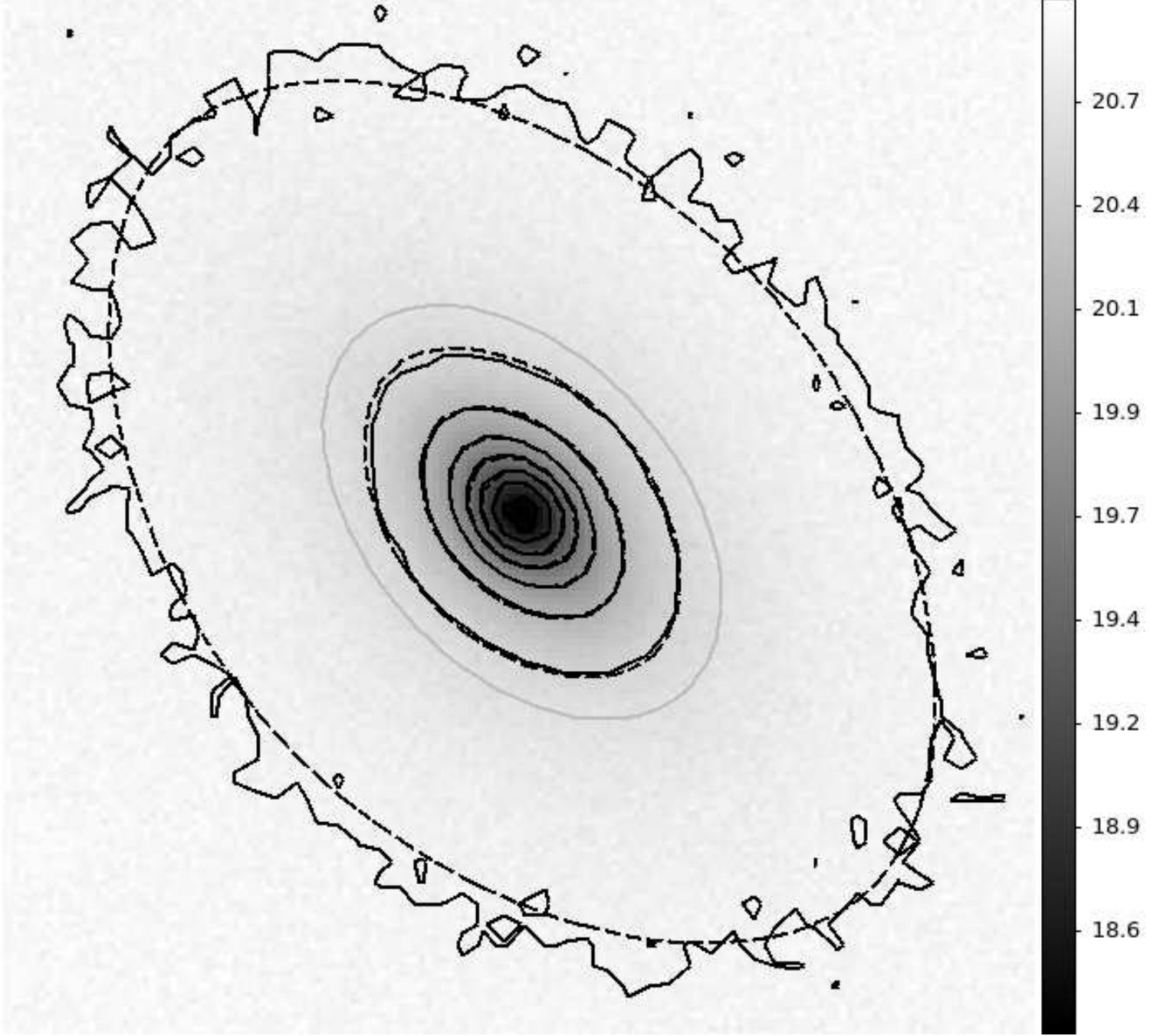}
\caption{Example mock image of the major axis projection of the central galaxy from a typical L* group after 10 Gyr of simulation. The image  shows SDSS-equivalent r-band photometry down to the mean sky level, overlaying contours from the image itself (dashed, black) and the best fit GALFIT Sersic model (solid, black). The gray ellipse shows the effective radius but with no boxiness parameter altering its shape. The image is 29 kpc or 150 SDSS pixels across.
\label{fig:pipeline}
}
\end{figure}

Next, we create a FITS image out to $8R_{eff,est}$ around the central galaxy. The image is smoothed by a point spread function (PSF) with a full-width at half-maximum (FWHM) of 1.43 arcseconds, typical for SDSS r-band observations \citep{SDSSEDR,SDSSDR7}. Galaxies are imaged at a mock redshift of $z_{obs}=0.025$, typical for the SDSS spectroscopic sample used in \citep[e.g.][]{HydBer09a,NaiAbr10}. \figref{pipeline} shows an example image of a typical galaxy. The pixel scale is identical to that used by SDSS, 0.396 arcsec/pixel. Most importantly, we add a sky background with both a mean surface brightness and variations comparable to SDSS observations. In the r band, the mean sky value is 20.86 and variations are Gaussian distributed with a standard deviation of $2.65\%$, equivalent to the SDSS asinh zero-flux magnitude of 24.80 (which itself was chosen to be approximately 1-sigma of typical sky noise). We also create maps of the projected dark matter distribution using the same pixel scale (but no PSF).

\begin{figure*}
\includegraphics[width=0.24\textwidth]{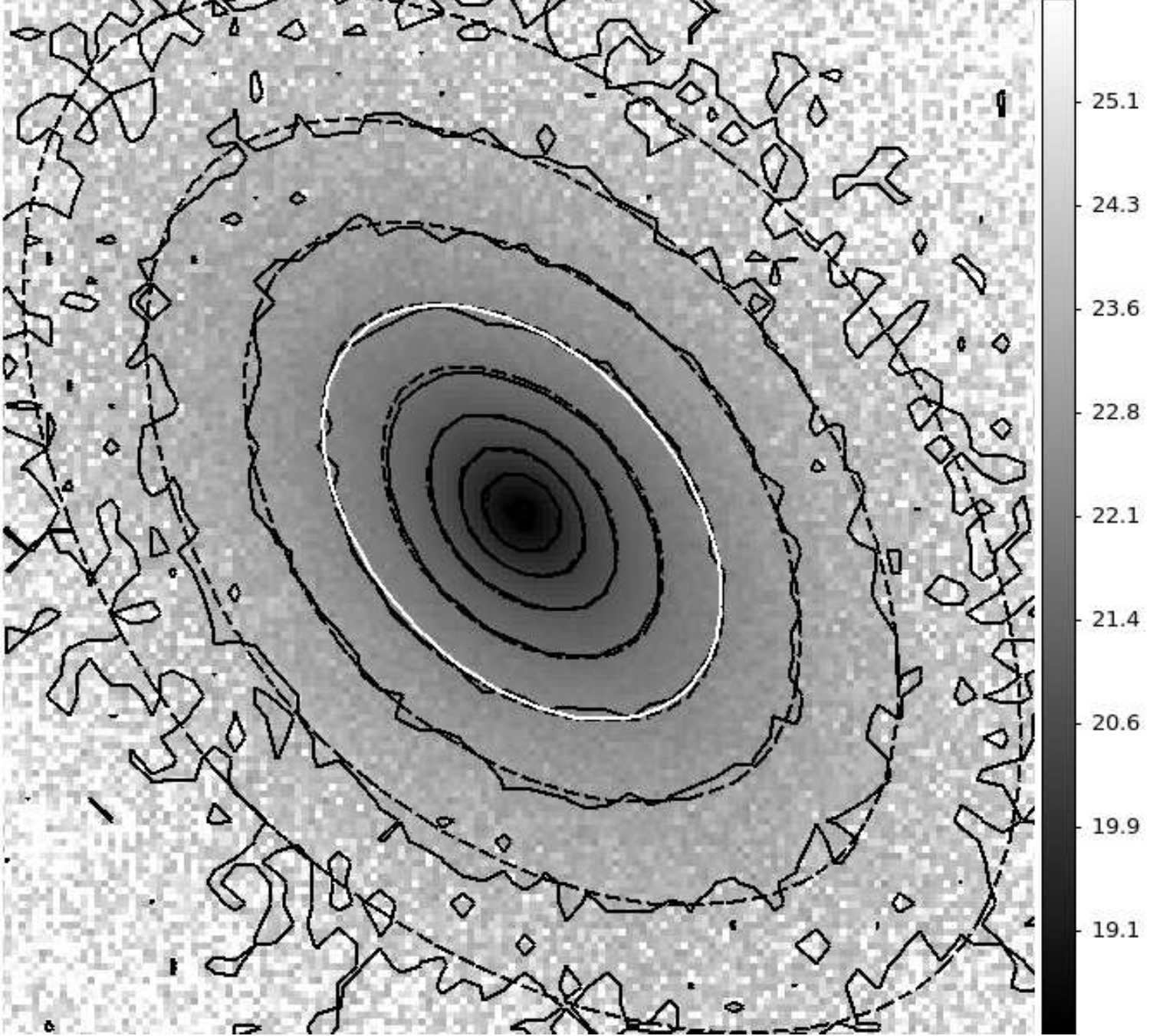}
\includegraphics[width=0.24\textwidth]{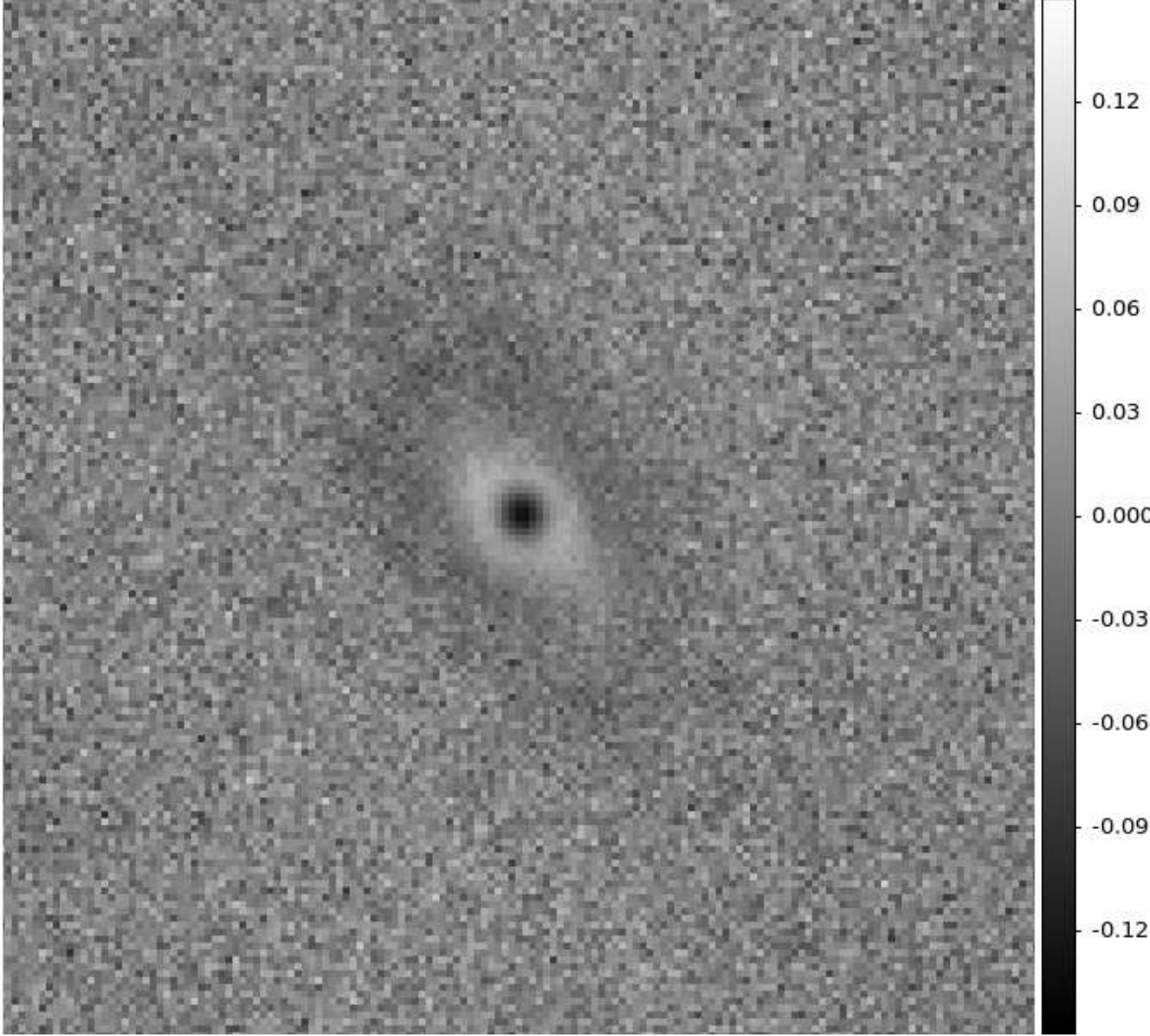}
\includegraphics[width=0.24\textwidth]{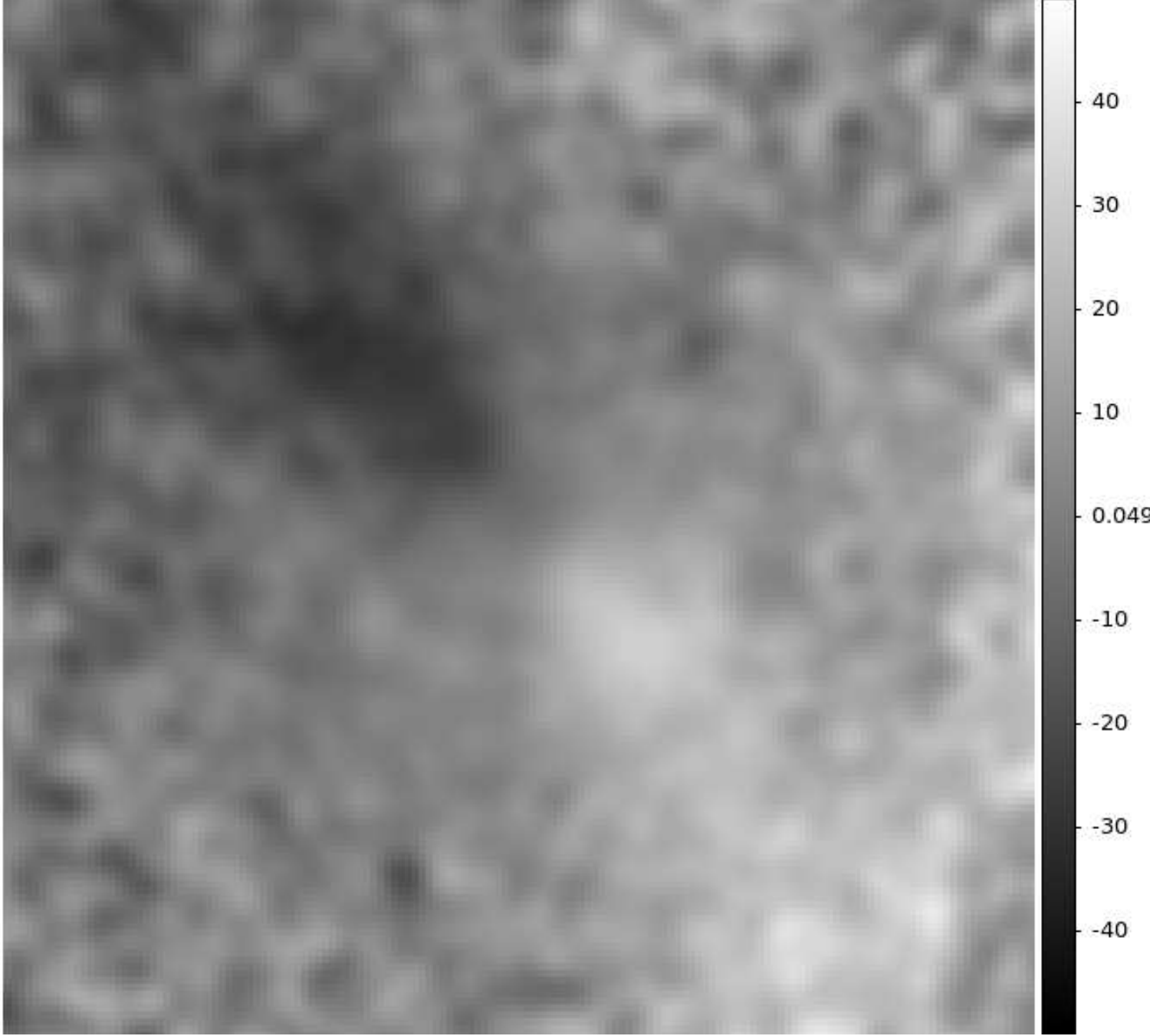}
\includegraphics[width=0.24\textwidth]{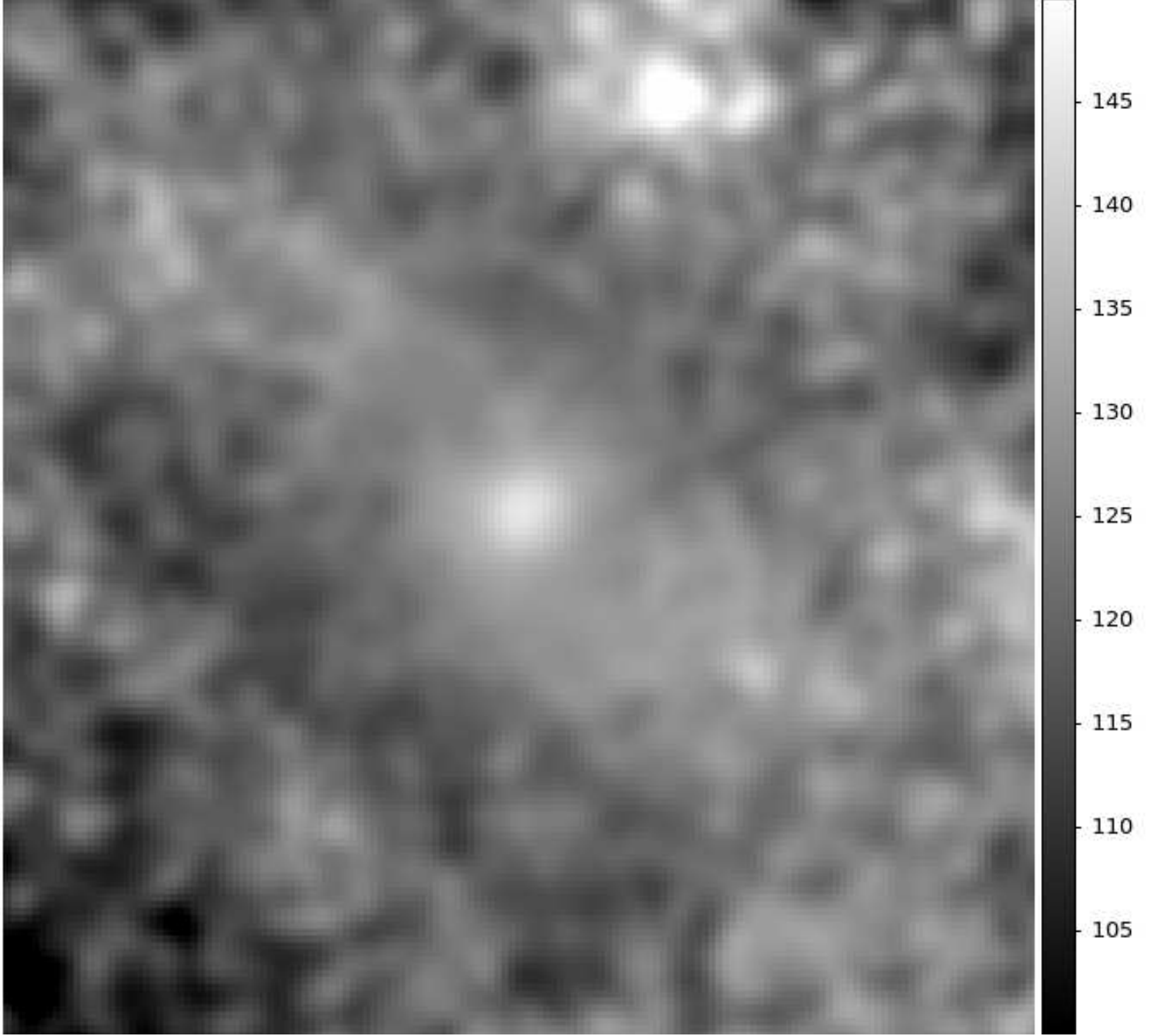}
\caption{Example images of a typical galaxy. The leftmost image is as \figref{pipeline} but now sky subtracted. The remaining panels show, from left to right, the GALFIT Sersic model residuals ((image - model)/image), and velocity and velocity dispersion maps in units of \kms. All images are 29 kpc or 150 SDSS pixels across, as in \figref{pipeline}.
\label{fig:pipeline2}
}
\end{figure*}

In addition to photometry, we create kinematical maps of the first four moments of the luminosity-weighted velocities of particles in each pixel (velocity, r.m.s. velocity dispersion $\sigma$, and v3 and v4). Although we do smooth these maps by the same PSF and use the same pixel scale as the photometry, we do not add a sky background or any instrument-specific noise. We do not perform any fitting to the kinematic quantities, choosing r.m.s. velocity dispersions rather than fitting any profiles, and so the kinematical maps remain largely instrument-agnostic beyond the choice of pixel scale and PSF. The maps can then be used both to measure central velocity dispersions and spatially resolved kinematics, comparing to SDSS and Atlas3D respectively.

Finally, we create an error map for the photometry, which will be used to perform $\chi^{2}$ minimization in fitting profiles to the galaxies. The error is the square root of the luminosity in each pixel multiplied by some constant factor, which scales the signal-to-noise ratio across the image. The constant itself is simply related to the image exposure time, given a certain zero-point equivalent to 1 count per second (for SDSS r-band, this is about 26.7) and mean sky variation. This scheme contrasts with, e.g., \citet{FelCarMay11}, and other simulations which use the square root of the number of particles in each pixel as the error. The per-pixel errors do not scale directly with the resolution of the simulation but should instead converge with increasing resolution. Similarly, setting a non-zero floor to the error map ensures that pixels with no signal are not ignored in the fit, which is necessary since the absence of a signal is meaningful.

For each galaxy, we create images in 10 randomly oriented but evenly spaced projections. These are the ten projections passing through opposite faces of a regular icosahedron, but arbitrarily rotated with respect to the central galaxy. We also use the three projections corresponding to estimates of the principal axes of the central galaxy. We fit every galaxy in the image with a single Sersic profile using GALFIT. Sufficiently large galaxies (including the central galaxy) fit a boxiness parameter (C0) as well, which allows for elliptical isophotes to vary from diamond-shaped (C0 $<$ 0) to rectangular- or box-shaped (C0 $>$ 0). For highly inclined disks with a bulge, this can also provide a better fit than an unmodified ellipse. The GALFIT fits also include a fixed sky background equal to the mean sky brightness. We do not allow for the sky value to vary, as doing so would result in over-fitting the sky, a common problem in observations. Since different surveys and even different data releases of the SDSS have employed various methods for fitting sky backgrounds, we opt to avoid the difficulty of reproducing each methodology and simply fit with the known mean sky value. This does not, however, remove the pixel-to-pixel variation in sky brightness, which sets the effective limiting surface brightness in the image.

We use the GALFIT fits to create a sky- and satellite-subtracted image of the central galaxy in each frame. This image is used to measure various quantities, including alternative non-parametric half-light radii. We also use GALMORPH to fit a single de Vaucouleurs profile to this sky-substracted image. This provides a direct comparison to the methodology used in \citet{HydBer09b}, with the caveat that our use of GALFIT to fit the profiles of satellite galaxies may not match the exact methodology employed in masking nearby sources in SDSS or other surveys. 

\subsection{Photometric and Kinematic Measures}

Sizes and luminosities of the central remnants are measured several different ways. The preferred luminosity measure is the total luminosity within the deconvolved model image of the central galaxy, roughly equivalent to model magnitudes in SDSS and other surveys. For comparison, we also measure several other sizes and luminosities, including non-parametric Petrosian radii \citep{Pet76} (see \citet{SDSSDR2} for the SDSS implementation and \citet{GraDriPet05} for analysis thereof). A thorough analysis of the suitability of these measures is presented in \appref{analysis_testing}.

Kinematical maps are used to measure the velocity distributions - mean velocities (for rotation measures), dispersions, and higher order moments. Generally, we use central dispersions within 1/8 $R_{eff}$ and rotation measures within $R_{eff}$. Velocity dispersions in the central remnants do vary radially, generally dropping from peak central values. Integral field surveys such as Atlas3D can measure dispersions out to 0.5 to 1 $R_{eff}$, whereas fiber dispersions from SDSS are measured within fixed angular diameters, and hence variable fractions of $R_{eff}$. Aperture corrections are often applied to fiber dispersions to convert them to a fixed fraction of $R_{eff}$, with 1/8 $R_{eff}$ a typical choice for SDSS observations \citep{HydBer09a}. However, we find that central dispersions are nearly identical to effective dispersions (within 1 $R_{eff}$), with most galaxies lying on a linear relation and only a handful of outliers, so aperture corrections are not necessary for the simulations.

The central velocity dispersions in simulations can be artificially depressed by softening of the gravitational potential. We mask out the central 300 pc to compensate, and measure central dispersions within 1/8 $R_{eff}$ where possible. For the few galaxies where 1/8 $R_{eff}$ is smaller than 300 pc, we enlarge the aperture by factors of 1/8 $R_{eff}$ until a reliable estimate is obtained.

We have also considered the kinetic energy measure $S=\sqrt{\sigma^2+v^2}$, or equivalently $S=\sigma \times \sqrt{1+(v/\sigma)^2}$. This is a more accurate measure of the stellar kinetic energy for galaxies with significant rotation. However, most simulated galaxies do not have sufficient rotational support for this correction to be significant, and there are not yet any large samples of galaxies with published dispersions and $v/\sigma$ to compare to. 

\section{Results}
\label{sec:results}

The main results presented in this paper are the morphologies and kinematics of central group galaxies. Although we do fit satellites as well, this is mainly to exclude their contribution from the profile of the central galaxies. Few satellite galaxies are sufficiently well resolved to recover sizes and Sersic indices accurately, but we only require their total luminosities to be recovered and subtracted from the central galaxy's profile.

Unless otherwise noted, all radii measured with elliptical annuli are $\sqrt{(a \times b)}$, where a and b are the major and minor axis lengths.

\subsection{Observational Comparisons}

Our results are compared to three published data sets for nearby galaxies. The Atlas3D survey \citep{CapEmsKra11}\citetext{hereafter A3D} is a volume-limited integral field unit survey of the kinematics of 260 nearby early-type galaxies. A3D provides kinematical maps \citep{EmsCapKra11} with a pixel size about twice as large as that of SDSS. This is mitigated by the larger aperture and very low redshifts (z$<$0.01) of the sample as compared to our nominal mock sample redshift (0.025). Sersic profile fits are also available from \citet{KraAlaBli12}, with photometry from a variety of sources but typically comparable to or better than SDSS.

\citet{SimMenPat11}\citetext{hereafter S+11} published three different profile fits for over a million SDSS galaxies. We use the single Sersic decompositions for direct comparison and the free Sersic (bulge) plus exponential (disk) decompositions for diagnostic purposes. Although these fits were performed with a different code - GIM2D \citep{SimWilVog02} - the fitting procedure is similar to our GALFIT fits. We select galaxies with spectroscopic redshifts $0.01 < z < 0.3$ to ensure availability of a reliable V$\mathrm{_{max}}$ volume correction term. We use the logarithmic median velocity dispersion between two sources - the SDSS DR7 and Princeton measurements (both included in DR7, \citet{SDSSDR7}). Stellar masses are based on the MPA-JHU DR7 catalog \footnote{Available at http://www.mpa-garching.mpg.de/SDSS/DR7/}, using fits to the multi-band photometry.

Detailed visual classification of of nearly 6,000 early-type galaxies from SDSS with z$<$0.1 is provided by \citet{NaiAbr10}\citetext{hereafter N+10}. Volume corrections are applied with the standard 1/V$\mathrm{_{max}}$ weighting scheme \citep{Sch68}. Profile fits from this catalog include Petrosian sizes from the SDSS pipeline \citep{SDSSEDR} and Sersic fits from S+11. Although the original catalog of N+10 contained over 14,000 galaxies, eliminating bad fits and unmatched/misclassified objects provides just over 11,000 galaxies, of which nearly 5,000 are early-types.

We exclude all SDSS galaxies with extreme velocity dispersions ($\mathrm{\sigma < 20}$ \kms or $\mathrm{\sigma > 400}$ \kms) or effective radii smaller than 0.3 kpc. Where visual classifications are available (A3D, N+10) we select galaxies with Hubble T-types \citep{deV59} less than 0 as early-types. T-types less than -3 are included in the elliptical sample while the remainder are classified as S0s. The majority of the S+11 sample does not have visual classifications, other than the small subset classified by N+10. We adopt a series of empirical cuts similar to those of \citet{DutConvdB11} to identify early-type galaxies, testing these against the N+10 subset. The early-type sample contains galaxies with $\mathrm{n_{s} > 1}$, and -- from the disk plus free $\mathrm{n_{s}}$ fits -- r-band bulge to total luminosity ratio $B/T_{r} > 0.4$, disk inclination less than 63 degrees, and bulge $\mathrm{r_{eff} > 0.5 kpc}$. Early-types must also have a spectroscopic eclass value less than -0.1 (see \citet{YipConSza04}, but note that the sign convention in SDSS is opposite), which selects galaxies with spectra consistent with a passive population. This early-type sample is subdivided into an elliptical subset, which imposes further cuts based on the single Sersic fits: g-band image smoothness S2 $\mathrm{<0.08}$, or g-band image smoothness S2 $\mathrm{<0.12}$ and $\mathrm{B/T_{r} < 0.6}$. These cuts are similar to those suggested by \citet{SimCloDes09} to select early-type galaxies from morphology alone, but also serve to decrease contamination from S0s and early-type spirals in the elliptical sample. All galaxies not classified as early-type but meeting the dispersion and $R_{eff}$ cut are identified as spirals.

\begin{table}
\caption{Morphological types of cuts used for S+11 sample}
\begin{tabular}{cccc}
\hline 
S+11 Subsample & N+10 Es & N+10 S0s & N+10 Spirals \\
\hline
    Es & 1874 & 1095 & 93 \\ 
    S0s & 98 & 350 & 323 \\ 
    Spirals & 193 & 1265 & 5272 \\ 
	Unclassified & 13 & 54 & 604 \\
	Total & 2178 & 2764 & 6292 \\
\hline
\end{tabular}
\tablecomments{Each row lists the breakdown of visual morphological classifications from N+10 of each of the subsamples from S+11, which are based on empirical cuts on various parameters rather than visual classification. Empirical cuts generally produce complete but impure samples of ellipticals and spirals, with substantial contamination by S0s.}
\label{tab:morphcuts}
\end{table}

The samples obtained by applying these cuts to the N+10 catalog are listed in \tabref{morphcuts}. The elliptical sample is 86\% complete. While it is only 61\% pure, the contamination mainly comes from S0s and not spirals. No cuts appear to be able to reliably classify S0s, which contaminate both elliptical and spiral samples. In principle, we could instead use the S+11 cuts on the N+10 sample rather than relying on visual classifications at all; however, visual classifications are repeatable and fairly robust (see \citet{NaiAbr10} for comparisons to previous classifications), and as seen in A3D, there are significant differences in rotational support between the elliptical and S0 population \citep{KraAlaBli12}, even if no automated morphological classification can separate them.

Additional weightings are necessary to compare these catalogs to our own simulations, which probe a range of about 5 in absolute magnitude and have a nearly flat luminosity function. We produce r-band luminosity functions for each sample, then weight by the ratio of the simulated luminosity function to the observed one. Elliptical and S0 subsamples use all simulated galaxies versus E/S0 classifications from observed catalogs - i.e., we do not morphologically classify simulated merger remnants. This weighting procedure turns each observational sample into a catalog with equal numbers of galaxies at each luminosity, directly comparable to our simulations. Although the weightings are not vital for tight scaling relations like the fundamental plane, they are necessary for fair comparisons of weaker correlations and histograms marginalizing over luminosity.

\subsection{Morphology}
\label{subsec:morphology}

As detailed in \subsecref{analysis_methods}, the central galaxies are fit with a single Sersic profile. Each profile has six free parameters (in addition to the two coordinates for the centre of the galaxy): the Sersic index $\mathrm{n_{s}}$, an effective half-light radius $r_{eff}$, a surface brightness at this radius, an ellipticity $\epsilon$, a position angle, and a boxiness parameter C0 modifying the shape of the ellipse from diamond (negative C0) to box-shaped (positive C0). Ellipticals have long been known to be best fit by larger Sersic indices than disks, to have small ellipticities and several correlations between size, luminosity and Sersic index. Any satellite galaxies in the image are also fit with a single Sersic profile.

\subsubsection{Sersic Indices}

\begin{figure*}
\includegraphics[width=0.33\textwidth]{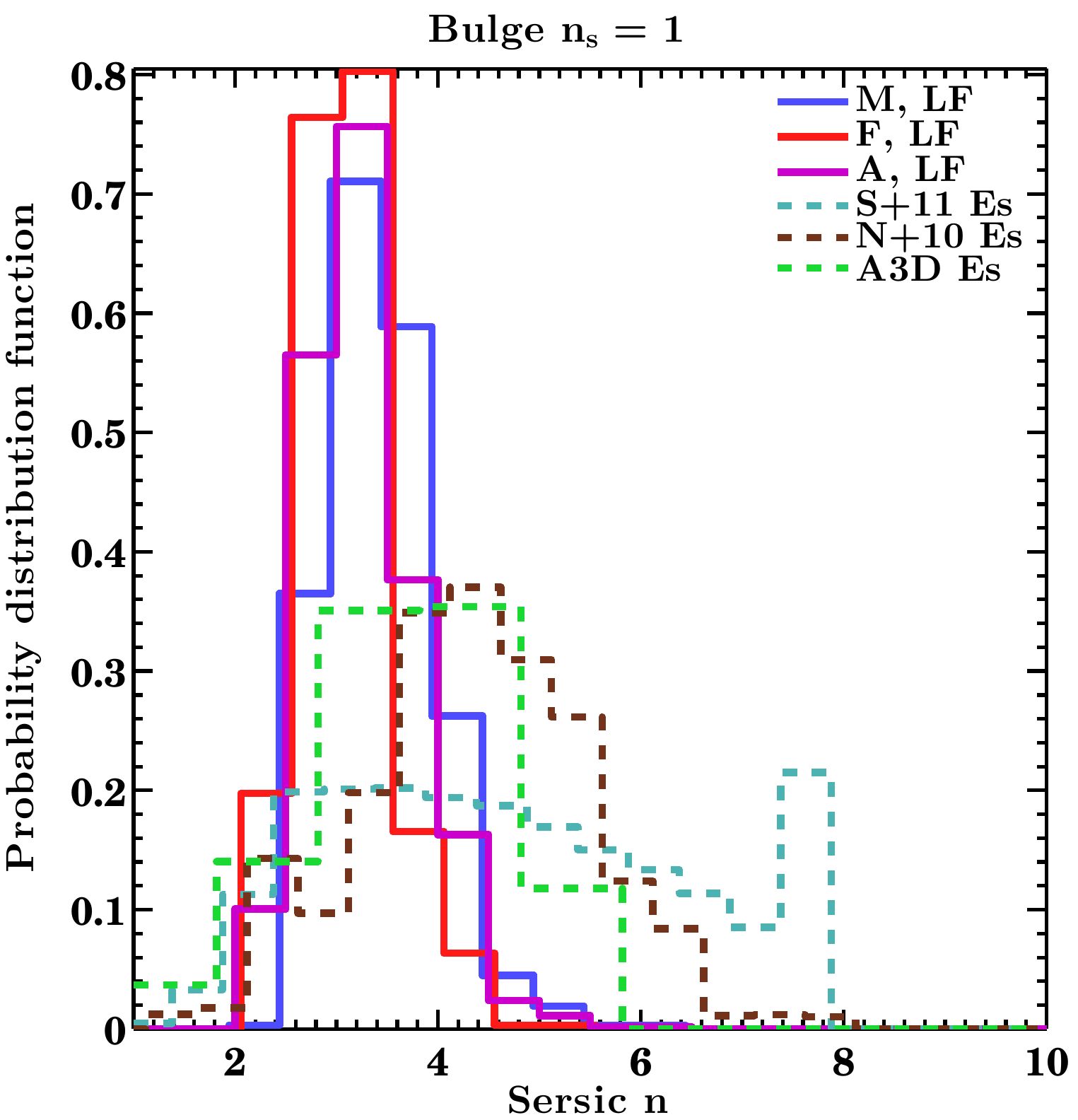}
\includegraphics[width=0.33\textwidth]{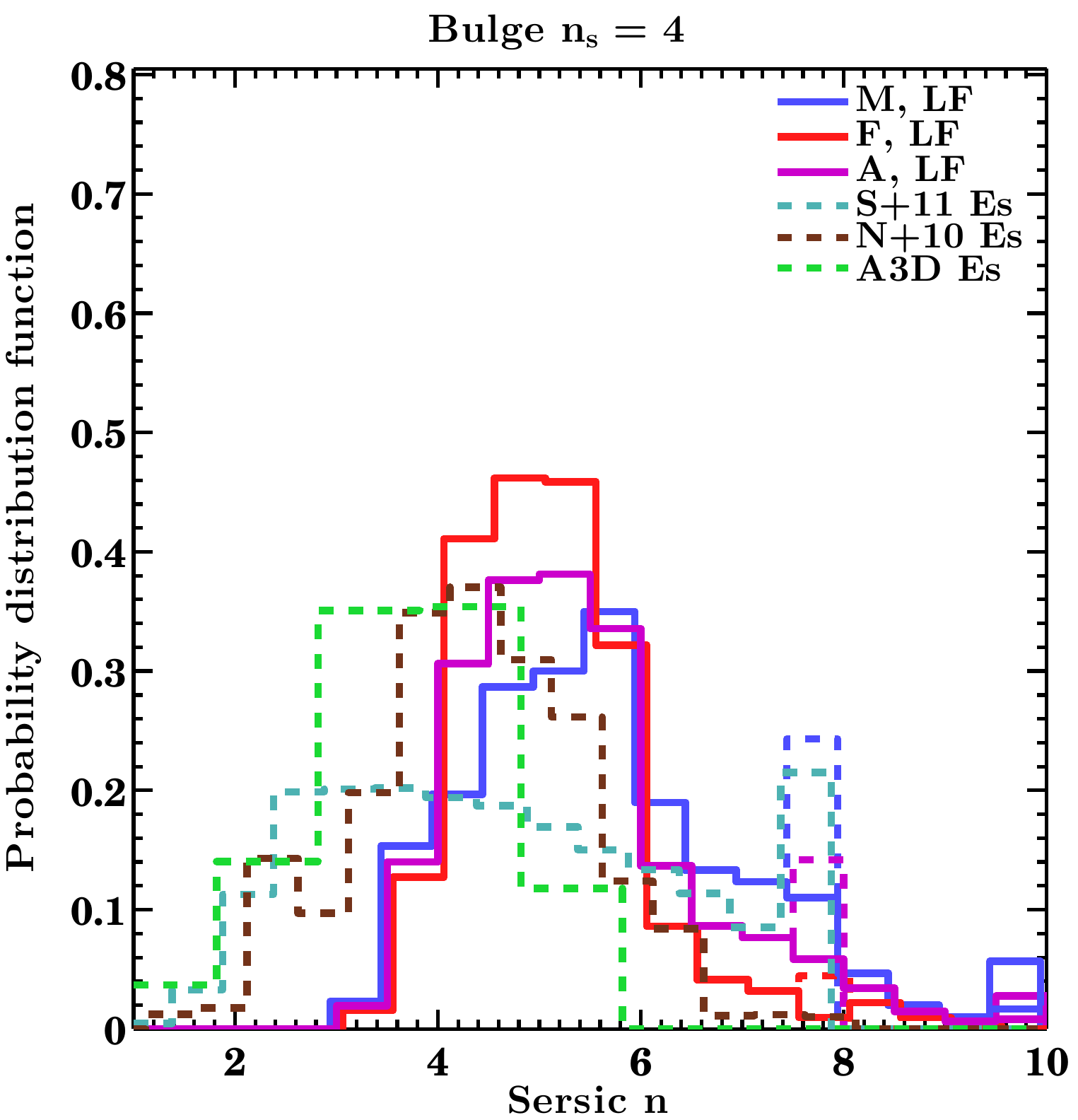}
\includegraphics[width=0.33\textwidth]{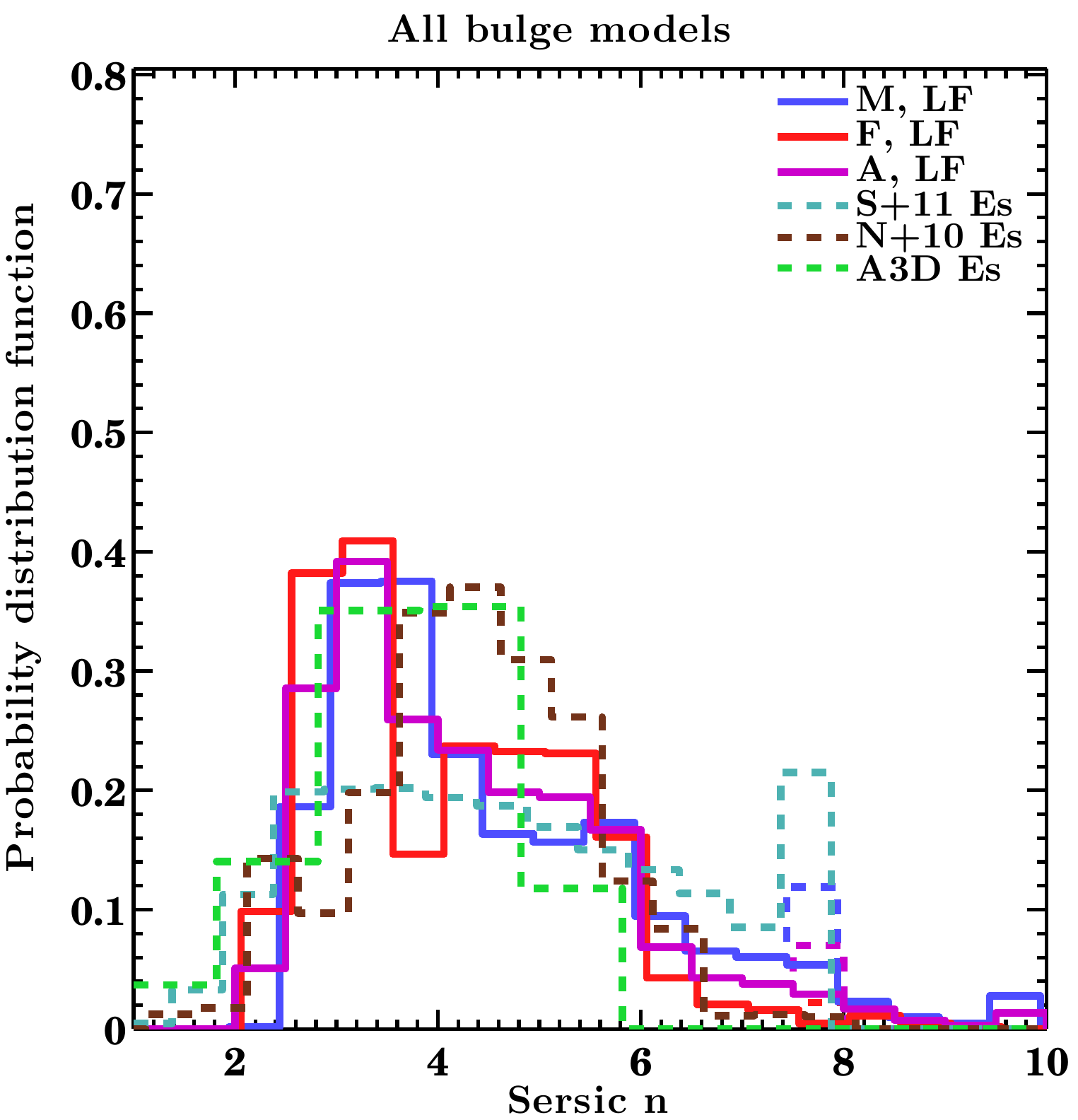}
\includegraphics[width=0.33\textwidth]{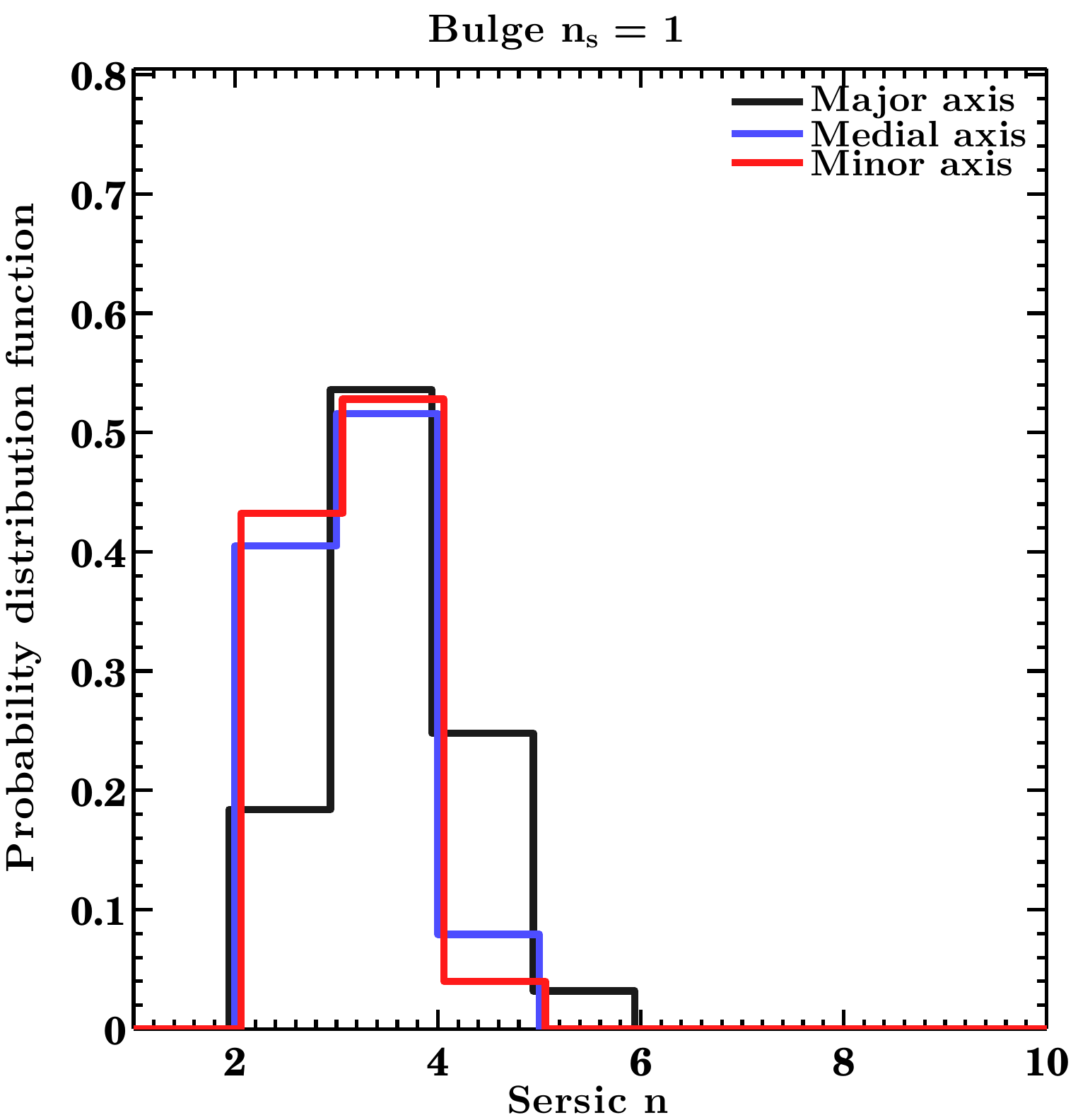}
\includegraphics[width=0.33\textwidth]{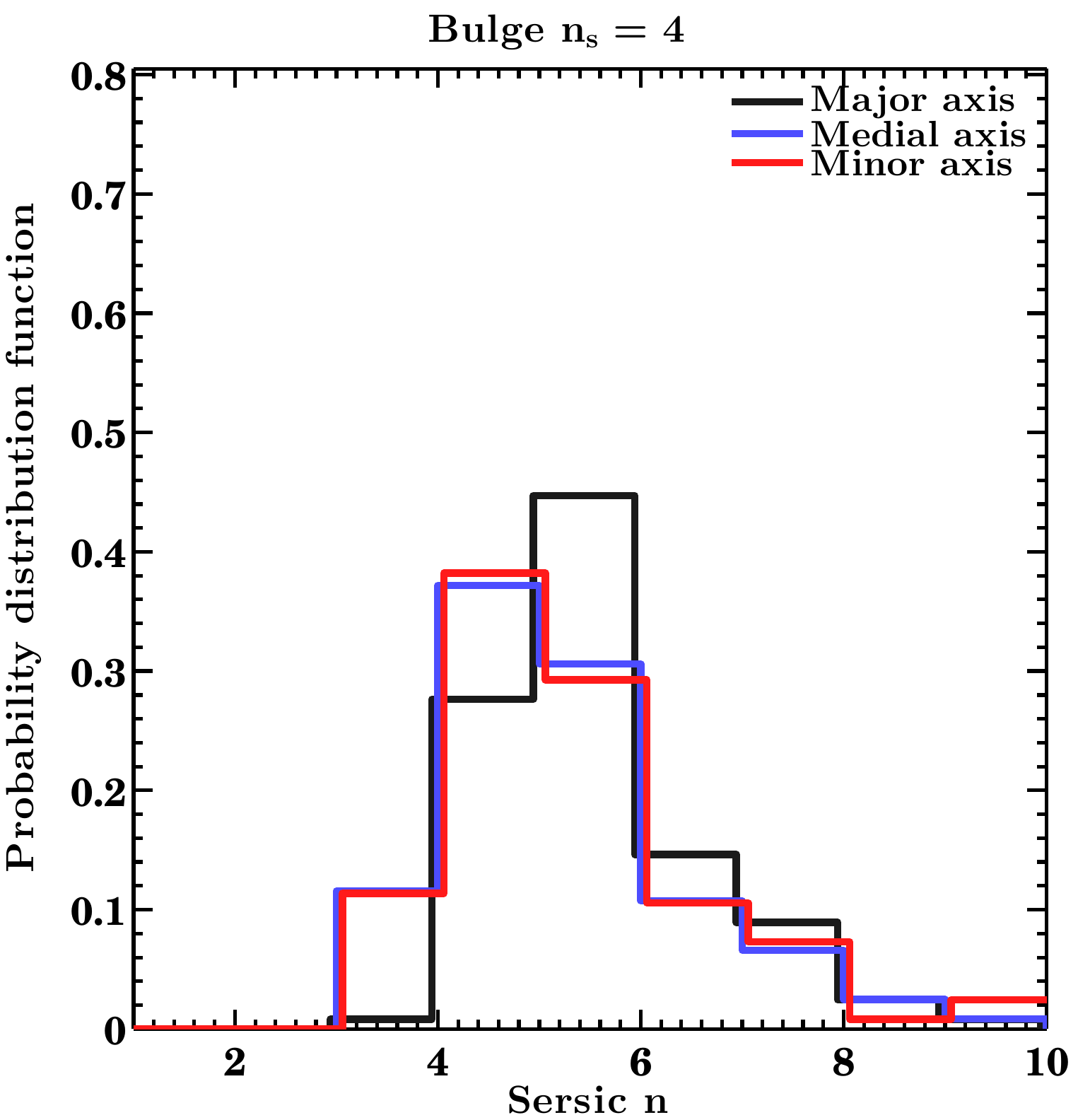}
\includegraphics[width=0.33\textwidth]{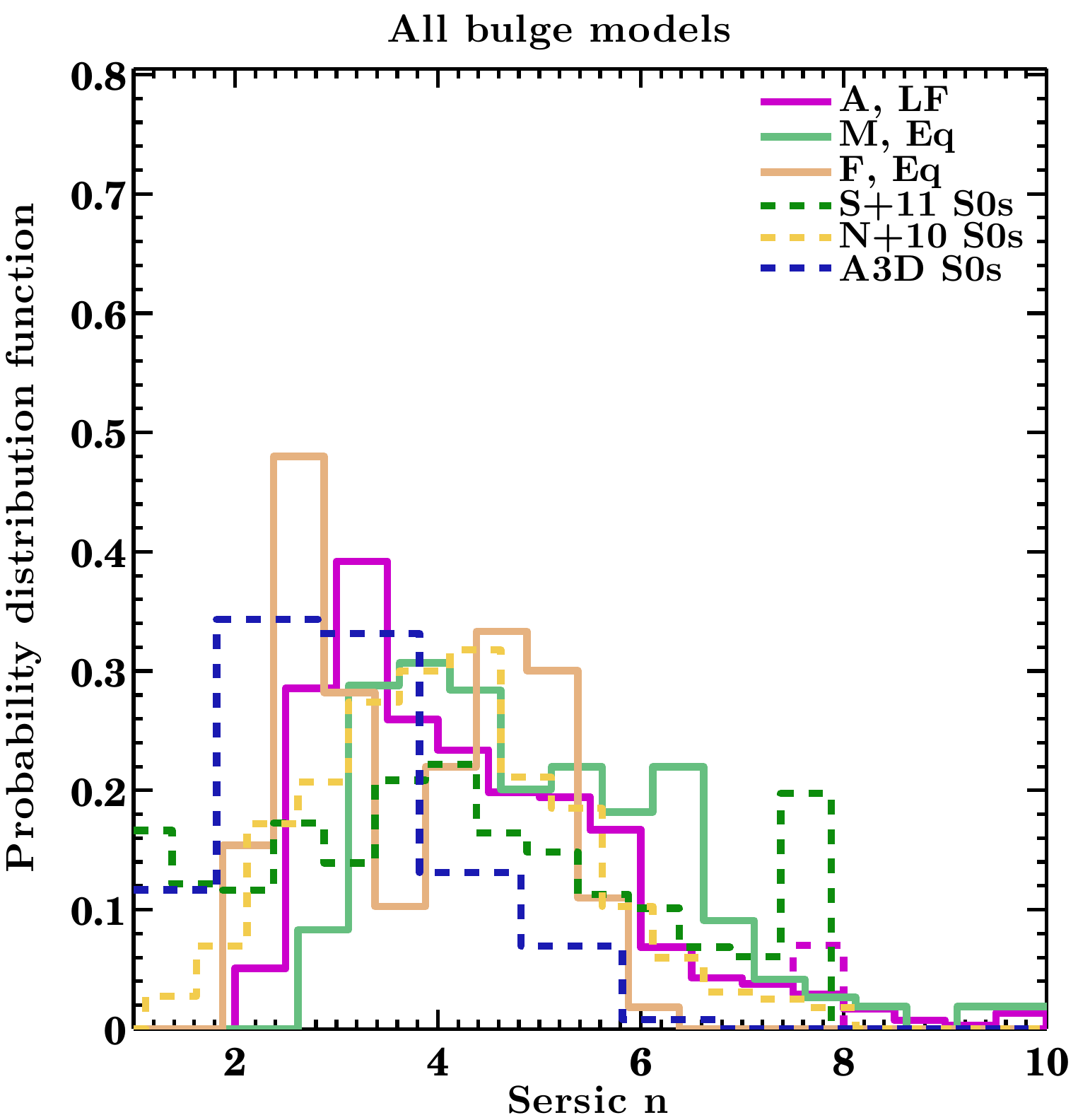}
\caption{Sersic indices of central ellipticals. The top row shows comparisons to observed elliptical galaxies. The bottom right panel shows distributions for equal mass mergers as well, as compared to observed S0 galaxies. The bottom left and middle panels show principal axis projections only. Histograms are offset slightly on the x-axis to prevent overlap. Simulation data are shown with several maximum $\mathrm{n_{s}}$ - 8 (dashed line) for comparison with N+10 and S+11, and no limit (solid line) for the simulations themselves. B.n$_{s}$=4 mergers match observed ellipticals best, but a range of bulge types appears to be required to reproduce the observed distributions.
\label{fig:sersicn_hist}}
\end{figure*}

~\figtextref{sersicn_hist} shows various histograms of the Sersic index distribution for the B.n$_{s}$=1, B.n$_{s}$=4 and B.n$_{s}$=all samples. Each individual bulge type produces a narrow distribution of Sersic indices. The B.n$_{s}$=4 sample's distribution is narrower and peaked at a larger value of $\mathrm{n_{s}}=5$ than the observational distributions. The B.n$_{s}$=1 sample's peak at $\mathrm{n_{s}}=3$ is significantly lower than those of N+10 and A3D ellipticals, and the distribution is narrower still than that of B.n$_{s}$=4. The combined B.n$_{s}$=all sample's $\mathrm{n_{s}}$ distribution is nearly bimodal due to this separation and approximately twice as broad as B.n$_{s}$=1 alone. By contrast, most observed distributions are unimodal, although the S+11 distributions show a larger peak of high $\mathrm{n_{s}}$ galaxies, which is only reproduced in the B.n$_{s}$=4 sample. There is also a hint of bimodality in the S0 distributions, which we have diminished by setting a lower limit of $\mathrm{n_{s}}=1$. The peak of the S0 distribution is best reproduced by the B.n$_{s}$=1, but, as will be demonstrated in \subsubsecref{rotation}, the remnants' rotational support is far lower than that of typical S0s. None of the simulation samples can reproduce the width of the observed S0 distributions.

Although each of the B.n$_{s}$=1 and B.n$_{s}$=4 samples are individually a poor fit to the elliptical data - particularly being too narrow of a distribution - the naive linear combination of the two (B.N$_{s}$=all) provides a better match. The B.n$_{s}$=all sample is also a better match to the elliptical distributions than the S0, the latter of which tend to smaller Sersic indices. While it is not a particularly realistic distribution - assuming that half of the groups in the universe contain galaxies with only exponential bulges while the other half contain de Vaucouleurs bulges - we will elaborate on the implications for more realistic bulge profile distributions in \secref{discussion}.

The difference in Sersic index between the Many- and Few-merger subsamples is small in the L.F.-sampled case but is maximized at about 0.5 for equal-mass mergers. Furthermore, the distributions of the Few-, equal-mass merger remnants in the different bulge samples are sufficiently narrow that the combined B.n$_{s}$=all, Few-merger subsample is distinctly bimodal. Thus, it appears that multiple mergers are sufficient to broaden the distributions of remnant Sersic indices, but sampling progenitors from a realistic luminosity function can accomplish the same purpose, even with relatively few mergers.

\begin{figure*}
\includegraphics[width=0.49\textwidth]{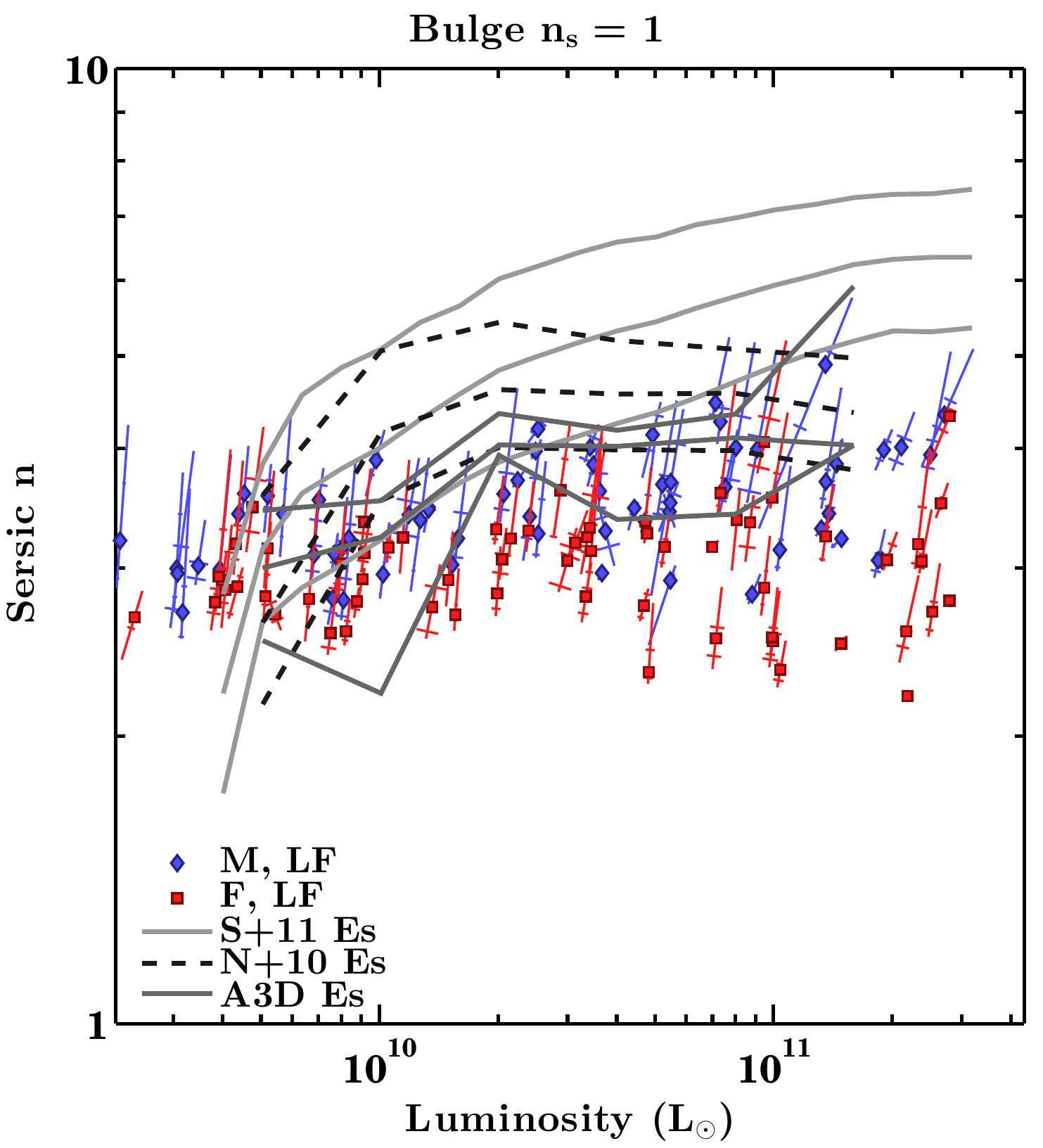}
\includegraphics[width=0.49\textwidth]{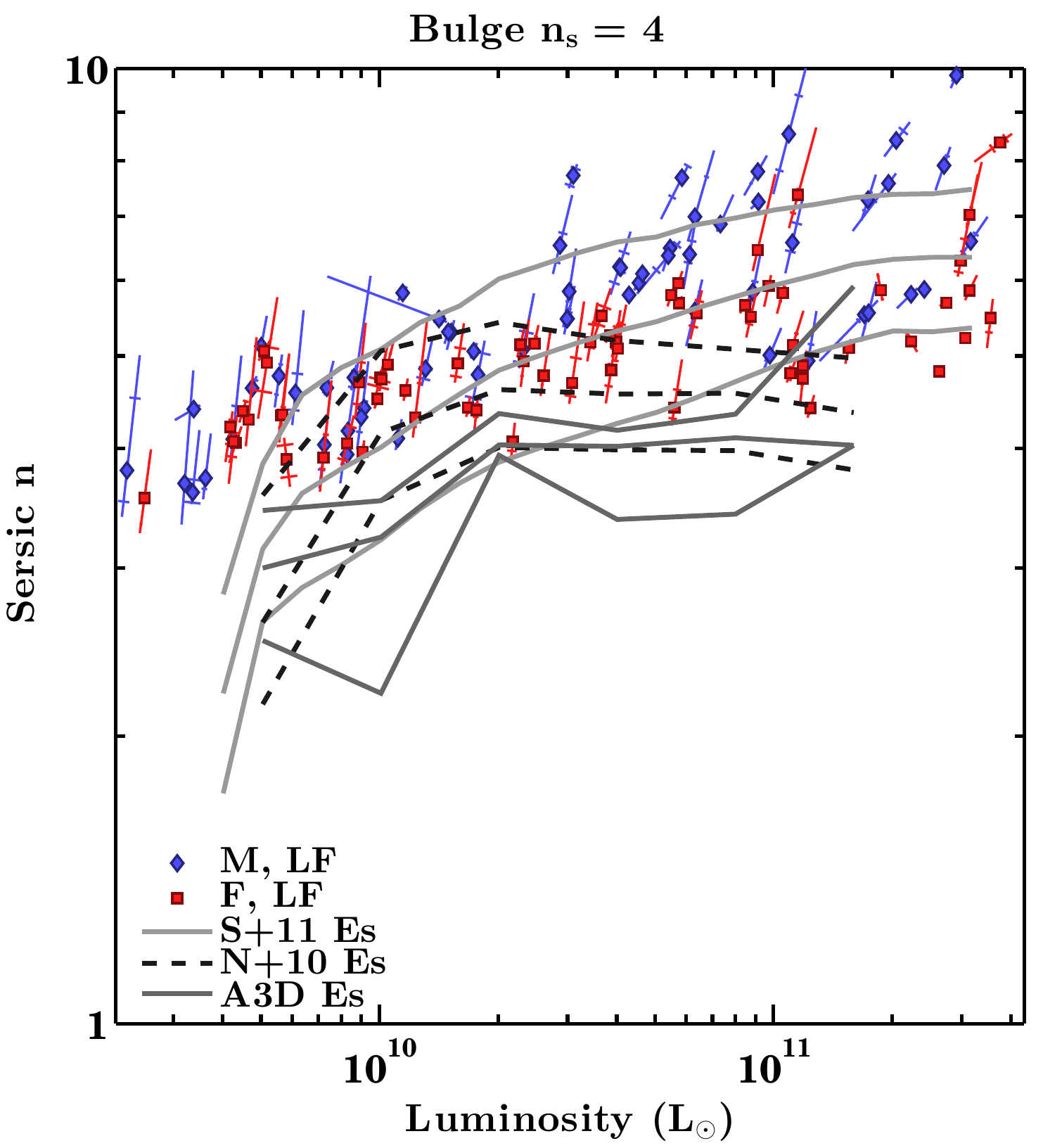}
\caption{Sersic indices of central ellipticals by galaxy luminosity. Classical-bulge mergers (right) have larger $\mathrm{n_{s}}$ for the same initial conditions and show a strong dependence of $\mathrm{n_{s}}$ on galaxy luminosity, as with observed ellipticals from S+11 but unlike exponential bulge mergers (left) and N+10 ellipticals. Different projections of the same group are shown as lines of best fit for clarity, with a single point marking the median projection. The length of the line shows the range of values from 10 random projections. Perpendicular lines cross at the 25th and 75th percentiles, with a length equivalent to the r.m.s. dispersion of points perpendicular to the line of best fit.
\label{fig:l_sersicn}}
\end{figure*}

Major axis projections of central remnants have systematically larger Sersic indices than the medial or minor axis projections (bottom left and middle panels of \figref{sersicn_hist}), with the peak of the distribution shifted by about 1. Medial and minor axes have nearly identical distributions, even though their ellipticities and semi-major axes are not necessarily the same. As \figref{l_sersicn} shows, the variation in Sersic index for a single galaxy over different viewing angles is not usually much larger than one (and often smaller), so projections aligned near the major axis appear to produce the largest $\mathrm{n_{s}}$ profiles.

Only B.n$_{s}$=4 mergers produce a correlation between luminosity and $\mathrm{n_{s}}$, as shown in \figref{l_sersicn}. This is partly a result of more massive ellipticals being produced by more mergers. In both B.n$_{s}$ samples, Many-merger remnants tend to have larger $\mathrm{n_{s}}$ at fixed luminosity. However, in the B.n$_{s}$=4 sample, even the Few-merger subsample shows a small positive slope in Sersic index, whereas the trend is flat or even slightly negative for B.n$_{s}$=1. The overall trend is dependent both on the initial bulge profile and the number of mergers. A positive dependence of merger rate on halo mass is a prediction of $\Lambda$CDM (e.g. \cite{HopCroBun10}). Exponential bulges, however, are simply not concentrated enough to create merger remnants with $\mathrm{n_{s}} > 4$, even with repeated merging. Thus, luminous ellipticals are unlikely to be the product of only exponential bulge mergers.

The degree of agreement between simulations and observations is difficult to judge, since the observational samples do not completely agree. The N+10 $\mathrm{n_{s}}$-L relation appears to flatten above $\mathrm{10^{10}L_{\odot}}$. This could be due to the larger redshift range of the S+11 sample; however, we find that GALFIT-derived Sersic fits to mock images at higher redshift tend to fit lower Sersic indices, so this systematic trend would have to be reversed in observed ellipticals to explain the shift. 

\subsubsection{Ellipticities}

\begin{figure*}
\includegraphics[width=0.33\textwidth]{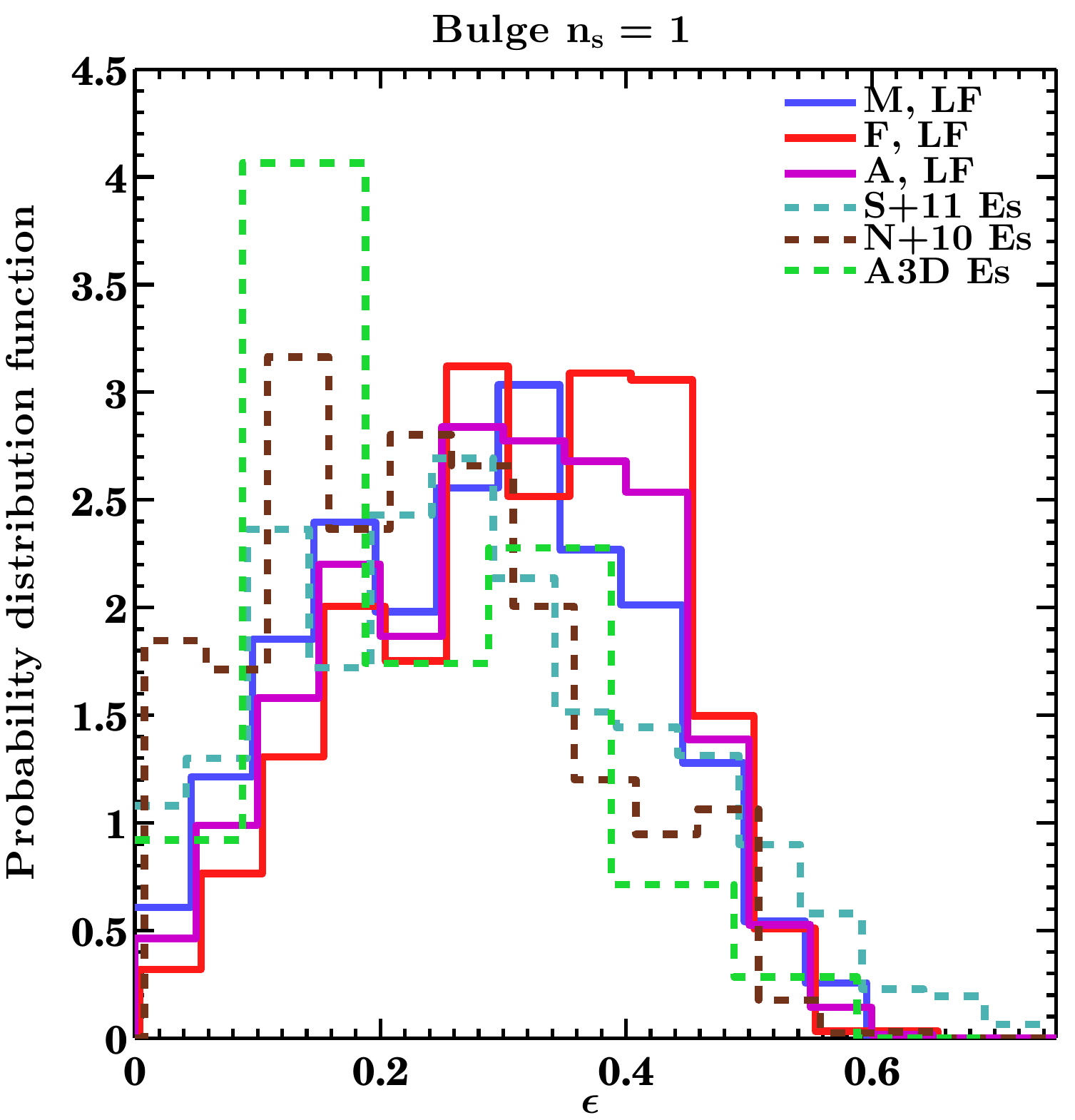}
\includegraphics[width=0.33\textwidth]{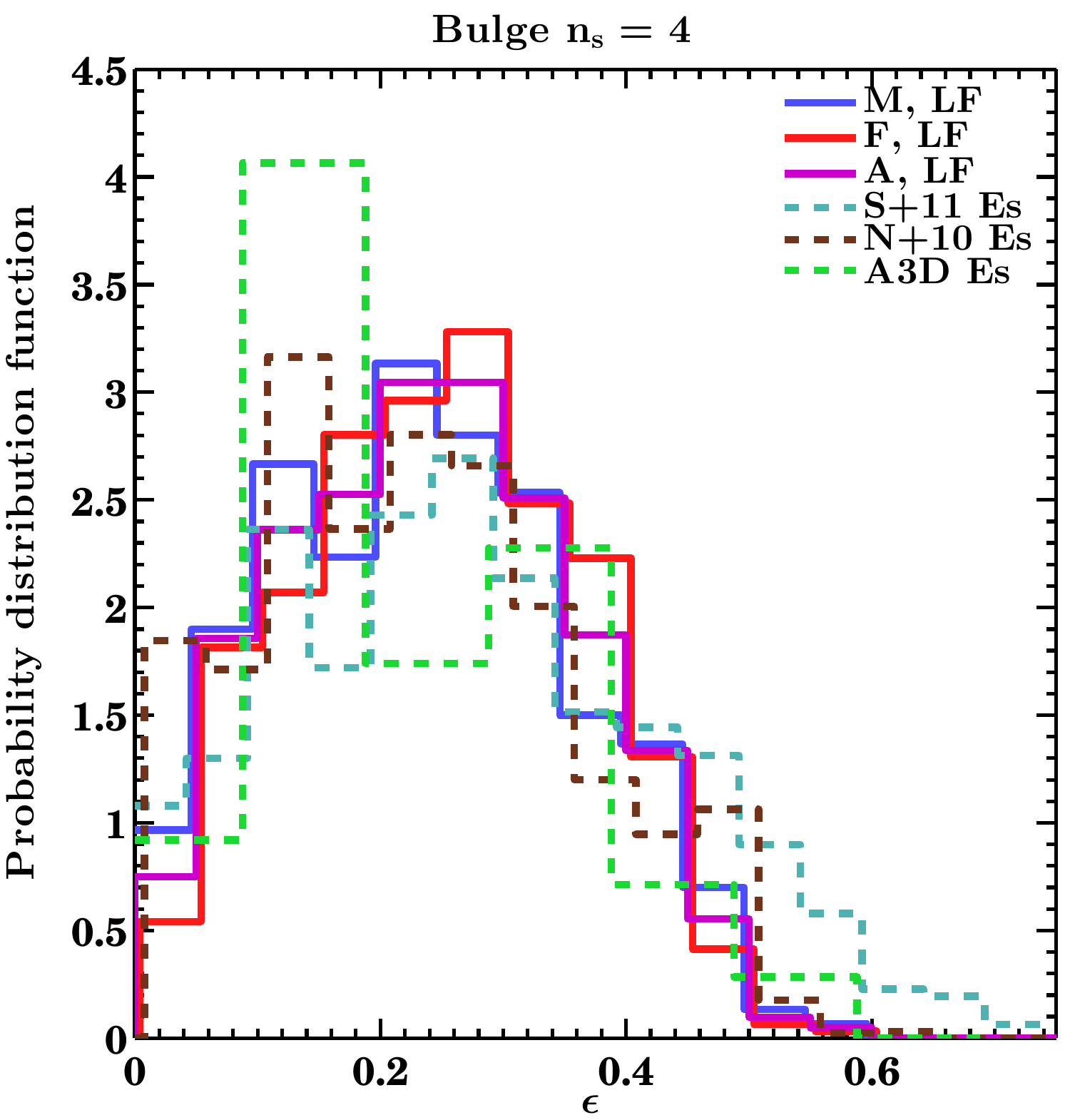}
\includegraphics[width=0.33\textwidth]{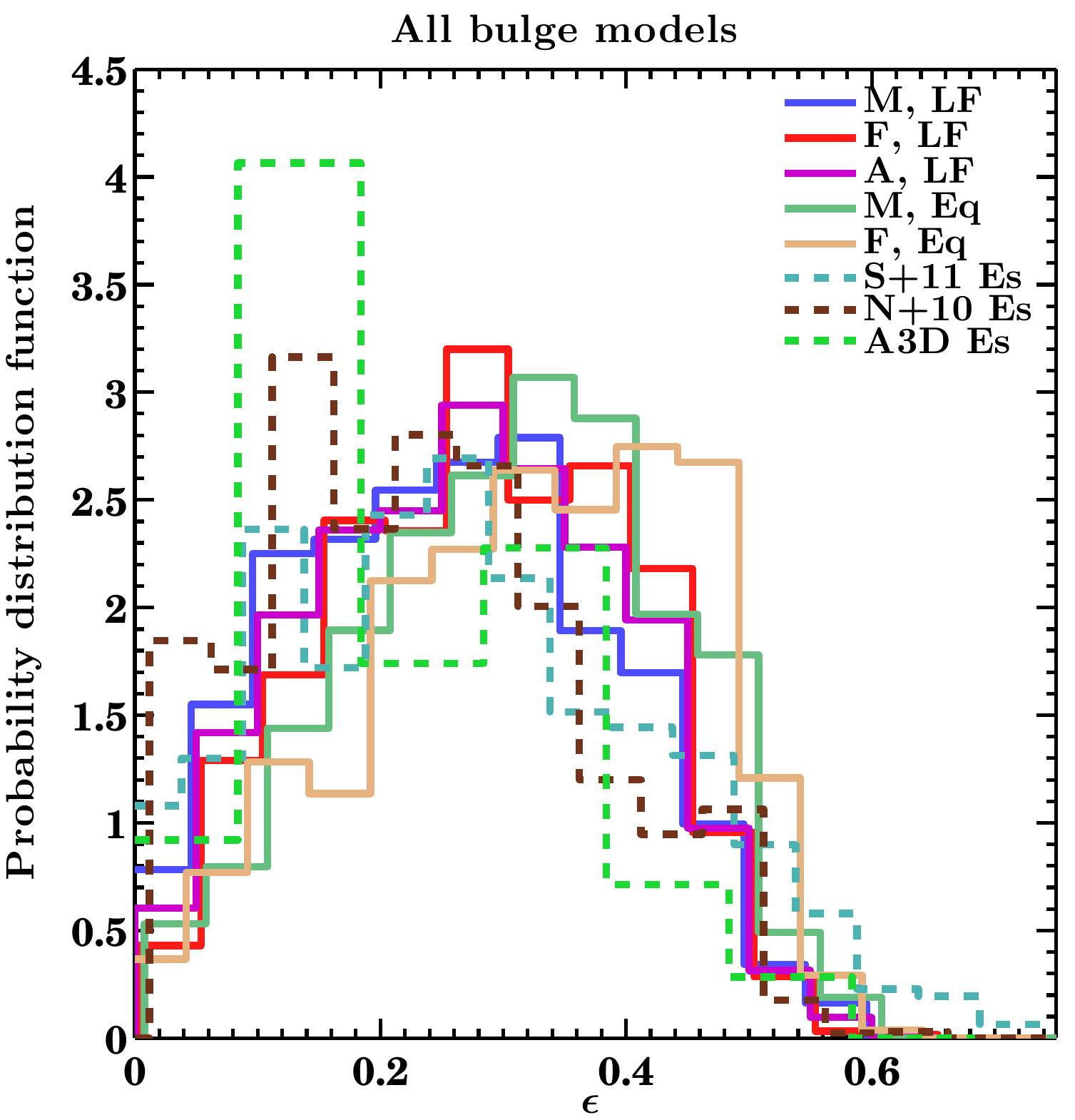}
\includegraphics[width=0.33\textwidth]{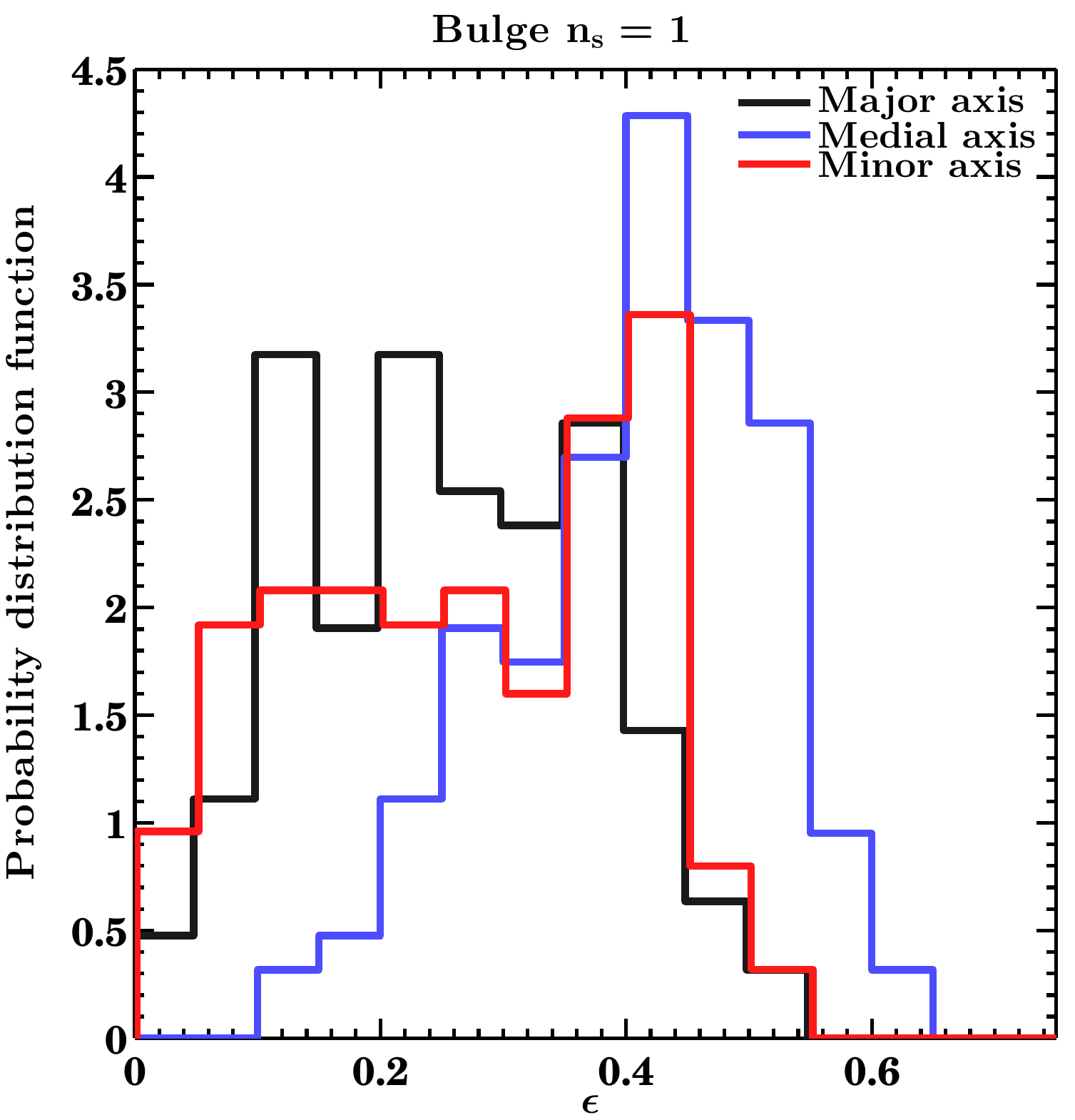} \includegraphics[width=0.33\textwidth]{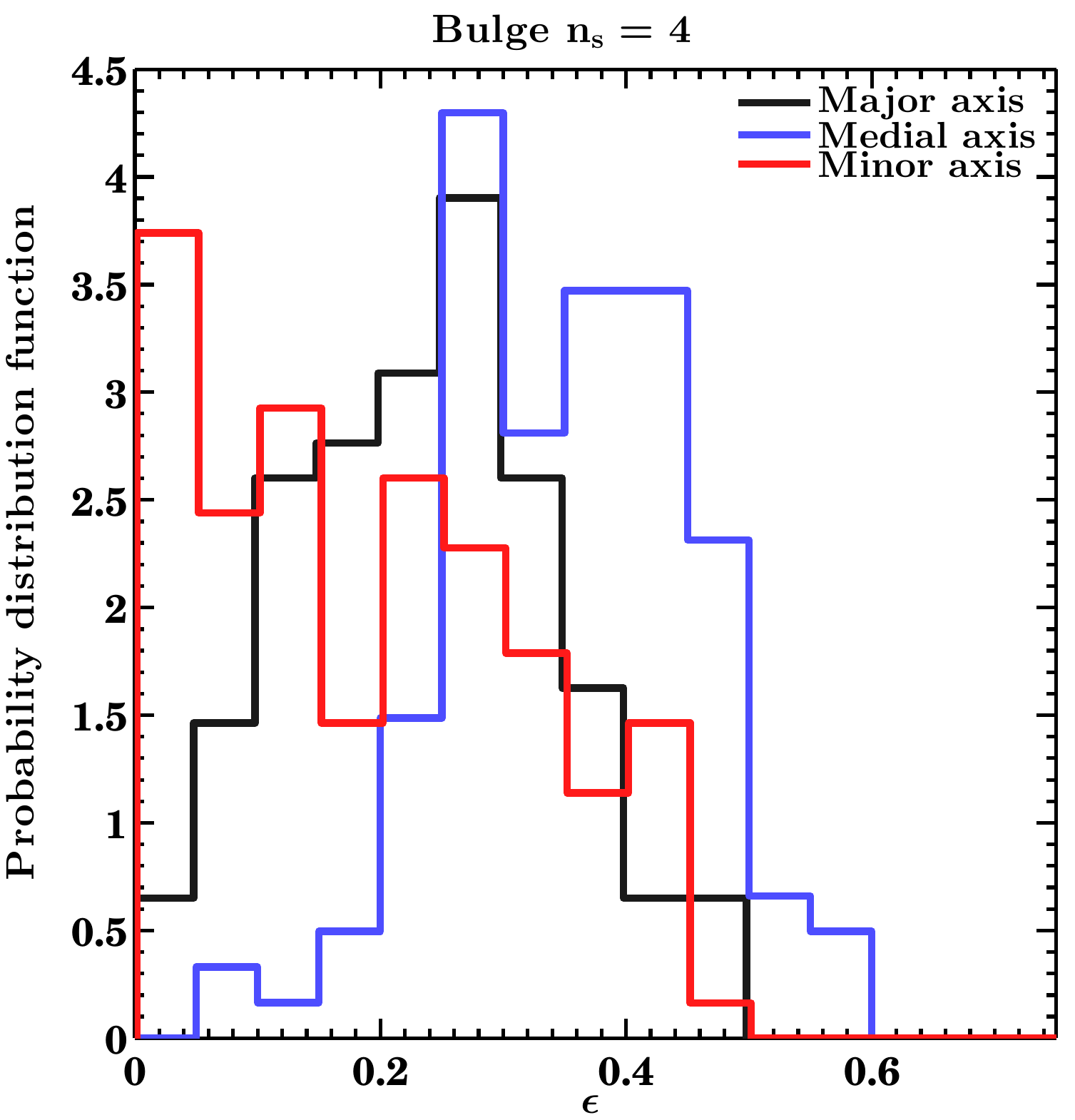}
\includegraphics[width=0.33\textwidth]{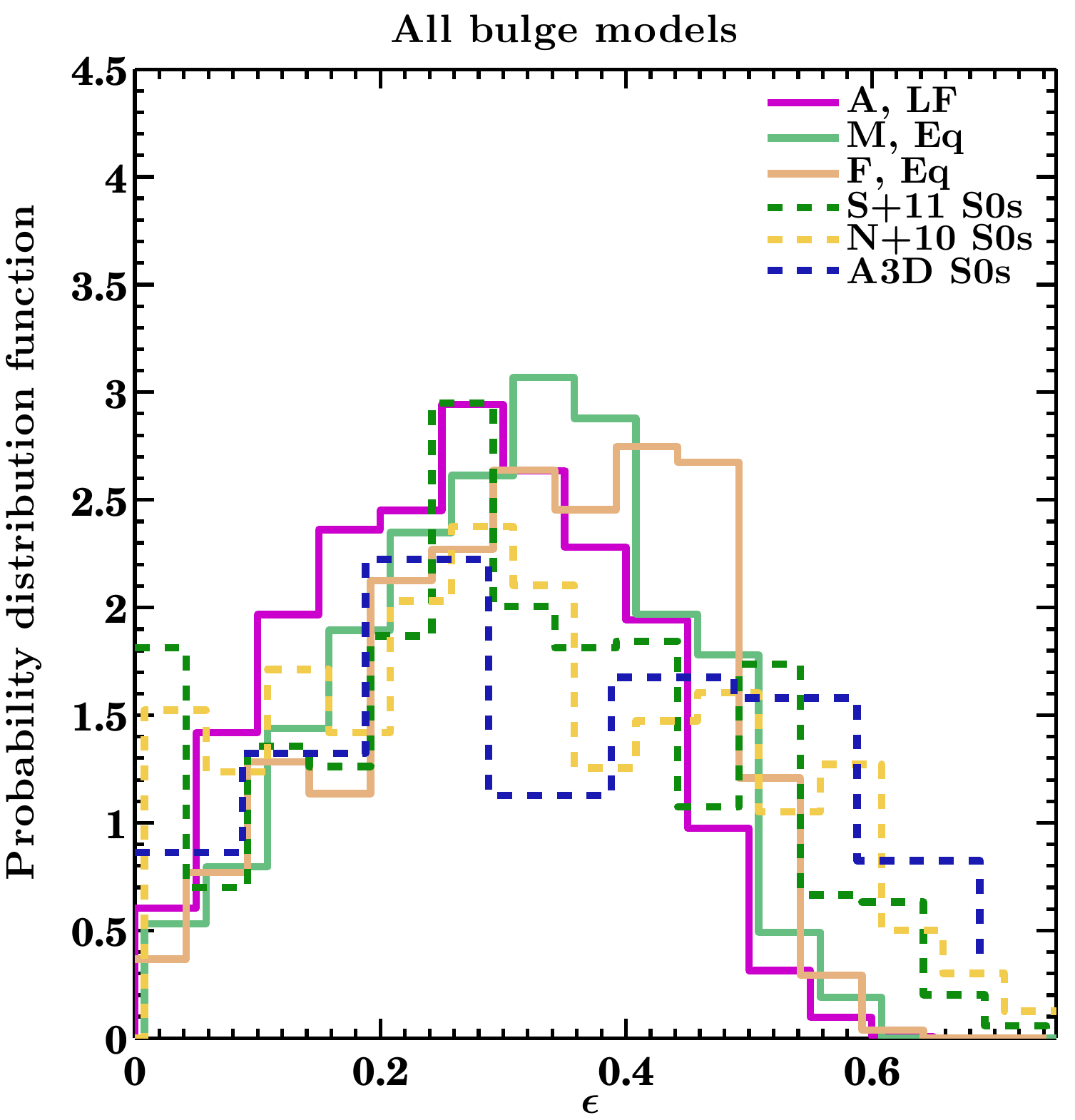}
\caption{Ellipticities of central ellipticals. Top panels show randomly oriented, evenly spaced projections of various models compared with observed ellipticals. Bottom left and middle panels show principal axis projections only, while the bottom right panel shows equal mass mergers and compares with observed S0s. Ellipticities of central galaxies are largely consistent with the observed distributions, though slightly more flattened on average. Simulated galaxies are intrinsically triaxial, with the minor axis projection being the roundest on average. The remnants are rounder than observed S0s and lack a tail of highly flattened ($\epsilon>$ 0.5) objects. 
\label{fig:e_hist}}
\end{figure*}

Ellipticities of the remnants are on the average slightly larger than observed elliptical samples but lower and more sharply peaked than S0s. ~\figref{e_hist} shows that there is only a small difference between the Many- and Few-merger subsamples, while there is about a 0.05 shift towards rounder remnants from the B.n$_{s}$=4 to B.n$_{s}$=1 samples. On the whole the distributions are not unreasonable, lying closer to observations of ellipticals than of S0s, while lacking the tail of highly elliptical shapes found in S0s. Although the Many-merger remnants are slightly rounder on average than the Few-merger, the difference is not large even in equal-mass mergers. This is somewhat surprising, considering that the progenitor galaxy orbits are nearly isotropic and should tend to produce spheroidal remnants as the number of mergers increases. We will elaborate on this point further in \secref{discussion}.

The intrinsic ellipticities of the remnants along the principal axis projections are also shown in the bottom left and middle panel of \figref{e_hist}. The distributions are consistent with the remnants being triaxial, with the median value in each projection being both different than the others and greater than zero. The smallest axis ratios are found for the minor axis projection, which would be the case for ellipsoids closer to prolate than oblate. Most galaxies have a medial axis ellipticity of around 0.4, with few being rounder than 0.2, indicating that almost all galaxies have a significantly shorter minor axis than the major axis.

\begin{figure*}
\includegraphics[width=0.49\textwidth]{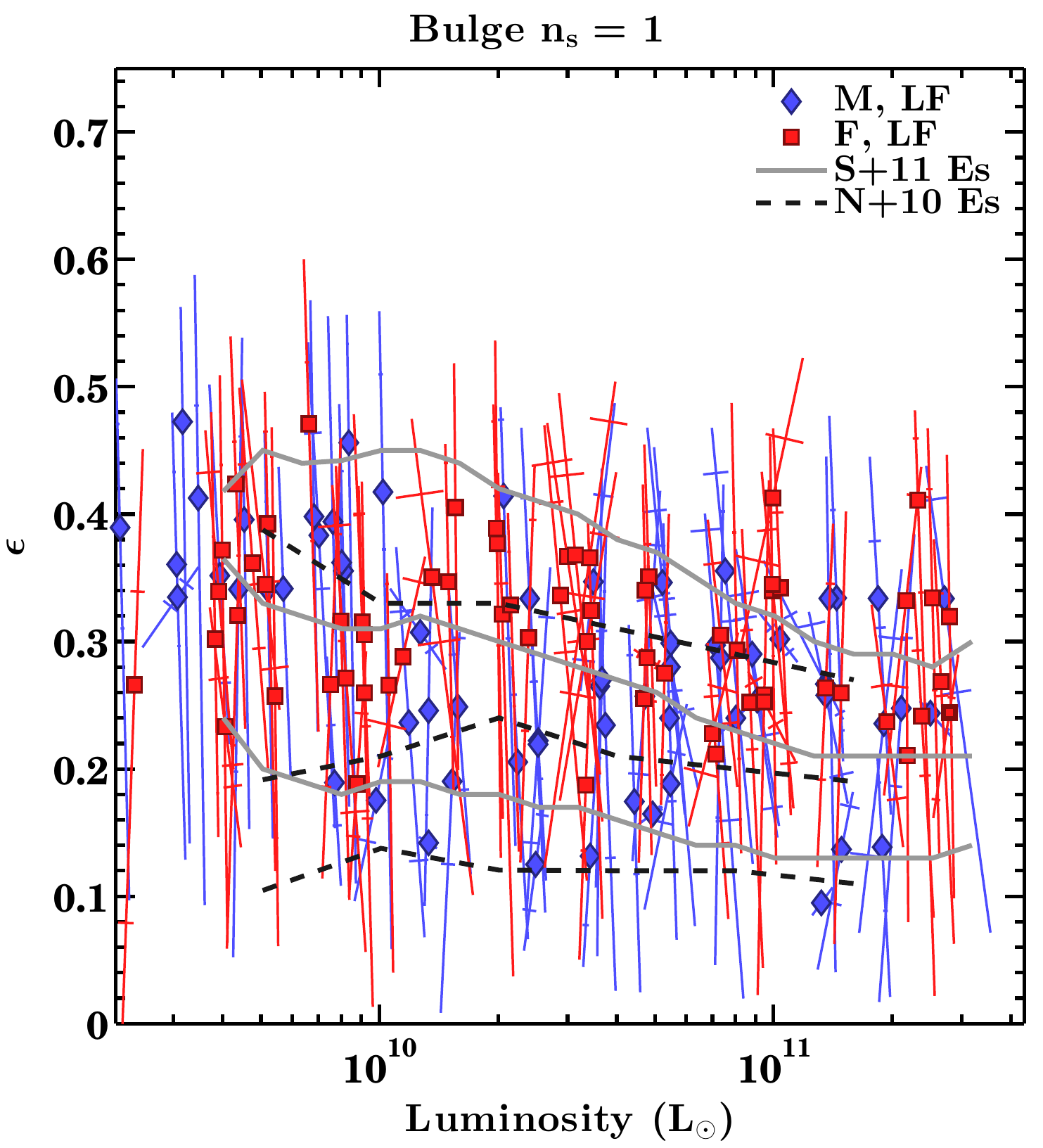}
\includegraphics[width=0.49\textwidth]{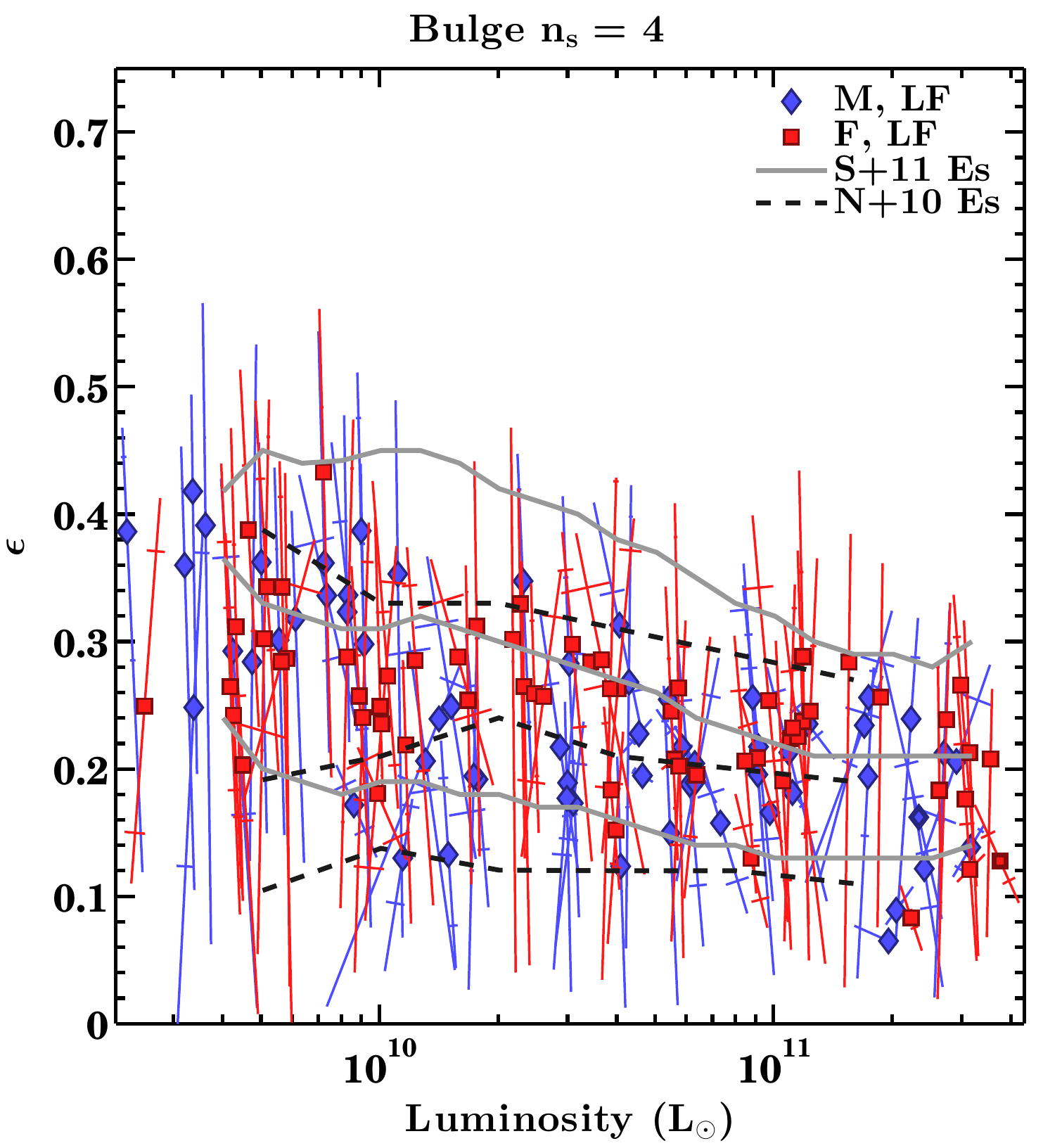}
\caption{Ellipticities of central ellipticals by galaxy luminosity. Median ellipticities tend to decrease slightly with luminosity, both in simulations and observations, while B.n$_{s}$=4 mergers are slightly rounder on average. Projection effects for single galaxies are quite large, with spreads of 0.2-0.3 in ellipticity being common. Line types are as in \figref{l_sersicn}.
\label{fig:l_ell}}
\end{figure*}

In addition to having larger Sersic indices, brighter galaxies trend toward smaller ellipticities and rounder shapes (\figref{l_ell}). This trend might be expected for more luminous galaxies with many mergers. If the orbits of the merging galaxies are isotropically distributed, the resulting remnant should be close to spherical. Such a trend is present in the simulations, although it appears stronger for the B.n$_{s}$=4 sample. Much of the scatter in the relation appears to be due to projection effects of the inherently triaxial simulated galaxies, although median ellipticities show significant scatter as well. The B.n$_{s}$=1 sample also appears to have few very round ($\epsilon < 0.1$) remnants, especially at low luminosities.

\subsection{Scaling Relations}
\label{subsec:scalerel}

\subsubsection{Size-Luminosity/Stellar Mass and Kormendy Relations}

\begin{figure}
\includegraphics[width=0.49\textwidth]{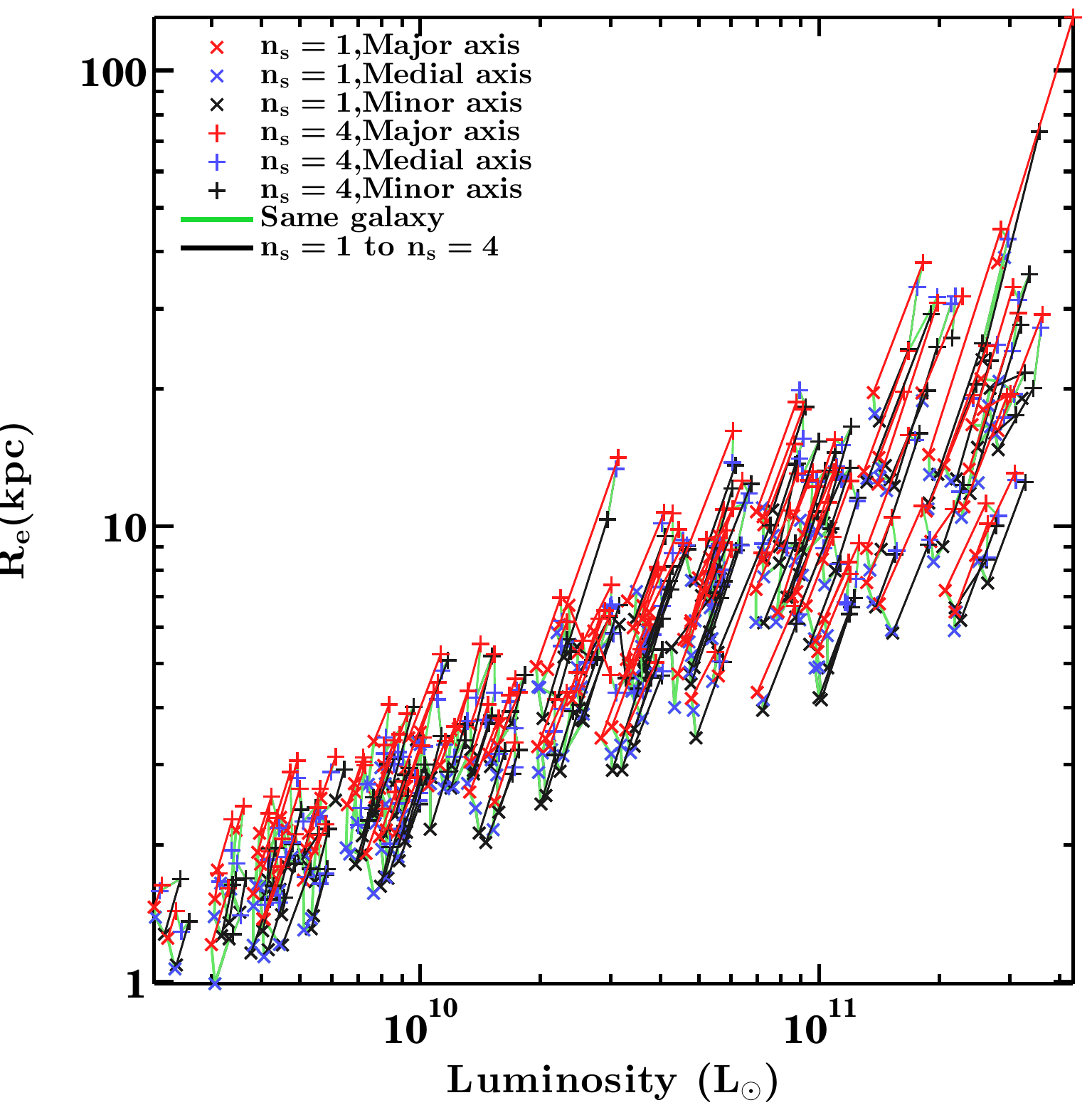}
\caption{The size-luminosity relation of merger remnants after 10 Gyr. Each point shows one of the principal axis projections. Light (green) lines connect different projections of the same galaxy. Darker (red and black) lines connect the same projection for groups with different progenitor bulge profiles but otherwise identical initial conditions. The light (green) lines can be viewed as contributions to scatter in the relation from projection effects, while the darker lines show differences from progenitor galaxies.
\label{fig:size_lum}}
\end{figure}

~\figref{size_lum} shows the Sersic model size-luminosity relation for principal axis projection of simulated galaxies after 10.3 Gyr, connecting otherwise identical groups with different spiral bulges. All relations have very small scatter. Part of the scatter is caused by the B.n$_{s}$=1 sample having smaller sizes (a real effect) and lower luminosities (partly a real effect, but largely systematic, as will be shown in \appref{analysis_testing}). Regardless, both projection effects and different progenitor bulge profiles contribute to the scatter in the relation. 

\begin{figure*}
\includegraphics[width=0.50\textwidth]{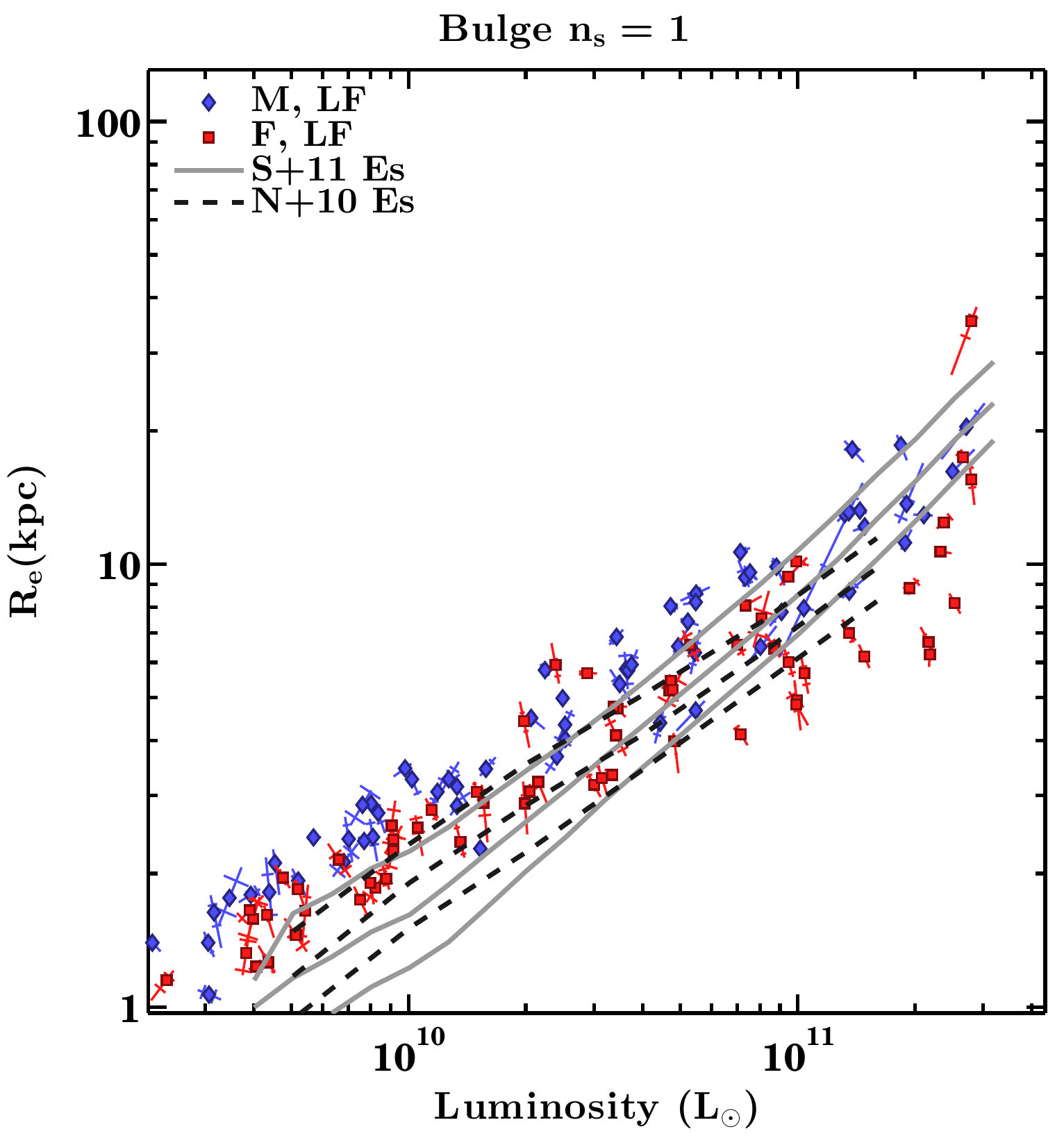}
\includegraphics[width=0.50\textwidth]{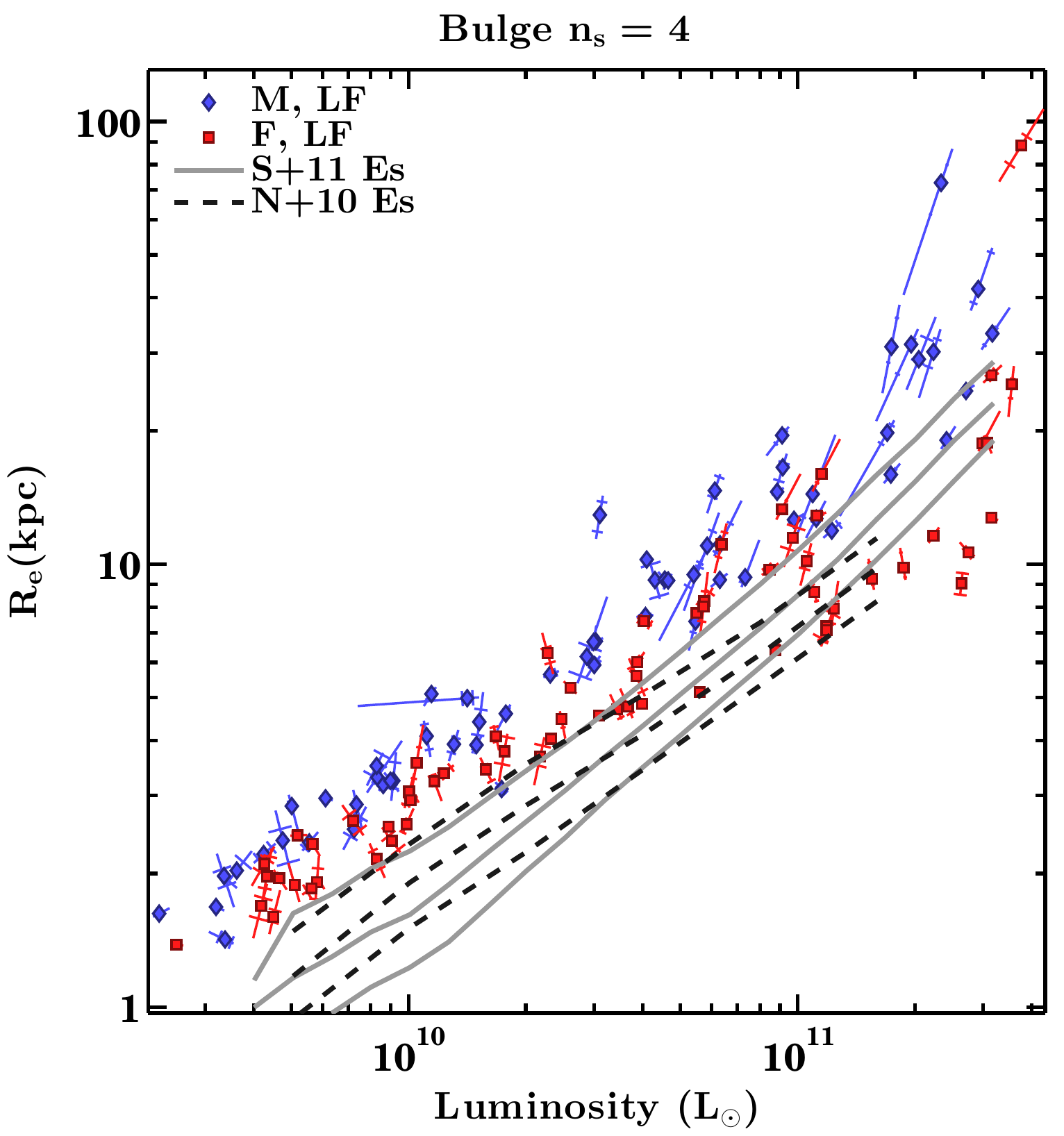}
\caption{Sersic model size-luminosity relation of central ellipticals after 10 Gyr. The simulated relations have small scatter and a similar slope to observations but are offset slightly, being too large at fixed luminosity. Many merger galaxies appear to be a better fit for luminous ellipticals, whereas few mergers match low luminosity ellipticals better. Line types are as in \figref{l_sersicn}.
\label{fig:sizelum}}
\end{figure*}

\begin{table}
\caption{Sersic model size-luminosity relations}
Simulations: Ten equally-spaced projections, randomly oriented \\
\begin{tabular}{ccccc}
\hline 
B.$\mathrm{n_{s}}$ & Subsample & Slope & Intercept & R.M.S. \\
\hline
1 & Unweighted & 0.53 $\pm$ 0.01 & -4.89 $\pm$ 0.07 & 0.10 \\ 
1 & Weighted & 0.58 $\pm$ 0.01 & -5.36 $\pm$ 0.06 & 0.10 \\ 
1 & Many & 0.55 $\pm$ 0.01 & -5.08 $\pm$ 0.05 & 0.07 \\ 
1 & Few & 0.50 $\pm$ 0.01 & -4.67 $\pm$ 0.08 & 0.10 \\ 
\hline
4 & Unweighted & 0.61 $\pm$ 0.01 & -5.66 $\pm$ 0.10 & 0.12 \\ 
4 & Weighted & 0.67 $\pm$ 0.01 & -6.20 $\pm$ 0.07 & 0.12 \\ 
4 & Many & 0.65 $\pm$ 0.01 & -5.95 $\pm$ 0.08 & 0.09 \\ 
4 & Few & 0.55 $\pm$ 0.01 & -5.11 $\pm$ 0.11 & 0.11 \\ 
\hline
All & Unweighted & 0.58 $\pm$ 0.01 & -5.32 $\pm$ 0.06 & 0.12 \\ 
All & Weighted & 0.63 $\pm$ 0.01 & -5.92 $\pm$ 0.06 & 0.12 \\ 
All & Many & 0.62 $\pm$ 0.01 & -5.69 $\pm$ 0.06 & 0.10 \\ 
All & Few & 0.54 $\pm$ 0.01 & -4.96 $\pm$ 0.08 & 0.12 \\ 
\hline
\end{tabular}

Principal axis projections, unweighted \\
\begin{tabular}{ccccc}
\hline 
B.$\mathrm{n_{s}}$ & Projection & Slope & Intercept & R.M.S. \\
\hline 
1 & Major axis & 0.55 $\pm$ 0.02 & -5.12 $\pm$ 0.19 & 0.10 \\ 
1 & Minor axis & 0.52 $\pm$ 0.02 & -4.70 $\pm$ 0.21 & 0.10 \\ 
4 & Major axis & 0.60 $\pm$ 0.02 & -5.53 $\pm$ 0.24 & 0.11 \\ 
4 & Minor axis & 0.58 $\pm$ 0.02 & -5.25 $\pm$ 0.29 & 0.11 \\ 
\hline
\end{tabular}

Observations \\
\begin{tabular}{cccccc}
\hline 
Cat. & Type & Weight & Slope & Intercept & R.M.S. \\
\hline 
S+11 & E & N & 0.85 $\pm$ 0.00 & -8.34 $\pm$ 0.01 & 0.12 \\ 
S+11 & E & Y & 0.80 $\pm$ 0.00 & -7.82 $\pm$ 0.02 & 0.12 \\ 
\hline
N+10 & E & N & 0.66 $\pm$ 0.01 & -6.40 $\pm$ 0.10 & 0.09 \\ 
N+10 & S0 & N & 0.63 $\pm$ 0.01 & -6.03 $\pm$ 0.11 & 0.12 \\
N+10 & E & Y & 0.65 $\pm$ 0.02 & -6.29 $\pm$ 0.21 & 0.09 \\ 
N+10 & S0 & Y & 0.61 $\pm$ 0.02 & -5.82 $\pm$ 0.24 & 0.12 \\ 
\hline 
A3D & E & N & 0.72 $\pm$ 0.03 & -6.90 $\pm$ 0.35 & 0.12 \\ 
A3D & S0 & N & 0.57 $\pm$ 0.05 & -5.41 $\pm$ 0.24 & 0.12 \\ 
A3D & E & Y & 0.72 $\pm$ 0.04 & -6.97 $\pm$ 0.65 & 0.12 \\ 
A3D & S0 & Y & 0.60 $\pm$ 0.05 & -5.70 $\pm$ 0.44 & 0.12 \\ 
\hline
\end{tabular}
\tablecomments{Slopes are given in log space, i.e., for $\log(R_{eff})$ as a function of $\log(L)$. Simulation data are from analyses after 10.3 Gyr, including various subsamples of randomly oriented (but equally spaced) projections, as detailed in the text, as well as principal axis projections. Observational data for each catalog (Cat.) are 1/V$\mathrm{_{max}}$ corrected, with fits optionally weighted (Weight) or not by the difference between the simulated and observed luminosity functions. R.M.S. lists the r.m.s. orthogonal scatter of all points from the best-fit relation.}
\label{tab:size_lum}
\end{table}

\tabtextref{size_lum} lists best-fit Sersic model size-luminosity relations for simulations and observations alike, obtained by least-squares minimization of the orthogonal scatter. In all of the simulation samples, the scatter is relatively small at about 0.1 dex. The scatter does not appear to be mainly due to projection effects or combining progenitors. Fits to major axis projections have similar scatter to the ten equidistant but randomly aligned projections. Similarly, though some groups show projection-dependent sizes and luminosities, these variations are smaller than the scatter in median values, and are likely a result of the mild correlation between Sersic index and luminosity of projections of the same galaxy (evidenced in \figref{l_sersicn}). If sizes and luminosities are generally accurate to within 10-20\% or 0.04-0.08 dex, as suggested by our testing, then some of the scatter could be intrinsic. The scatter in the unweighted simulation data is comparable to that in observed ellipticals (slightly larger than N+10), while the slope is considerably shallower and the intercepts larger. 

Separate fits to the Many- and Few-merger subsamples show a large difference of 0.05 to 0.1 in slope. Also, as \figref{sizelum} demonstrates, the Many-merger subsample is larger at fixed luminosity than the few merger sample. Thus, the slope of the predicted relation can be maximized by giving a larger weight to luminous, Many-merger remnants (and a smaller weight to faint galaxies), while applying the opposite weighting to groups of relatively few galaxies, such that their weights are largest at low luminosities. We apply such a weighting in \tabref{size_lum} and find that it can steepen the slope of the size-luminosity relation further than even the Many-merger subsample alone, bringing it close to observed values for N+10 but still short of S+11 and A3D.

\tabtextref{size_lum} also lists values for observational data, with both 1/V$\mathrm{_{max}}$ corrections and optional weighting to match the luminosity function of the simulations. This weighting scheme only makes a significant difference in the S+11 sample - otherwise, most scaling relations are insensitive to weighting method, as one would expect if they are truly linear with uniform scatter. Some curvature may exist at the low- or high-luminosity extremes, but it is unclear whether it is real or systematic.

\begin{table}
\caption{Petrosian size-luminosity relations}
Simulations: Ten equally-spaced projections, randomly oriented \\
\begin{tabular}{ccccc}
\hline 
B.$\mathrm{n_{s}}$ & Subsample & Slope & Intercept & R.M.S. \\
\hline
1 & Unweighted & 0.50 $\pm$ 0.01 & -4.62 $\pm$ 0.06 & 0.10 \\ 
1 & Weighted & 0.54 $\pm$ 0.01 & -5.06 $\pm$ 0.05 & 0.10 \\ 
1 & Many & 0.52 $\pm$ 0.01 & -4.81 $\pm$ 0.06 & 0.08 \\ 
1 & Few & 0.48 $\pm$ 0.01 & -4.43 $\pm$ 0.09 & 0.09 \\ 
\hline
4 & Unweighted & 0.49 $\pm$ 0.01 & -4.47 $\pm$ 0.05 & 0.10 \\ 
4 & Weighted & 0.54 $\pm$ 0.01 & -4.99 $\pm$ 0.06 & 0.10 \\ 
4 & Many & 0.52 $\pm$ 0.01 & -4.72 $\pm$ 0.07 & 0.08 \\ 
4 & Few & 0.46 $\pm$ 0.01 & -4.22 $\pm$ 0.07 & 0.09 \\ 
\hline
All & Unweighted & 0.49 $\pm$ 0.00 & -4.50 $\pm$ 0.04 & 0.10 \\ 
All & Weighted & 0.54 $\pm$ 0.01 & -5.04 $\pm$ 0.04 & 0.10 \\ 
All & Many & 0.52 $\pm$ 0.00 & -4.76 $\pm$ 0.06 & 0.08 \\ 
All & Few & 0.47 $\pm$ 0.01 & -4.34 $\pm$ 0.06 & 0.09 \\ 
\hline
\end{tabular}

Observations \\
\begin{tabular}{cccccc}
\hline 
Cat. & Type & Weight & Slope & Intercept & R.M.S. \\
\hline 
N+10 & E & N & 0.62 $\pm$ 0.01 & -6.04 $\pm$ 0.07 & 0.08 \\ 
N+10 & S0 & N & 0.57 $\pm$ 0.01 & -5.53 $\pm$ 0.07 & 0.10 \\ 
N+10 & E & Y & 0.58 $\pm$ 0.01 & -5.61 $\pm$ 0.22 & 0.08 \\ 
N+10 & S0 & Y & 0.54 $\pm$ 0.01 & -5.14 $\pm$ 0.09 & 0.10 \\
\hline 
\end{tabular}
\tablecomments{Column definitions are as in \tabref{size_lum}.}
\label{tab:petro_size_lum}
\end{table}

The Petrosian $R_{50}$ size-luminosity relation (\tabref{petro_size_lum}) shows smaller scatter than Sersic sizes, despite the fact that uncorrected Petrosian half-light radii systematically underestimate the luminosities of pure Sersic profiles and simulated galaxies alike. This is especially true for the B.n$_{s}$=4 sample, which has slightly lower scatter than B.n$_{s}$=1 mergers, despite having greater systematic errors on $R_{eff}$ due to its larger mean $\mathrm{n_{s}}$. The slopes are still shallower than those observed in N+10, but the difference can shrink to less than 0.05 if considering weightings for both simulations and observations. The implications of these results will be discussed further in \secref{discussion}.

\begin{table}
\caption{Sersic size-stellar mass relations}
\begin{tabular}{cccccc}
\hline 
Cat. & Type & Weight & Slope & Intercept & R.M.S. \\
\hline
S+11 & E & N & 0.78 $\pm$ 0.00 & -7.95 $\pm$ 0.02 & 0.13 \\ 
S+11 & E & Y & 0.75 $\pm$ 0.00 & -7.53 $\pm$ 0.03 & 0.13 \\ 
\hline 
N+10 & E & N & 0.64 $\pm$ 0.01 & -6.39 $\pm$ 0.06 & 0.09 \\ 
N+10 & S0 & N & 0.57 $\pm$ 0.01 & -5.61 $\pm$ 0.09 & 0.12 \\ 
N+10 & E & Y & 0.60 $\pm$ 0.02 & -5.89 $\pm$ 0.20 & 0.09 \\ 
N+10 & S0 & Y & 0.48 $\pm$ 0.02 & -4.63 $\pm$ 0.17 & 0.13 \\ 
\hline
\end{tabular}
\tablecomments{Column definitions are as in \tabref{size_lum}.}
\label{tab:size_mstellar}
\end{table}

The best-fit relations between size and stellar mass for the S+11 and N+10 catalogs are listed in \tabref{size_mstellar}. The slopes are slightly shallower than those for the size-luminosity relations and closer to (but not quite matching) those predicted by the simulations, which do not have significant variations in the stellar mass-to-light ratio. Thus some of the tension between the slopes of the simulated and observed size-luminosity relations can be resolved by accounting for the variable stellar mass-to-light ratio of observed galaxies, which increases in more luminous observed ellipticals but is nearly constant by construction in the simulated remnants.

\begin{table}
\caption{Petrosian model size-stellar mass relation}
\begin{tabular}{cccccc}
\hline 
Cat. & Type & Weight & Slope & Intercept & R.M.S. \\
\hline
N+10 & E & N & 0.59 $\pm$ 0.01 & -5.97 $\pm$ 0.09 & 0.09 \\ 
N+10 & S0 & N & 0.50 $\pm$ 0.01 & -4.922 $\pm$ 0.08 & 0.11 \\ 
N+10 & E & Y & 0.52 $\pm$ 0.02 & -5.21 $\pm$ 0.21 & 0.09 \\ 
N+10 & S0 & Y & 0.41 $\pm$ 0.01 & -4.03 $\pm$ 0.11 & 0.12 \\ 
\hline
\end{tabular}
\tablecomments{Column definitions are as in \tabref{size_lum}.}
\label{tab:petro_size_mstellar}
\end{table}

The Petrosian size-stellar mass relation shows slightly shallower slope, as with Sersic models. In fact, the slope and scatter of the weighted simulations (0.52 and 0.09) are within the quite small bootstrap errors (0.01) of the weighted observations (0.51 and 0.09), while the intercept is higher (-4.81 versus -5.08) but still also within the more generous error bars. Thus, it is entirely possible to match the slopes, and, to a lesser extent, the intercepts of the size-mass relation, depending on the fitting technique and sample weights. However, this alone does not justify either weighting scheme. The observational scheme is reasonable, since matching luminosity functions is necessary in order to make a fair comparison. The simulation scheme is not as well justified, since the number of mergers per group is somewhat arbitrary.

\begin{figure*}
\includegraphics[width=0.50\textwidth]{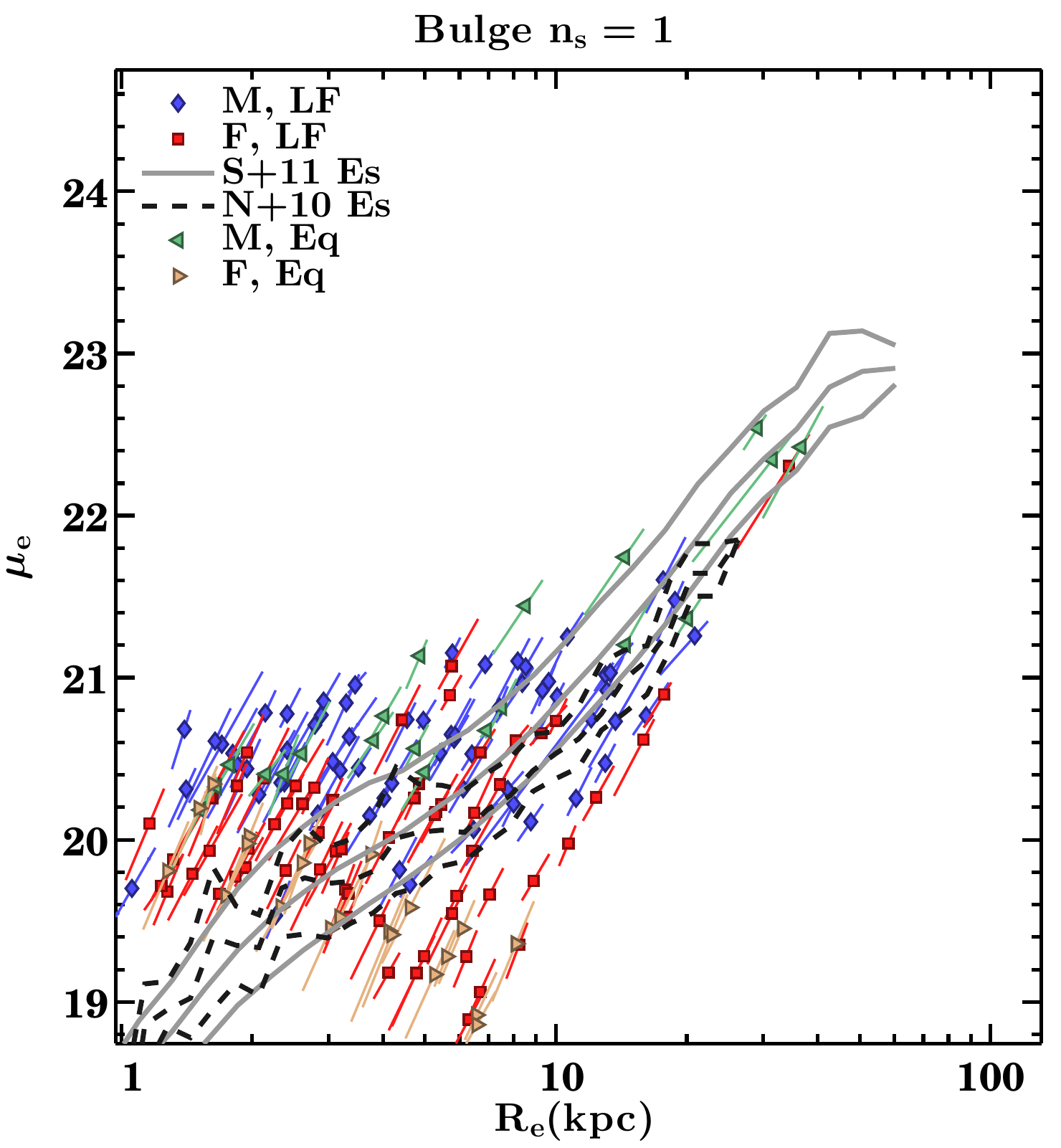}
\includegraphics[width=0.50\textwidth]{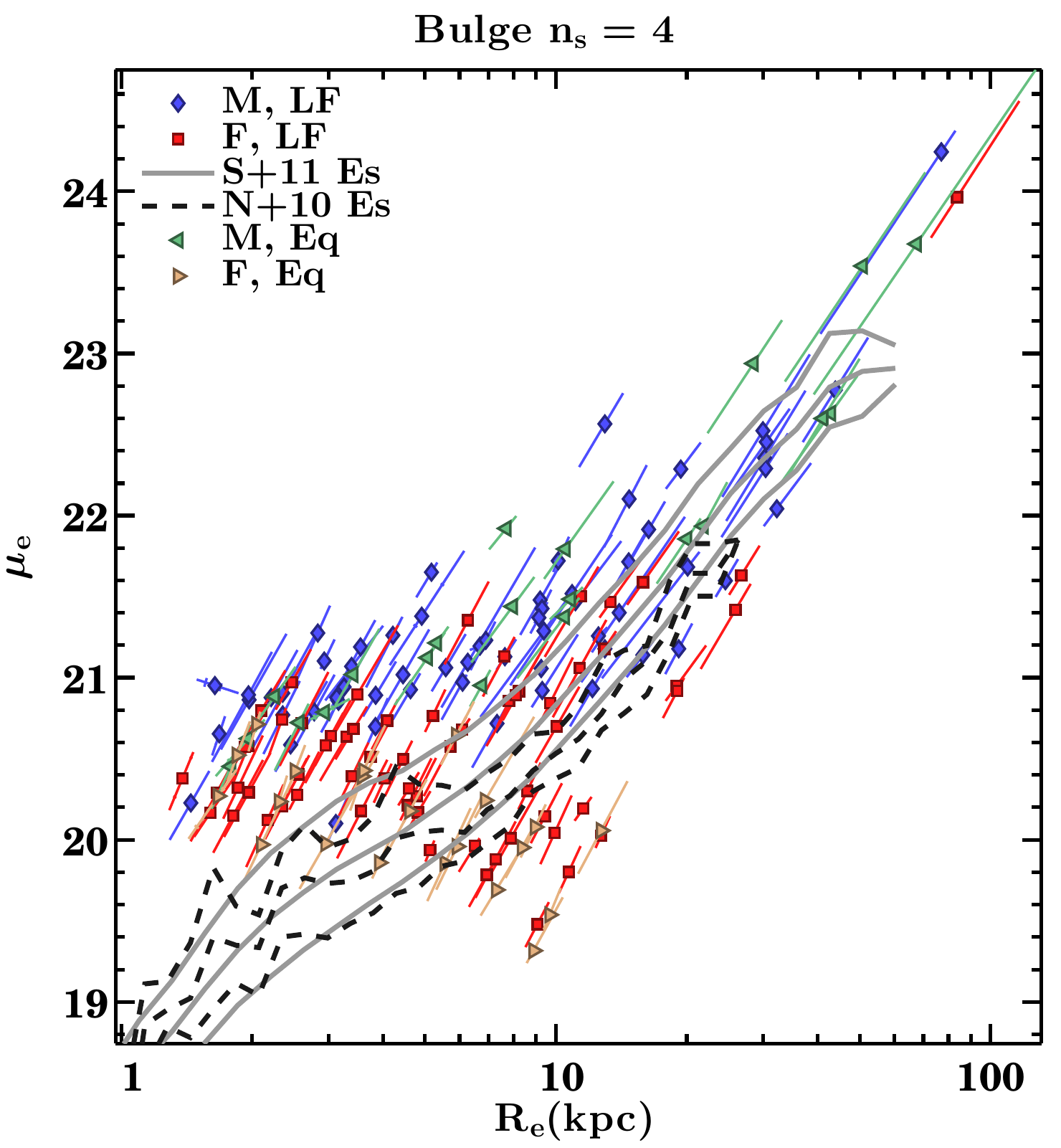}
\caption{The Kormendy relation of central ellipticals. Only B.n$_{s}$=4 simulations show a distinct Kormendy relation, but they are also too faint at a fixed size. The observed relation can be better reproduced if few mergers and B.n$_{s}$=1 are the preferred source of small galaxies and many/B.n$_{s}$=4 produced large ellipticals. Line types are as in \figref{l_sersicn}.
\label{fig:kormendy}}
\end{figure*}

The Kormendy relation \citep{Kor77}, shown in \figref{kormendy}, has large scatter and shallow slope, especially for the B.n$_{s}$=1 relation, which is nearly flat. None of the observed relations are quite linear. While the kink at small sizes is likely a systematic artifact, the curvature near 5-6 kpc appears more robust and also more significant than the equivalent curvature in the size-luminosity relation. As in the size-luminosity relation, it appears as if the simulated galaxies are either too faint for their size or too large to be so faint. Interestingly, the relation for large ellipticals appears to asymptote towards the slope of constant luminosity ($\mathrm{d\log(\mu_{e})/d\log(R_{eff})}$ = 5), which suggests that bright ellipticals can grow significantly in size without adding a large amount of stellar mass. In fact, many BCGs have exceptionally large effective radii and faint mean surface brightnesses. However, most of the similar simulated remnants in \figref{size_lum} are mergers of many equal-mass spirals, rather than luminosity function-sampled remnants - without this M-Eq subsample, the simulated Kormendy relation is rather weak, especially for the B.n$_{s}$=1 sample.

\subsubsection{Faber-Jackson Relation}

\begin{figure*}
\includegraphics[width=0.49\textwidth]{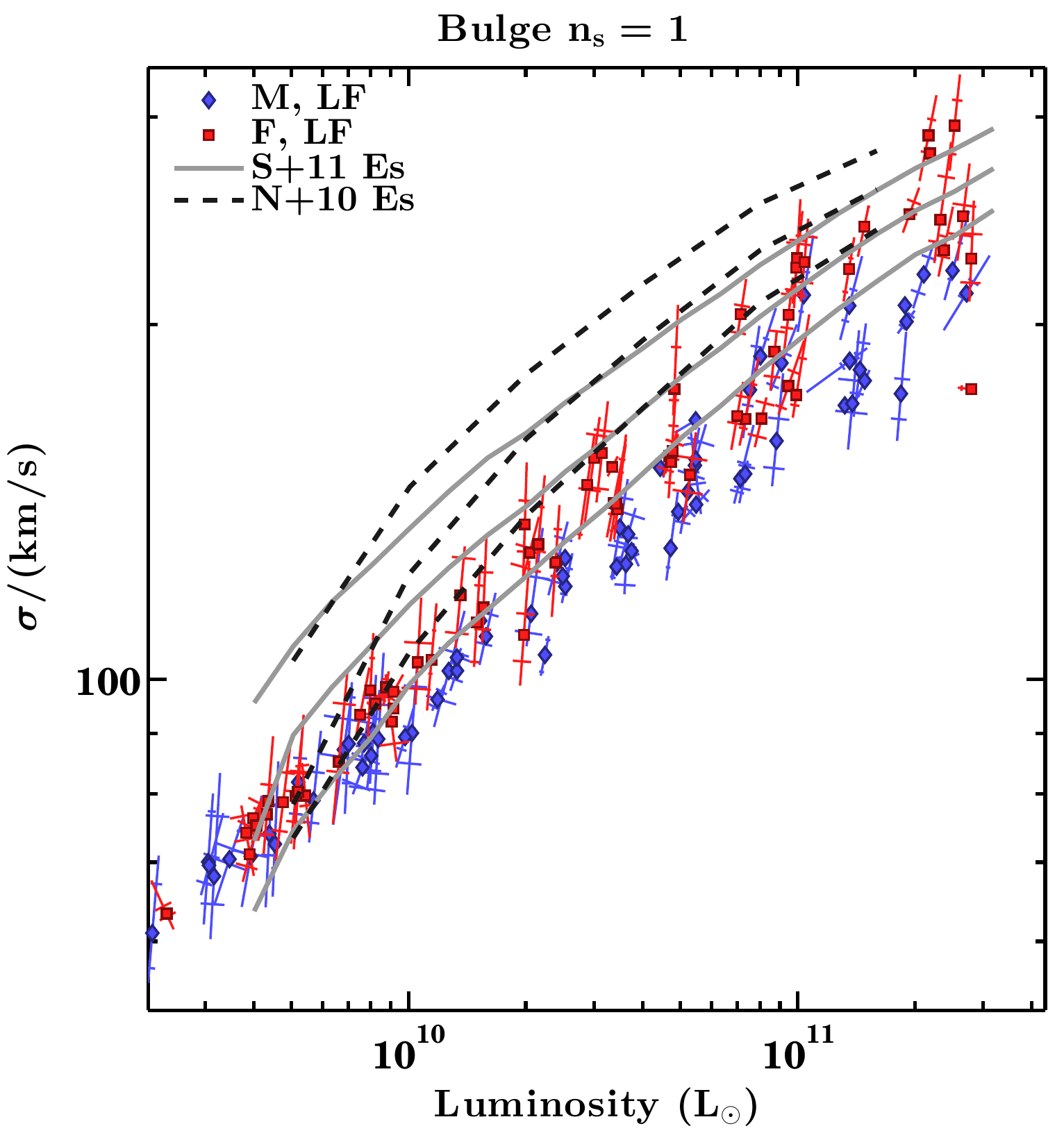}
\includegraphics[width=0.49\textwidth]{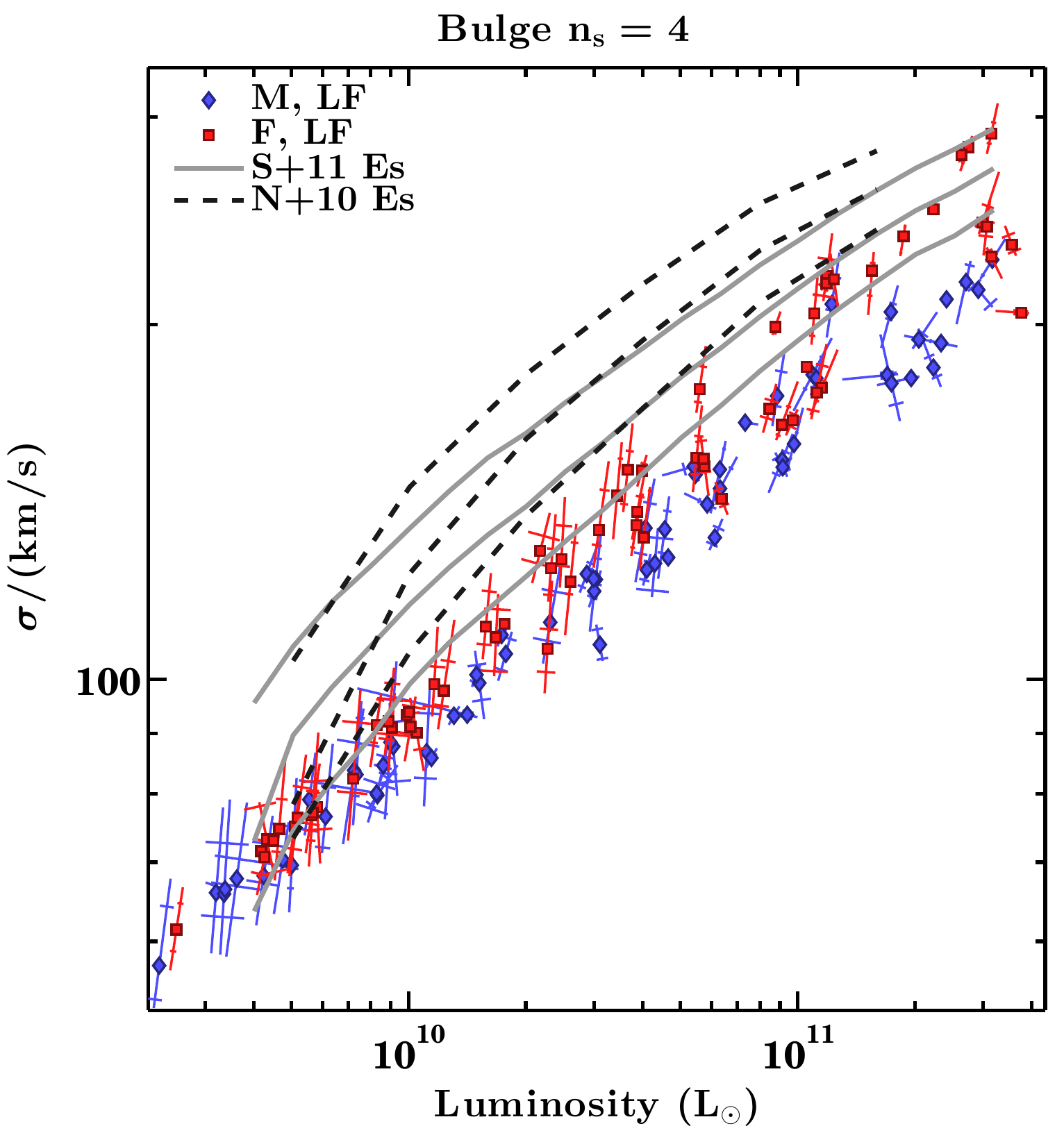}
\caption{The Faber-Jackson relation. Both simulation samples follow a similar slope to observations but are offset to lower dispersions, regardless of which subsample is used. The scatter in simulations is tighter than observations, since the observed relations show 25th, 50th and 75th percentiles. Line types are as in \figref{l_sersicn}.} 
\label{fig:fj}
\end{figure*}

\begin{table}
\caption{Sersic model Faber-Jackson relations}
Simulations: Ten equally-spaced projections, randomly oriented \\
\begin{tabular}{ccccc}
\hline 
B.$\mathrm{n_{s}}$ & Subsample & Slope & Intercept & R.M.S. \\
\hline
1 & Unweighted & 0.28 $\pm$ 0.00 & -0.85 $\pm$ 0.02 & 0.04 \\ 
1 & Weighted & 0.27 $\pm$ 0.00 & -0.68 $\pm$ 0.02 & 0.05 \\ 
1 & Many & 0.27 $\pm$ 0.00 & -0.68 $\pm$ 0.03 & 0.03 \\ 
1 & Few & 0.30 $\pm$ 0.00 & -0.98 $\pm$ 0.04 & 0.04 \\ 
\hline
4 & Unweighted & 0.29 $\pm$ 0.00 & -0.90 $\pm$ 0.02 & 0.04 \\ 
4 & Weighted & 0.27 $\pm$ 0.00 & -0.74 $\pm$ 0.02 & 0.04 \\ 
4 & Many & 0.27 $\pm$ 0.00 & -0.75 $\pm$ 0.02 & 0.03 \\ 
4 & Few & 0.30 $\pm$ 0.00 & -0.99 $\pm$ 0.03 & 0.04 \\ 
\hline
All & Unweighted & 0.28 $\pm$ 0.00 & -0.86 $\pm$ 0.02 & 0.04 \\ 
All & Weighted & 0.27 $\pm$ 0.00 & -0.71 $\pm$ 0.02 & 0.05 \\ 
All & Many & 0.27 $\pm$ 0.00 & -0.72 $\pm$ 0.02 & 0.04 \\ 
All & Few & 0.30 $\pm$ 0.00 & -0.98 $\pm$ 0.03 & 0.04 \\ 
\hline
\end{tabular}

Principal axis projections, unweighted \\
\begin{tabular}{ccccc}
\hline 
B.$\mathrm{n_{s}}$ & Projection & Slope & Intercept & R.M.S. \\
\hline 
1 & Major axis & 0.27 $\pm$ 0.01 & -0.70 $\pm$ 0.06 & 0.04 \\ 
1 & Minor axis & 0.30 $\pm$ 0.01 & -1.01 $\pm$ 0.06 & 0.04 \\ 
4 & Major axis & 0.27 $\pm$ 0.01 & -0.69 $\pm$ 0.05 & 0.04 \\ 
4 & Minor axis & 0.30 $\pm$ 0.01 & -1.12 $\pm$ 0.06 & 0.04 \\ 
\hline
\end{tabular}

Observations \\
\begin{tabular}{cccccc}
\hline 
Cat. & Type & Weight & Slope & Intercept & R.M.S. \\
\hline 
S+11 & E & N & 0.27 $\pm$ 0.00 & -0.63 $\pm$ 0.01 & 0.08 \\ 
S+11 & E & Y & 0.30 $\pm$ 0.00 & -0.92 $\pm$ 0.02 & 0.08 \\ 
\hline
N+10 & E & N & 0.28 $\pm$ 0.01 & -0.67 $\pm$ 0.10 & 0.07 \\ 
N+10 & S0 & N & 0.36 $\pm$ 0.01 & -1.57 $\pm$ 0.09 & 0.10 \\ 
N+10 & E & Y & 0.37 $\pm$ 0.01 & -1.67 $\pm$ 0.31 & 0.08 \\ 
N+10 & S0 & Y & 0.48 $\pm$ 0.03 & -2.85 $\pm$ 0.20 & 0.11 \\  
\hline
\end{tabular}
\tablecomments{Column definitions are as in \tabref{size_lum}.}
\label{tab:fj}
\end{table}

The Faber-Jackson (velocity dispersion-luminosity, \citet{FabJac76}) relation is shown in \figref{fj}, with best fits tabulated in \tabref{fj}. The simulated relations have slopes fairly close to the observations, though the intercepts are significantly lower. The turnover or curvature at low velocity dispersions ($<$100 \kms) is likely not entirely real, since such low dispersions are near the spectrograph's resolution limit and unlikely to be reliable \citep{SDSSDR8}. The luminosity function weightings make a significant difference in slope for the N+10 sample, which is likely due to this same curvature. The scatter appears to be mostly due to projection effects at the low-luminosity end but increases at high luminosities, where the Many- and Few-merger samples appear to diverge. The most robust conclusions from the data are that the slope for the S0 sample is significantly steeper than that for ellipticals, which in turn is slightly steeper than the canonical slope of 0.25, depending on the weighting scheme used. The scatter in the simulated relations is also significantly lower than in the observations, even when both bulge samples are combined.

\begin{table}
\caption{Velocity dispersion-stellar mass relations}
\begin{tabular}{cccccc}
\hline 
Cat. & Type & Weight & Slope & Intercept & R.M.S. \\
\hline 
S+11 & E & N & 0.26 $\pm$ 0.00 & -0.60 $\pm$ 0.01 & 0.07 \\ 
S+11 & E & Y & 0.28 $\pm$ 0.00 & -0.88 $\pm$ 0.01 & 0.07 \\ 
\hline    
N+10 & E & N & 0.29 $\pm$ 0.00 & -0.89 $\pm$ 0.05 & 0.06 \\ 
N+10 & S0 & N & 0.36 $\pm$ 0.01 & -1.74 $\pm$ 0.08 & 0.09 \\ 
N+10 & E & Y & 0.36 $\pm$ 0.03 & -1.70 $\pm$ 0.26 & 0.07 \\ 
N+10 & S0 & Y & 0.43 $\pm$ 0.01 & -2.45 $\pm$ 0.10 & 0.09 \\ 
\hline
\end{tabular}
\tablecomments{Column definitions are as in \tabref{size_lum}.}
\label{tab:sigma_mstellar}
\end{table}

Unlike the size-mass/luminosity relations, the velocity dispersion-stellar mass relation (\tabref{sigma_mstellar}) is hardly changed from the velocity dispersion-luminosity relation, although the scatter shrinks slightly. The velocity dispersion-stellar mass relation also deviates from the canonical Faber-Jackson relation slope of 0.25, showing a scaling closer to $\mathrm{\sigma \propto M_{*}^{0.3}}$.

\subsection{Rotational Support}
\label{subsubsec:rotation}

\begin{figure}
\includegraphics[width=0.50\textwidth]{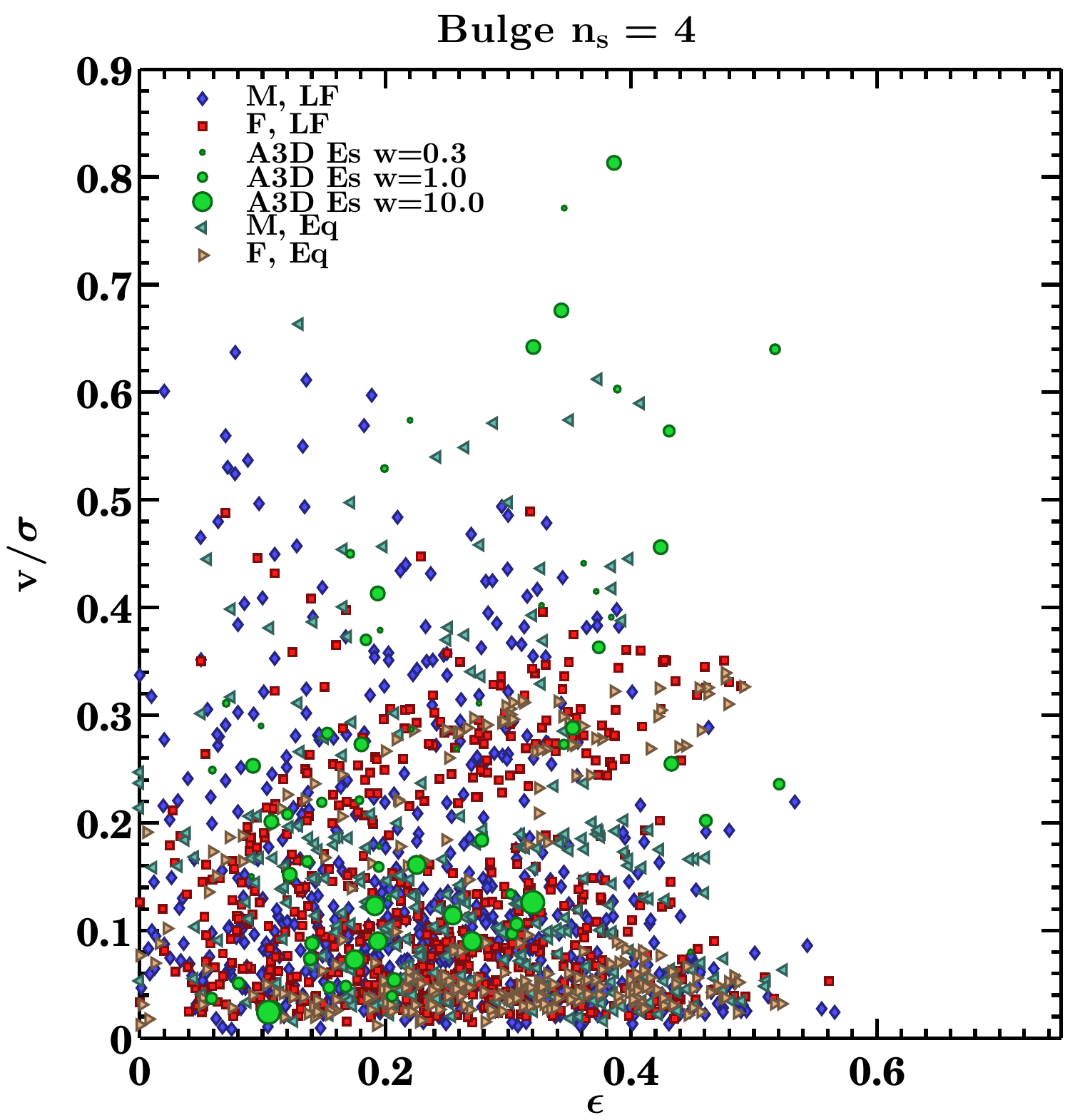}
\caption{Rotational support of simulated galaxies by classical $v/\sigma$ measure. Atlas3D ellipticals are shown with areas of points roughly corresponding to their relative weights on a logarithmic scale. Most simulated ellipticals are slow rotators, but some have modest rotational support.
\label{fig:e_vdivsigma}}
\end{figure}

We measure $v/\sigma$ as the luminosity-weighted average within $R_{eff}$, as used in IFU observations like Atlas3D \citep{CapEmsKra11}. Most simulated ellipticals are slow rotators (\figref{e_vdivsigma}). However, some projections show $v/\sigma$ as large as 0.35 and can be classified as fast rotators despite having been formed from dry mergers. \citet{CoxDutDiM06} found that dry binary mergers only form slow rotators ($v/\sigma <$ 0.1), whereas some group mergers are clearly capable of producing fast rotators. Nonetheless, the scarcity of remnants with $v/\sigma > 0.3$ strongly suggests that dissipation is necessary to form fast rotators, as will be elaborated further in \secref{discussion}.

\begin{figure*}
\includegraphics[width=1\textwidth]{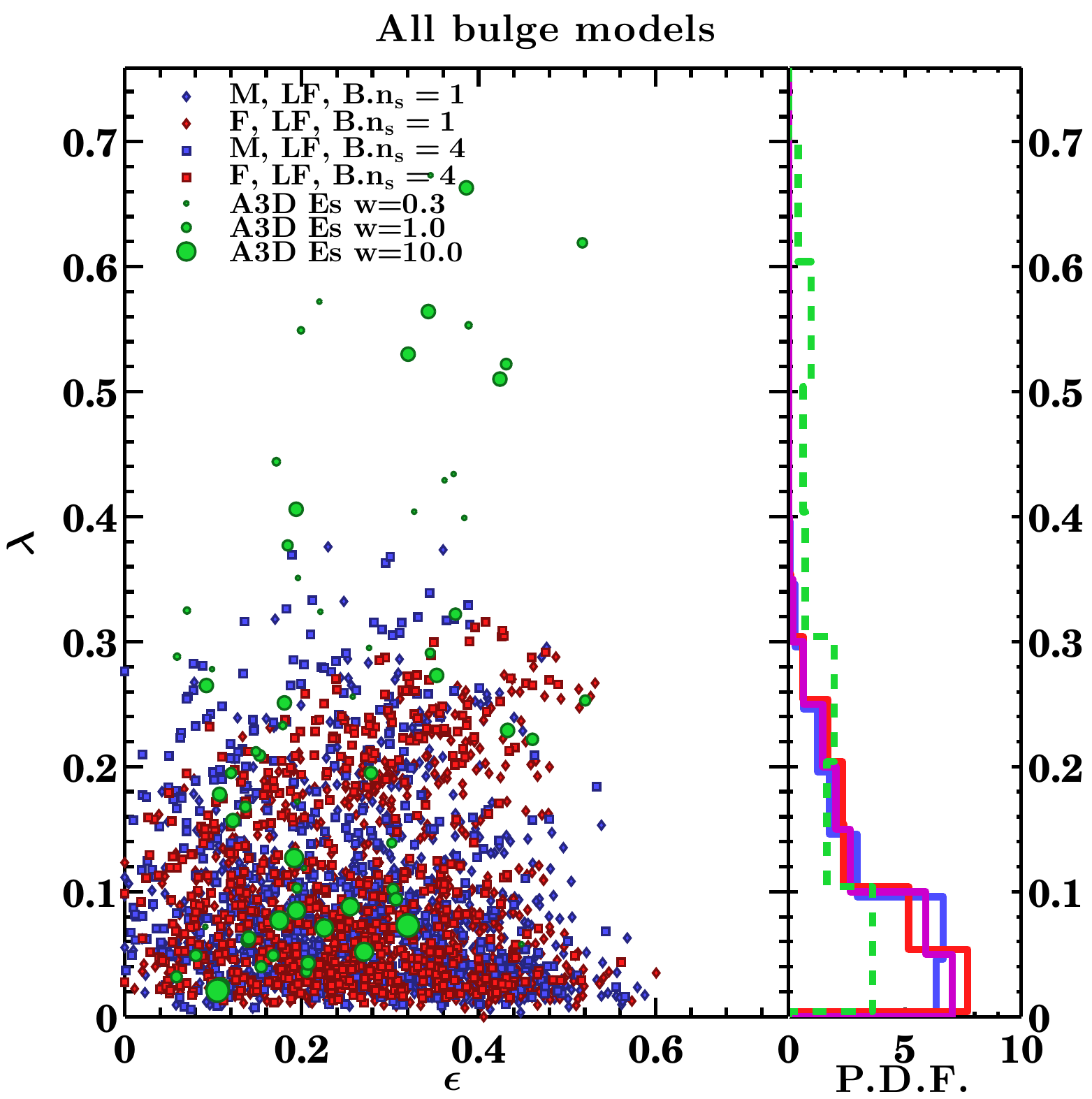}
\caption{Rotational support of simulated galaxies by dimensionless angular momentum measure $\lambda$. Observed data are shown with point sizes proportional to the logarithm of the relative weights to match the luminosity function of the simulated galaxies. Despite this weighting scheme, too many simulated galaxies have low rotational support, and none have very high support ($\lambda >$ 0.4).
\label{fig:e_L}}
\end{figure*}

We also measure rotation in \figref{e_L} by the more physically motivated measure $\lambda$ \citep{CapEmsKra11} - essentially a radially-weighted $v/\sigma$ tracing net projected angular momentum. While the distribution of rotational support is not wildly different from Atlas3D, there is a significant excess of slow rotators (especially flattened ones) and a complete absence of simulated galaxies with $\lambda > 0.4$. B.n$_{s}$=1 mergers and Many-merger remnants tend to be slightly slower rotators, but the differences in both cases are not large.

\figtextref{L_e_prax} shows rotational support for the principal axis projections. The minor axis projection shows minimal rotation, which is expected if there are no stable orbits about the major axis. In general, B.n$_{s}$=4 mergers are rounder despite having faster rotation for the same set of initial conditions. As with random projections, there appear to be two distinct tracks for galaxies, which is more readily apparent in the B.n$_{s}$=4 mergers. Most galaxies have a range of ellipticities in their major axis projections but only show modest increases in rotational support from the minor to medial axis projections. These appear as horizontal lines with a shallow slope near the bottom of the figure. A smaller subset of galaxies are nearly round in the minor axis projections, with very modest rotation ($\lambda < 0.1$), but are significantly flattened ($\epsilon > 0.2$ and rotationally supported ($\lambda > 0.2$) in major and medial axes alike. In fact, for most of these galaxies it appears as if the minor and medial axes are nearly identical, and so these galaxies are probably prolate spheroids. In this case, the distinction between major and medial axis projection is not very meaningful.

\begin{figure*}
\includegraphics[width=0.50\textwidth]{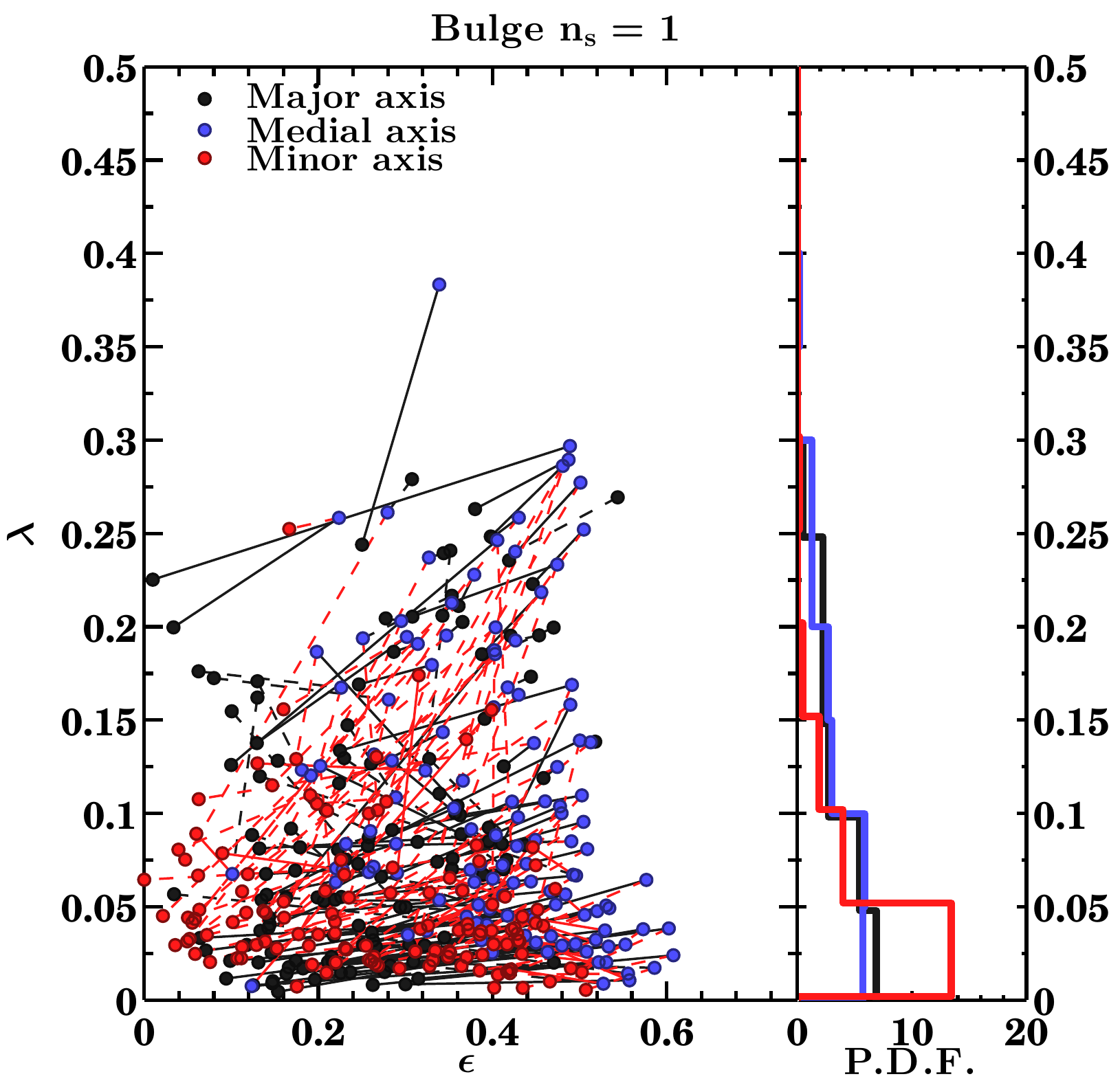}
\includegraphics[width=0.50\textwidth]{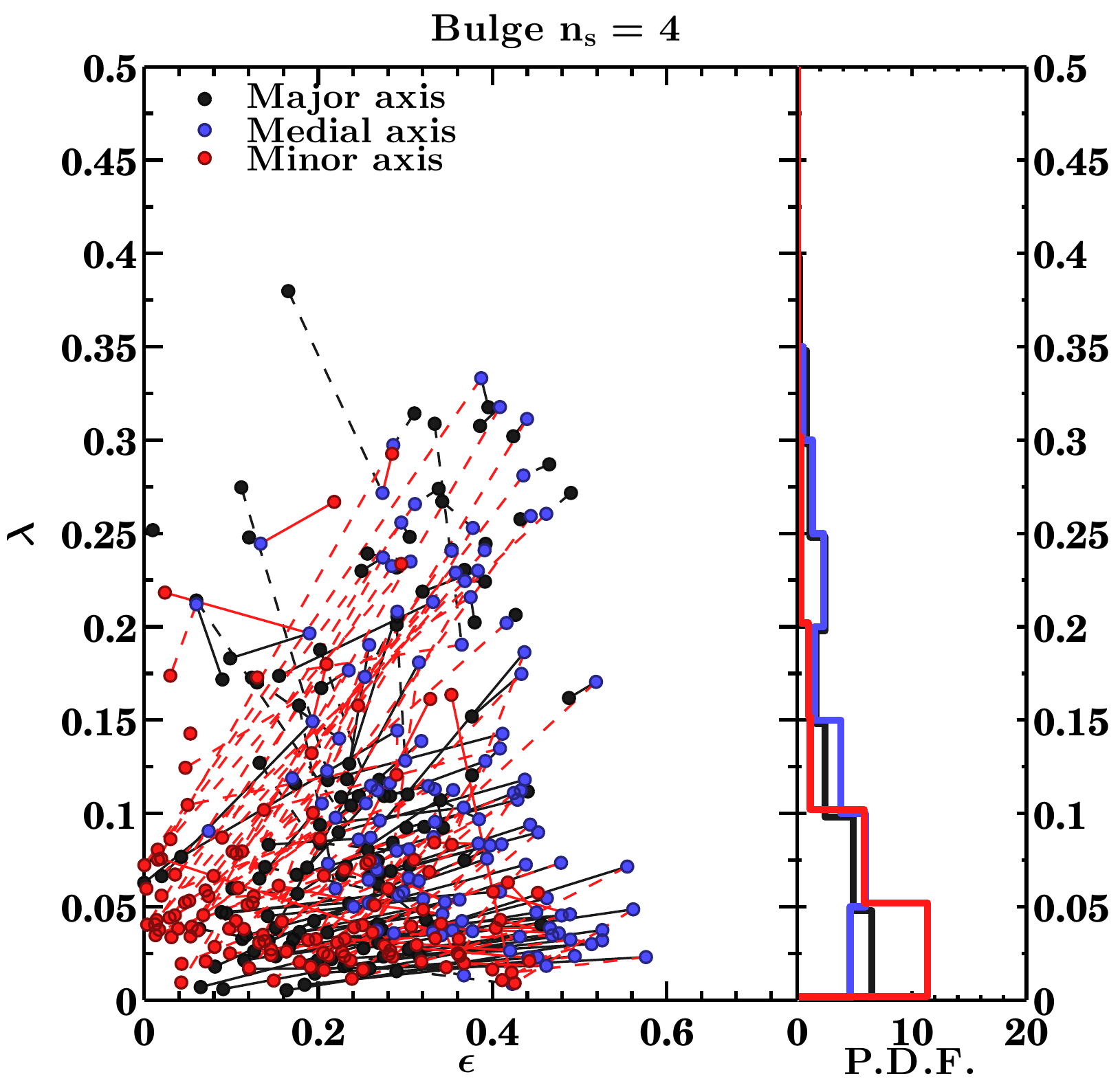}
\caption{Rotational support of principal axis projections of simulated galaxies by dimensionless angular momentum measure $\lambda$. Different projections of the same galaxy are connected by lines, with dashed lines if the second point is lower on the y-axis than the first. Minor axis projections show very little rotation, while medial and major axis projections have similar amounts of rotational support, with medial axis projections being slightly more flattened.
\label{fig:L_e_prax}}
\end{figure*}

\begin{figure}
\includegraphics[width=0.49\textwidth]{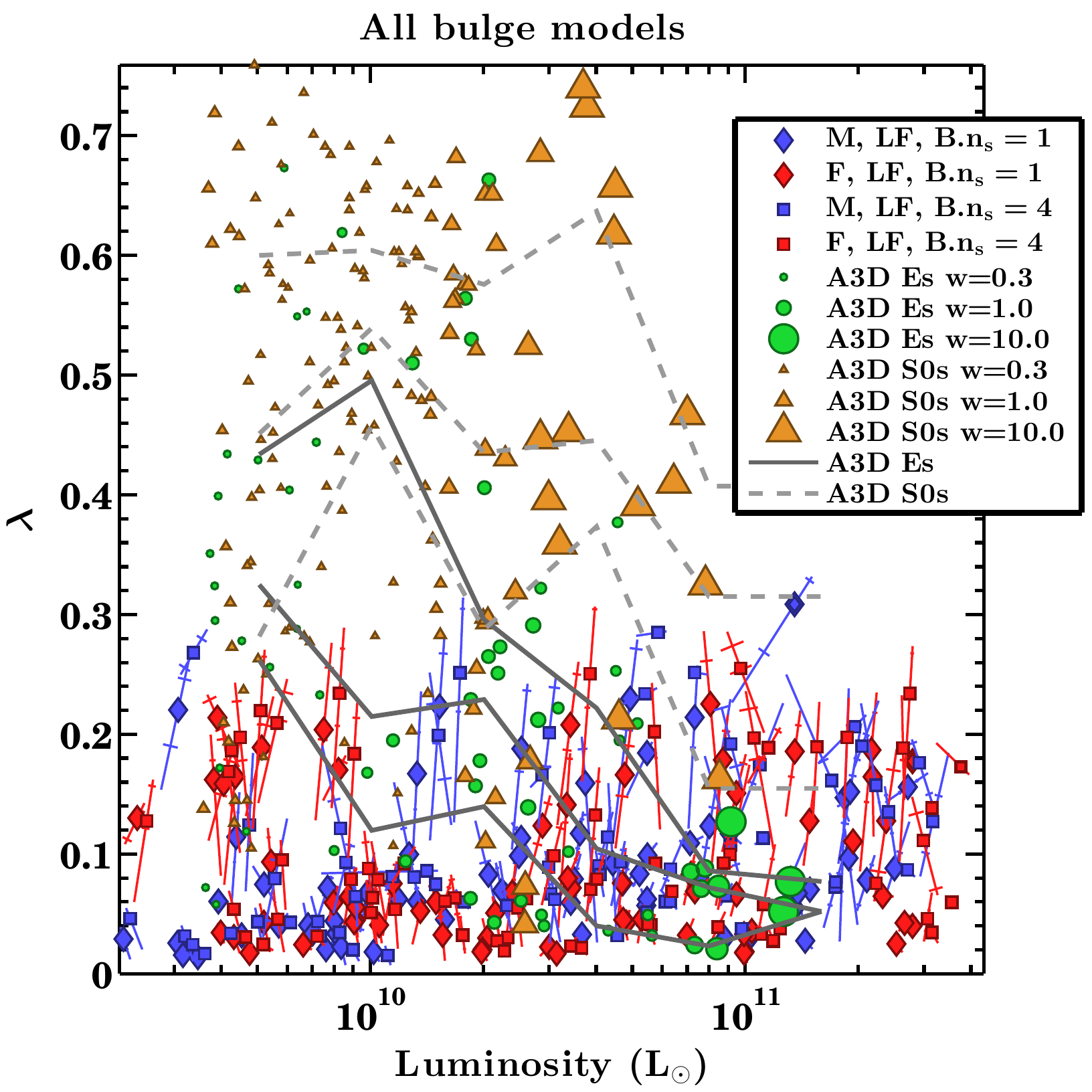}
\caption{Rotational support of elliptical galaxies by dimensionless angular momentum measure $\lambda$ as a function of luminosity. Observational data show a trend of lower rotational support at higher luminosity, whereas the simulated trend is nearly flat. Almost all S0s are faster rotators than simulated remnants.
\label{fig:l_L}}
\end{figure}

Rotational support decreases with increasing luminosity in A3D ellipticals but not in simulations, as shown in \figref{l_L}. This is largely due to the inability of dry mergers to produce fast rotating, faint ellipticals. Furthermore, even if the morphological properties of some remnants (particularly Sersic indices of B.n$_{s}$=1 remnants) are more consistent with S0s than ellipticals, observed S0s have far more rotational support than the vast majority of the simulated galaxies.

\begin{figure*}
\includegraphics[width=0.49\textwidth]{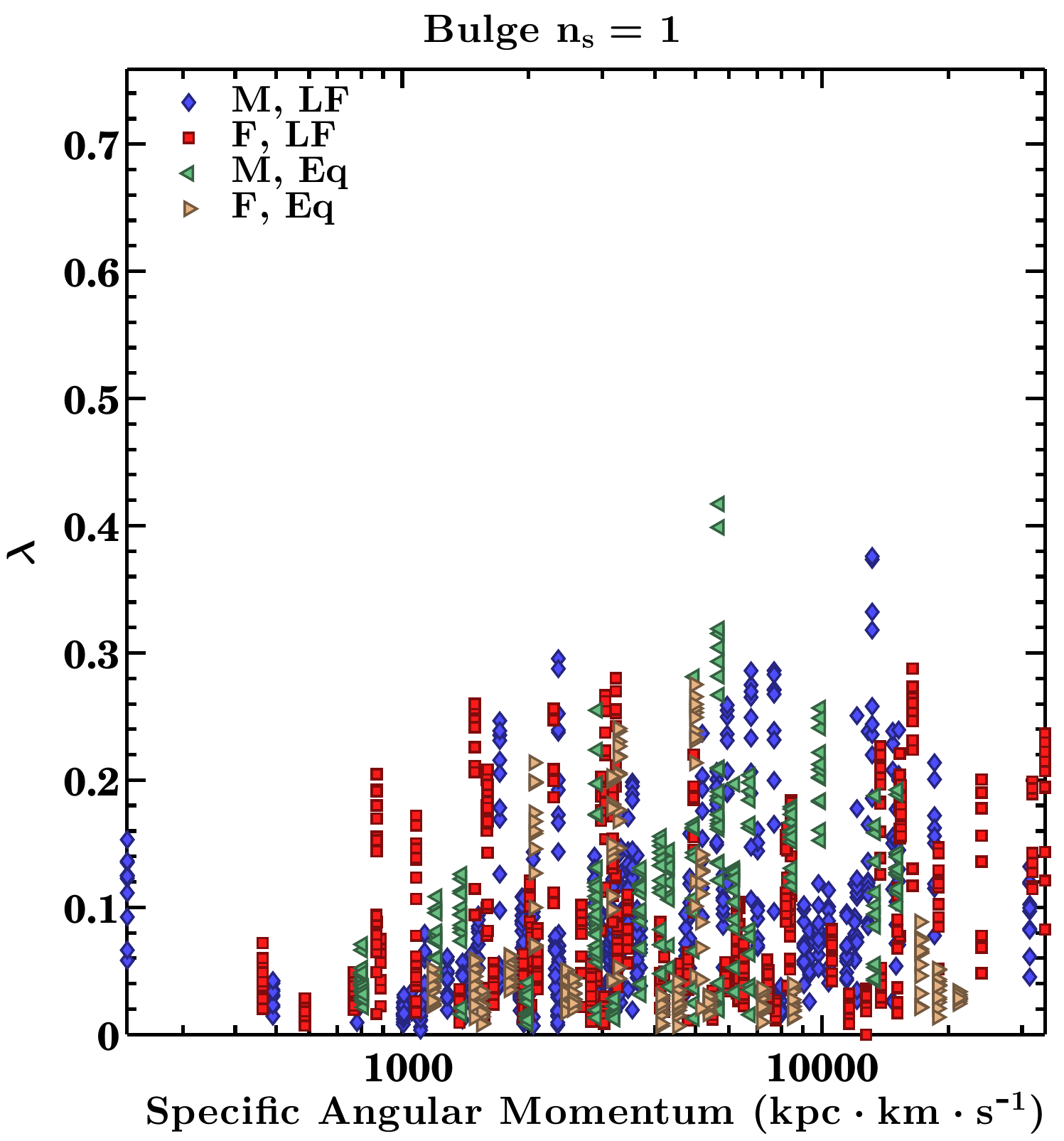}
\includegraphics[width=0.49\textwidth]{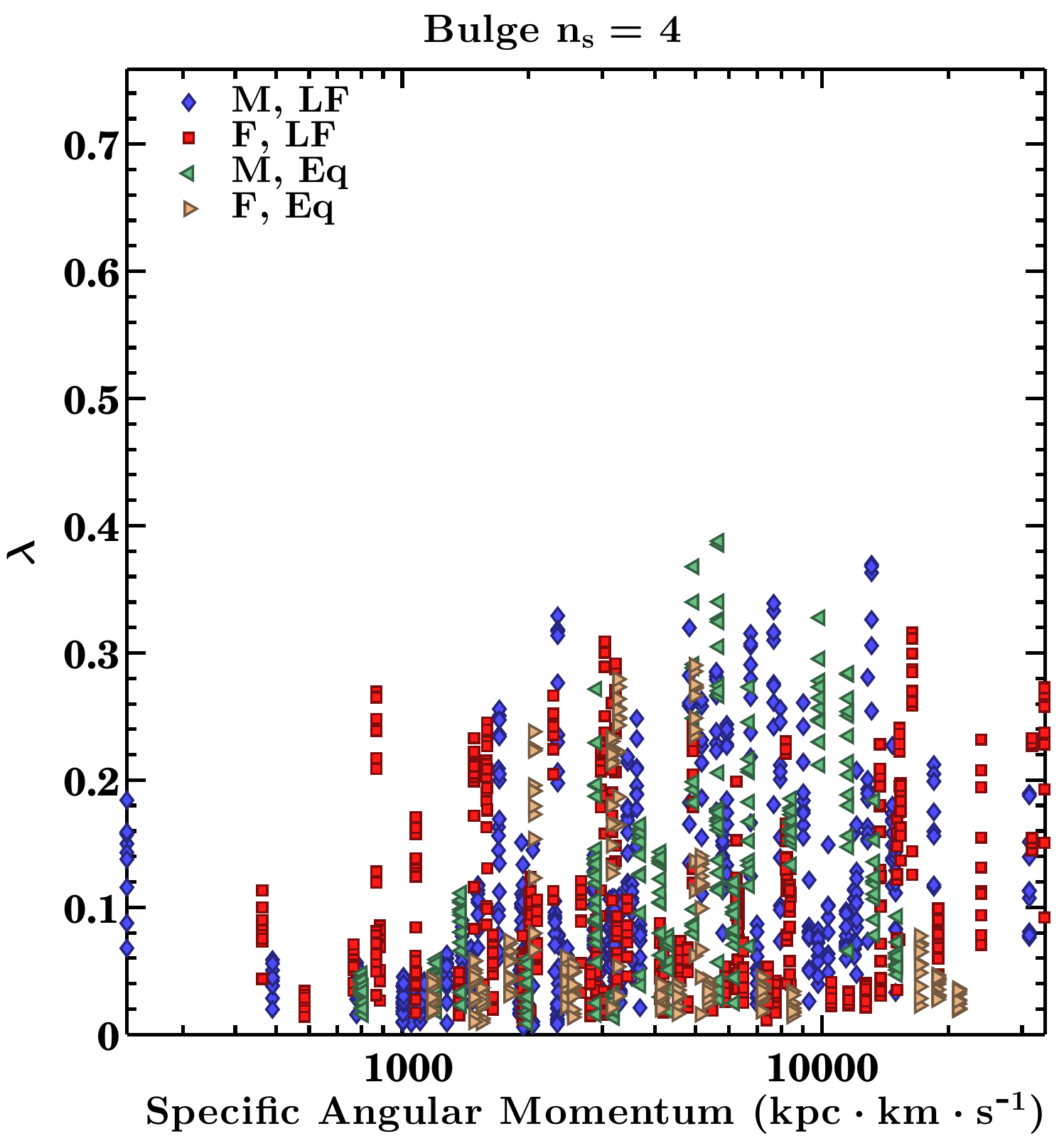}
\caption{Rotational support of simulated galaxies as a function of initial group orbital angular momentum per unit mass. For reference, M31's estimated orbit about the Milky Way with a tangential velocity of 30 \kms and a distance of 700 kpc would be amongst the smaller values in the sample.
\label{fig:angmom}} 
\end{figure*}

There does not appear to be any strong correlation between rotational support and number of mergers in \figref{e_vdivsigma}, or with total group mass or central galaxy luminosity. One might expect rotational support to at least correlate with net group specific angular momentum, assuming most of the halos merge and this angular momentum is conserved - however, \figref{angmom} does not show any such correlation. It appears that repeated, mostly isotropic mergers cannot produce very fast rotators, even if the group itself has some net orbital angular momentum in one or more satellite galaxies.

\section{Discussion}
\label{sec:discussion}

The main results of \secref{results} are that collisionless mergers in groups can produce central remnants with properties very similar to nearby elliptical galaxies. However, we do note several key differences between the simulation predictions and observed elliptical galaxies, not all of which are easily reconciled with dissipationless merging. We will also highlight how and why these results differ from previously published simulations.

\subsection{Morphology}

In \appref{analysis_testing}, it is shown that at the resolutions used in this study, luminosities, sizes and Sersic indices of spherical Sersic model galaxies can be recovered within about 5\%, usually underestimating the true values. For group merger remnants, Sersic fits typically recover luminosities and sizes to within 10\%, although luminosities tend to be more precisely recovered. By contrast, Petrosian radii systematically underestimate galaxy sizes and luminosities, negating the advantage of a non-parametric fit unless corrected for. Thus we conclude that single Sersic model fits are suitable for the simulated galaxies and can be compared directly to S+11 catalog fits, with the caveat that Sersic indices are the least robust parameter at low resolutions and are likely systematically underestimated. However, it is also true that in practice, Sersic fits can produce larger scatter on size than on luminosity, whereas Petrosian half-light radii appear to limit scatter in sizes - likely because they systematically underestimate the total luminosity of galaxies with large Sersic indices. Given these issues, our solution is to compare sizes between simulations and observations as fairly as possible, so that any systematic errors are likely to be shared between simulated and observed galaxies.

We have compared Sersic index and ellipticity distribution to single Sersic profile fits of local (A3D) and SDSS (N+10, S+11) galaxies. Neither B.$\mathrm{n_{s}}$ sample is a good fit to observed ellipticals alone, but the naive linear combination of the two is a better fit while remaining inconsistent with S0s. However, such a naive combination still produces a near-bimodal distribution, in contrast to the single peak typical of observed ellipticals. A more natural choice of progenitors would likely smooth out this bimodality. For example, groups with half of the spirals having exponential bulges and the other half de Vaucouleurs would likely produce remnants with intermediate properties, filling in the gap between the two peaks of single-progenitor distributions. A smooth, realistic distribution of bulge profiles and bulge fractions would likely flatten the peaks and further broaden the distribution of remnant Sersic indices.

Sersic indices of observed ellipticals are generally larger for more luminous galaxies, a trend reproduced by the simulated galaxies in \figref{l_sersicn}. \citet{HopHerCox09} predicted that the dissipationless component in mergers (including both binary spiral mergers and some re-mergers of the resulting remnants) should show only a weak increase with luminosity and have low median values of about $\mathrm{n_{s}}$=3, with scatter of about 1. We find similar results for the B.$\mathrm{n_{s}}$=1 sample, for which $\mathrm{n_{s}}$ is nearly constant with luminosity at a mean of 3 and with a range from 2 to 4. By contrast, the B.$\mathrm{n_{s}}$=4 sample not only has larger mean $\mathrm{n_{s}}$ at about 5, but the median $\mathrm{n_{s}}$ increases with luminosity by $\sim$0.5-1 per dex. This slope is close to that observed for N+10 and shallower than that in S+11, the discrepancy between these two samples having no obvious cause beyond probable contamination by S0s in the S+11 elliptical sample. Since the simulations have the same initial conditions other than their bulge profiles, this demonstrates that sufficiently concentrated progenitors can produce remnants with large $\mathrm{n_{s}}$ through dissipationless merging. Furthermore, bulge $\mathrm{n_{s}}$=4 mergers appear to be a better fit for luminous ellipticals, while bulge $\mathrm{n_{s}}$=1 remnants match the less luminous ellipticals. If progenitor bulge profiles scale with luminosity (i.e., luminous spirals have larger bulge $\mathrm{n_{s}}$ and merge to form luminous ellipticals), the scaling of elliptical Sersic index with luminosity can be matched more closely.

Both simulations and observations show a slight tendency for more luminous ellipticals to be rounder, especially above $10^{11}L_{\odot}$ (\figref{l_ell}). Again, the S+11 sample differs from N+10, in this case being more flattened on average - likely due to S0 contamination. Nonetheless, we also find similar luminosity-dependent behaviour as for Sersic indices, as the B.n$_{s}$=4 sample is a better match for bright, rounder ellipticals. The B.n$_{s}$=1 sample has a shallower slope and appears to be too flattened on average. The observed distribution does not obviously require a combination of both simulation samples in the same way as the Sersic index-luminosity relation does, but such a combination does not disagree with the S+11 relation either. The N+10 ellipticity-luminosity relation is considerably flatter and could be reproduced by the B.n$_{s}$=4 sample alone, with the scatter due to the substantial projection effects.

\subsection{Scaling Relations}

\subsubsection{Size-Luminosity Relation}

Despite having randomized initial conditions, the simulated galaxies typically produce tight size-luminosity relations, with slight dependence on which size measure is used and whether different B.n$_{s}$ samples are combined. The Sersic model relations are somewhat tighter ($\sim 0.1-0.12$ dex scatter) than those reported by \citet{NaiAbr11} ($\sim 0.12-0.15$ dex scatter). This is partly systematic, since \citet{NaiAbr11} used circular Sersic fits provided by \citet{BlaSchStr05}. Using the elliptical Sersic model fits of S+11 - which are more directly comparable to our own methodology and overlap with the N+10 sample - yields smaller scatter in the N+10 size-luminosity relation of $\sim 0.09$ dex. We also find slightly tighter scatter in the remnant Petrosian size-luminosity relation ($\sim 0.09$ dex), whereas the scatter for N+10 ellipticals remains largely unchanged whether Sersic or Petrosian model sizes are used.

The small scatter in the size-luminosity relation should allay concerns that stochastic merging processes cannot produce tight scaling relations. \citet{NipTreAug09} used simulations with multiple mergers of spheroidal galaxies to conclude that "a remarkable degree of fine tuning is required to reproduce the tightness of the local scaling relations with dry mergers". Instead, we find that mergers of many galaxies typically produce slightly tighter correlations than those with fewer galaxies, and the relations are tight regardless of which formation time is assumed for the groups (\appref{evolution}). No fine tuning in galaxy orbits, number of mergers or any other parameters are required to produce tight scaling relations. Moreover, the Faber-Jackson relation has even tighter scatter than the size-luminosity relation. Rather than scattering galaxies away from existing scaling relations, multiple mergers appear to converge remnants towards a common relation, a behavior somewhat like the central limit theorem. However, it is still true that dry mergers of spirals in groups produce remnants with larger sizes and smaller velocity dispersions at fixed mass or luminosity, a problem shared with mergers of spheroids \citep[e.g.][]{NipLonCio03,BoyMaQua06,NipTreBol09}. Also, the scatter does appear to increase slightly with luminosity. This could simply be a reflection of the wide range of galaxy and merger counts for the luminous groups, which may not match the true range of cosmological merger histories for galaxy groups.

We have tested mergers of spirals following a zero-scatter Tully-Fisher relation. The estimates for the scatter of merger remnant scaling relations can be considered lower limits, as they would likely have been higher had progenitors followed Tully-Fisher relations with intrinsic scatter and/or evolving slope and scatter. The observed Tully-Fisher relation does have significant scatter, even at low redshift (about 0.12 dex, from \citet{CouDutvdB07}), but the intrinsic scatter could be much lower. \citet{HopCoxHer08} estimate that a scatter of 0.1 dex in the Tully-Fisher relation contributes about 0.04 dex scatter in the Fundamental Plane scaling relation; comparable scatter added to the existing size-luminosity relation scatter of 0.10-0.12 dex would make little difference if added in quadrature.

While limiting scatter does not appear to be a challenge, in almost all cases the slope of the size-luminosity relation is shallower than observed and the intercept larger, so most galaxies are too large for their luminosities. The slopes of the remnant size-luminosity relations (typically $R \propto L^{0.5-0.6}$) are steeper than the progenitor spiral scaling relation ($R \propto L^{0.42}$) and the group scaling relation ($\rho$=constant, $R \propto L^{1/3})$. However, the remnants slopes are still shallower than those for the observations, which range from $R \propto L^{0.6}$ to $R \propto L^{0.8}$ depending on the observational sample and size measure. Encouragingly, the best matches are found between simulated remnants and N+10 ellipticals ($R \propto L^{0.66}$), the largest sample for which visual classifications are available.

The steeper slope of the S+11 elliptical Sersic size-luminosity relation (0.75 to 0.78) is of some concern. However, the elliptical classification for S+11 is based on empirical cuts on various parameters and results in significant ($\sim30\%$) contamination by S0s (\tabref{morphcuts}). The much smaller A3D elliptical sample also has a slightly larger slope than N+10, and the luminosity function weighting does not change the slope. Since A3D used a slightly different fitting methodology with a much smaller volume sample, it is not clear whether this discrepancy is significant.

The size-luminosity relation slopes for the simulated remnants are also steeper than that of $R \propto L^{\sim0.3}$ predicted for binary mergers remnants by \citet{HopHerCox09}. However, those simulations began with a spiral scaling relation of similar slope (0.3), and so the merging process did not steepen the size-mass relation. By contrast, we have shown that group mergers are capable of steepening the slope of the size-luminosity relation by $\sim 0.1-0.2$ from progenitors to merger remnants without dissipation.

Our models predict virtually no dependence of stellar mass-to-light ratio on luminosity - while the bulge and disk stellar mass-to-light ratios have different values, the fraction of disk stars within the effective radius varies little. However, luminous ellipticals do tend to have larger stellar mass-to-light ratios, so comparing to observed size-stellar mass relations lessens the discrepancies in the slopes by about 0.05 dex, depending on the sample and size measure. Such a dependence could be produced by more massive progenitor spirals having larger mean stellar mass-to-light ratios. We also did not include any scatter in the progenitor spiral Tully-Fisher relation or any scatter or luminosity dependence in bulge fractions. Extra scatter in either of these input galaxy properties would likely result in increased scatter in the remnant scaling relations. Any realistic luminosity dependence in the large M31 model bulge fraction would likely flatten the slope still further, since faint ellipticals would be produced by faint spirals with weak bulges. 

Dissipation is a tempting solution to the shallow size-luminosity relation slope problem. Dissipation should decrease sizes at fixed luminosities and preferentially shrink faint ellipticals if their progenitors had larger gas fractions, resulting in a remnant with a larger fraction of stars formed in a central starburst. Luminosity-dependent gas fractions have been proposed by \citet{RobCoxHer06,HopCoxHer08} as the source of the tilt in the fundamental plane scaling relation, a hypothesis which will be addressed in Paper II.

Another possible remedy to the shallower slopes of the simulated size-luminosity relations is to weight the contributions from various simulation subsamples differently. Applying a simple linear weighting scheme of favoring B.n$_{s}$=1 groups at low luminosity and B.n$_{s}$=4 at high luminosities yields a steeper slope than a uniform weighting and a closer match with observations. Such a weighting also produces steeper slopes than either the Few- or Many-merger relations alone and can be justified if more massive halos undergo more mergers. While average halo merger rates are not strongly mass dependent \citep{SteBulWec08,FakMaBoy10}, the groups we have simulated here would likely be those with higher than average merger rates.

Although these schemes could resolve the mismatch in slopes, none save dissipation are viable solutions to the problem that simulated remnants are generally too large at fixed luminosity (\figref{sizelum}). Our estimated stellar mass-to-light ratios are already quite low, so making small galaxies brighter appears to be out of the question. Numerical resolution effects are not large (\appref{numerics}). Barring a strong redshift dependence in the sizes of observed ellipticals, this discrepancy is real. As a result, the Kormendy relation (\figref{kormendy}) is poorly reproduced. Remnants are too faint at fixed sizes, so their effective surface brightnesses are also too faint by about a magnitude for small galaxies. The shallower slopes of the size-mass relation also translate into a weak simulated Kormendy relation (nearly non-existent in the case of the B.n$_{s}$=1 sample).

\subsubsection{Faber-Jackson Relation}

The Faber-Jackson relation of simulated galaxies shows even smaller scatter (0.04 dex) than their size-luminosity relation or any observed Faber-Jackson relation (typically 0.08 dex, as in \figref{fj} and \tabref{fj}). The simulated remnants also have slightly shallower slopes ($\sigma \propto L^{0.28}$) than the observations ($\sigma \propto L^{0.27-0.37}$), again depending on sample and weighting scheme. Curiously, the slope of the Faber-Jackson relation is nearly identical to that of the progenitor spiral Tully-Fisher relation ($V \propto L^{0.29}$), so multiple mergers appear to preserve the scaling of orbital velocity with mass while converting ordered rotation into random motions. This is despite the fact that the virial ratio in each group varies significantly, and so galaxy orbits within each group are not scaled uniformly the same way that stellar orbits within galaxies are.

In virtually all cases, the slope of the remnant Faber-Jackson relation is steeper than the canonical value of 0.25 (or L$\propto \sigma^{4}$). However, the observed relations show similar deviations and there is no compelling reason why ellipticals should follow this canonical relation. Indeed, the simulations of \citet{BoyMaQua06} predict scalings as steep as M$\propto \sigma^{12}$ for major mergers with very small pericentric distances, so such deviations from the canonical relation are not unexpected.

In most samples, the simulations have smaller dispersions than observed galaxies of the same luminosity. No weighting scheme can resolve this mismatch in the intercept of the Faber-Jackson relation, which is of similar magnitude (but opposite sign) as the offset in the intercept of the size-luminosity relation. Increasing the stellar mass-to-light ratios of the simulations would make galaxies of the same dispersion fainter but would worsen the match to the size-luminosity relation by making small remnants even fainter. Dissipation appears to be necessary here - central starbursts have been shown to increase velocity dispersions and shrink effective radii compared to purely dissipationless mergers \citep{HopHerCox09}. However, it is not clear whether a mass-dependent gas fraction would preserve the slope or flatten it. The mild curvature in the observational relations may be a systematic effect at low dispersions, although we have attempted to minimize such systematics by including two independent dispersion measurements. On the other hand, the simulated relations are insensitive to the choice of velocity dispersion measure (central or effective; including rotational support or not) and most observed relations are insensitive to various weighting schemes.

\subsubsection{Time or Redshift Dependence}

All of the results presented above apply to simulations analyzed after 10.3 Gyr, assuming an initial formation redshift of z=2.0 - a redshift at which pure dry mergers of disks are not likely to be common. However, the first merger in the group typically only occurs after another 1-2 Gyr. The scaling relations of remnants after 5 and 7.7 Gyr are similar to those at 10.3 Gyr, as shown in \appref{evolution}, and so similar conclusions would be reached by assuming that the first merger occurred at z=0.5, when mergers were more likely to be dry or gas-poor. At face value, this also implies that the evolution in scaling relations is minimal; however, we caution that all of the groups are effectively the same age, so this prediction does not include any evolution from varied ages and assembly histories of real group galaxies.

\subsection{Spiral Progenitors and Their Bulges}

In the case of Sersic index distributions and scaling relations, it is tempting to consider whether a combination of progenitor bulge types (and possibly bulge fractions) could resolve the tensions with observations. To examine this further it is useful to ask what the distributions of bulge Sersic index and bulge fraction are for spirals as a function of luminosity.

\begin{figure}
\includegraphics[width=0.49\textwidth]{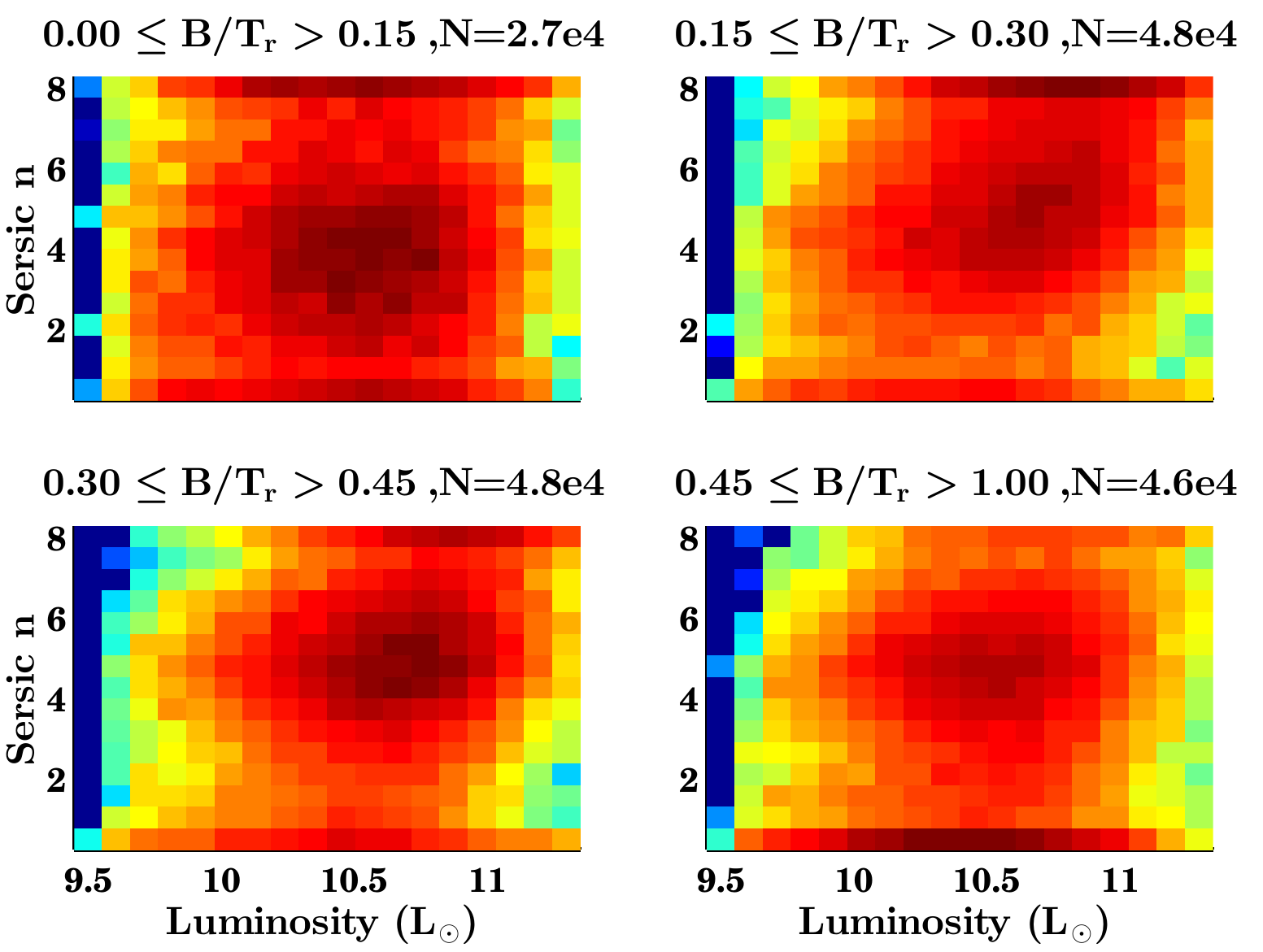}
\caption{Relations between total luminosity and best-fit bulge Sersic index as a function of r-band bulge-to-total luminosity ratio in S+11 spiral galaxies. Plots show probability density on a logarithmic scale, with dark red the highest and blue the lowest densities. Only galaxies with a distinct bulge component are included. All galaxies have an F-test probability that a bulge component is not required for a good fit of less than $0.32$. Each galaxy is weighted by 1/V$\mathrm{_{max}}$ to correct for incompleteness.
\label{fig:spiralbulges2}}
\end{figure}

\begin{figure}
\includegraphics[width=0.49\textwidth]{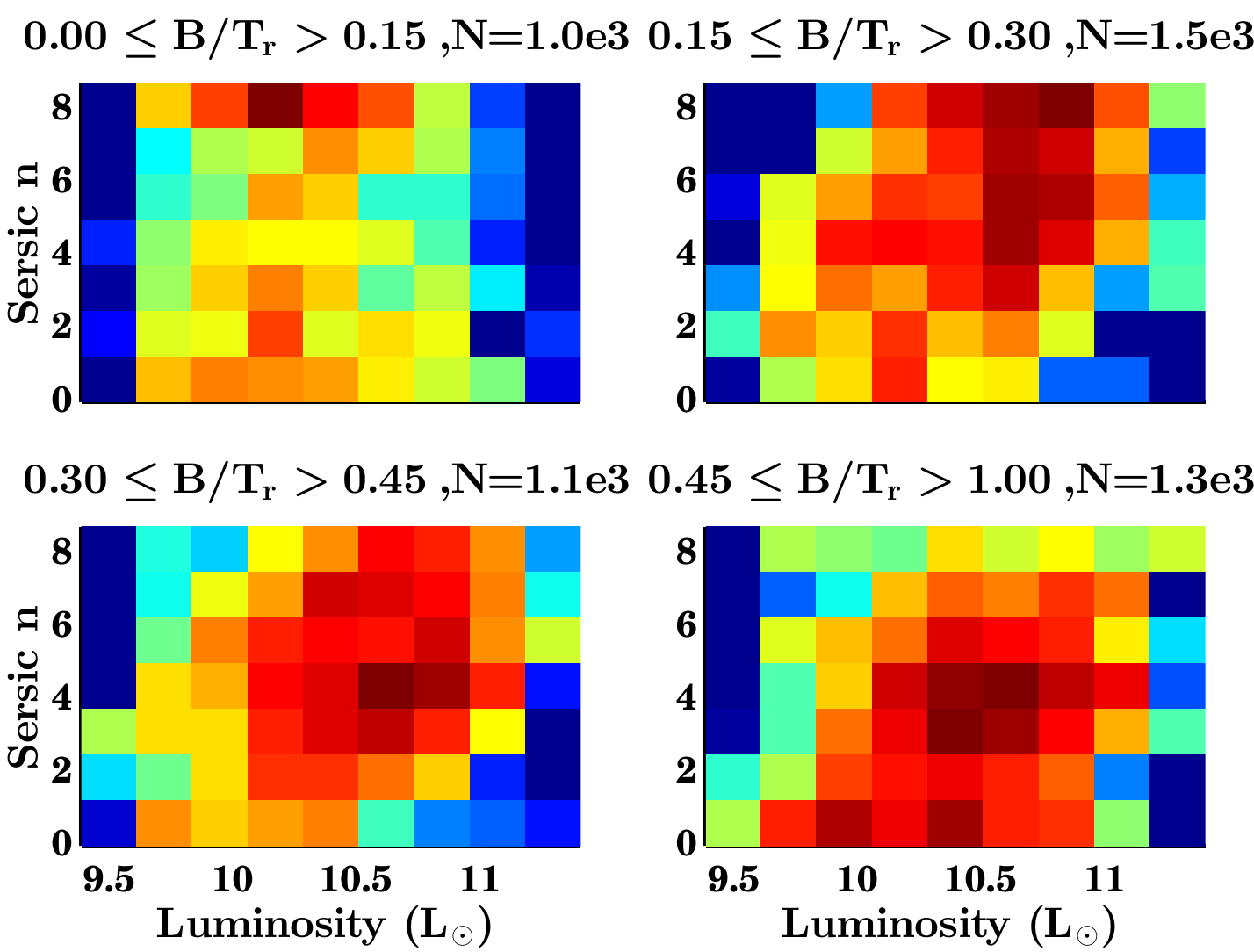}
\caption{Spiral galaxy bulge properties as in \figref{spiralbulges2}, but now for visually classified N+10 spirals. Although the statistics are barely sufficient, there does appear to be a weak correlation between luminosity and bulge Sersic index for spirals with large bulge fractions, more so than in \figref{spiralbulges2}.
\label{fig:spiralbulges2n}}
\end{figure}

Not all S+11 spirals have a distinct bulge, nor are most images of sufficient quality to accurately measure bulge properties, so we consider the subset for which a bulge plus disk fit is required - that is, those with an F-test probability that a de Vaucouleurs bulge is not required is less than 0.32. This is about half of the spiral sample. The proportion for which a free Sersic bulge is required over a de Vaucouleurs bulge is much smaller, so we do not limit the sample any further. \figref{spiralbulges2} shows the probability densities of bulge Sersic indices as a function of galaxy luminosity, split into different bulge fraction bins. In all bins, classical ($\mathrm{n_{s}}$=4) bulges are at least a local maximum, although extreme bulge Sersic indices ($\mathrm{n_{s}}$=0.5 and $\mathrm{n_{s}}$=8, which are the lower and upper limits for S+11) are often the most common. The dependence on luminosity is not very strong, but in most bulge fraction bins, fainter spirals are slightly more likely to have low Sersic index bulges than high.

The S+11 spiral sample is known to be contaminated by S0s. In \figref{spiralbulges2n}, we instead use the much smaller but visually classified sample of spirals from N+10. This smaller sample does slight evidence for correlation between luminosity and bulge Sersic index, at least for more bulge-dominated spirals. Also, the large fraction of bulge Sersic indices below 1 is greatly diminished, suggesting that those could be primarily S0 contaminants in the S+11 sample, or possibly more poorly resolved, higher-redshift spirals which appear in S+11 but not N+10. In either case, both samples contain substantial fractions of spirals with large bulge fractions.

The M31 model used in our simulations has a large bulge mass fraction (0.33) and luminosity fraction (0.5). Such fractions are not uncommon, even at low luminosities. de Vaucouleurs bulges are also quite common, whereas exponential bulges are at least not exceptionally rare, especially for bulge-dominated spirals. Even if groups of spirals have broad distributions of bulge  profiles, as in \figref{spiralbulges2}, their median values could also lie close to the limiting cases of exponential or de Vaucouleurs in our simulations. Also, a wide distribution of bulge profiles is indeed a realistic solution to the problem of single-progenitor mergers producing remnants with narrow Sersic index distributions. Real mergers in groups would likely produce wider, less bimodal distributions of Sersic indices than the single-progenitor simulations.

\subsection{Rotational Support}

The abundance of fast-rotating, faint ellipticals is at odds with the simulation predictions. However, as Figs. \ref{fig:e_vdivsigma} and \ref{fig:e_L} show, multiple mergers can product remnants with moderate rotational support. This contrasts with the results of \citet{CoxDutDiM06} that dissipationless binary mergers only produce slow-rotating remnants with $\mathrm{v/\sigma < 0.15}$ - all the more so because \citet{CoxDutDiM06} measured major axis rotation curves, whereas our simulations (and A3D) average over $R_{eff}$. Nonetheless, our simulations are unable to produce any remnants with $\mathrm{\lambda > 0.35}$. The simulated remnants show little or no change in rotational support as a function of luminosity (\figref{l_L}), unlike observations, and do not produce any of the fast-rotating, moderately luminous S0s found in A3D. There is also an abundance of flattened remnants with minimal rotation, unlike in Atlas3D. \citet{CoxDutDiM06,BoiEmsBou11} and others have shown that significant rotation can be easily produced in gas-rich mergers. Dissipation is likely necessary to produce some ellipticals, particularly faint ones, and most likely a large fraction of S0s - if S0s are formed through mergers. However, it should emphasized that many of the simulated galaxies are consistent with the properties of some A3D galaxies, particularly bright ellipticals, so dissipation may not be necessary in all cases.

\section{Conclusions}
\label{sec:conclusions}

We have investigated the hypothesis that elliptical galaxies can form through collisionless mergers of spiral galaxies by creating a sample of numerical simulations of such mergers and comparing the results directly with observations of local ellipticals. We draw the following key conclusions:

\begin{enumerate}
\item For a given fixed bulge type, central remnants have narrow distributions of Sersic indices, with mergers of spirals with exponential bulges producing less concentrated remnants ($\sim \mathrm{n_{s}}$=3) than classical-bulge merger remnants ($\sim \mathrm{n_{s}}$=5). Although classical-bulge mergers alone are a better fit than exponential-bulge mergers, a combination of progenitor bulge profiles is required to reproduce observed Sersic index distributions.
\item Classical-bulge mergers produce a correlation between luminosity and Sersic index, whereas exponential bulge mergers do not. The observed correlation is best reproduced if exponential bulge mergers preferentially produce faint ellipticals and classical bulge mergers produce bright ellipticals.
\item Every simulation sample produces tight scaling relations, with approximately 0.1 dex scatter for the size-mass relation and 0.04 dex scatter in the Faber-Jackson relation. Thus, even multiple dry mergers can produce ellipticals with exceptionally tight scaling relations. However, the scatter estimates represent a lower limit, because the progenitor spirals in our simulations follow a fixed, zero-scatter Tully-Fisher relations. The scatter in the remnants scaling relations would likely increase (though not necessarily significantly) if the progenitor scaling relations had larger intrinsic scatter or evolved with redshift.
\item The remnant size-luminosity relation typically has a shallower slope ($R \propto L^{0.5-0.6}$) than observed relations ($R \propto L^{0.6-0.8}$), depending on the sample and weighting scheme used. The simulated slope is also steeper than that of the progenitor spiral size-luminosity relation ($R \propto L^{0.42}$), suggesting that mergers can steepen the size-luminosity relation.
\item As a consequence of the shallower slopes and larger intercepts of the simulated size-luminosity relation, the simulated Kormendy relation is shallower than observed - nearly flat for exponential bulge mergers - and has larger scatter.
\item The remnant Faber-Jackson relation has a slightly shallower slope ($\sigma \propto L^{0.28}$) than most of the observed relations ($\sigma \propto L^{0.27-0.37}$), but is virtually unchanged from the progenitor spiral Tully-Fisher relation, $V \propto L^{0.29}$.
\item The slopes of the scaling relations can be better reproduced if massive ellipticals are produced by many mergers and less massive by fewer mergers, or if stellar mass is compared instead of luminosity.
\item The intercepts of the size-mass and Faber-Jackson relations can be individually matched by adjusting the stellar mass-to-light ratios of the galaxies; however, each relation requires adjustment in the opposite sense (remnants of a fixed luminosity being too large and having too low of a velocity dispersion), so it is not possible to match both intercepts simultaneously.
\item Multiple mergers can produce remnants with modest rotational support (v/$\sigma >$ 0.1); however, most remnants are slow rotators, and there is no correlation between luminosity and v/$\sigma$, whereas such a correlation is found in Atlas3D ellipticals.
\end{enumerate}

These results demonstrate that many of the properties of elliptical galaxies are consistent with their emergence through multiple dry mergers of spiral galaxies. Perhaps most importantly, these properties also differ significantly from those of remnants formed through binary dry mergers of spirals, as reported in previous studies. This not only adds to an increasing body of evidence supporting the case for multiple mergers \citep[e.g.][]{BouJogCom07,NaaJohOst09,TruFerdeL11,HilNaaOst12} but also demonstrates that such mergers can produce tight scaling relations - in some cases tighter than observed ellipticals - as long as the progenitor spirals are drawn from a realistic luminosity function and scaled appropriately. 

Several major concerns remain for a purely dissipationless formation scenario for elliptical galaxies. The first is the limited amount of rotational support in the merger remnants and the absence of any correlation between rotation and luminosity. The second is the large sizes (and low velocity dispersions) of faint ellipticals, which result in a shallow size-luminosity relation and poorly reproduced Kormendy relation. While this second point could be resolved without dissipation (e.g. by merging more compact disks at high redshift), dissipation does appear to be necessary to produce fast-rotating ellipticals. Dissipation could also solve the second problem, as central starbursts would produce more compact remnants with higher dispersions.

Perhaps the greatest challenge for dry mergers lies in matching the tilt of the fundamental plane with respect to the virial relation. Previous work has suggested that dissipational processes are the cause of this tilt and that dry mergers cannot produce any tilt \citep{RobCoxHer06,HopCoxHer08}. This point will be addressed in Paper II of this series \citep{TarDubYee13b}.

\section{Acknowledgments}
\label{sec:acknowledgments}

D.T. would like to thank B. Abraham, L. Bai, D. Krajnovic, T. Mendel and P. Nair for fruitful discussions and for providing data used herein, as well as the anonymous referee for helpful suggestions. D.T. acknowledges the support of Ontario Graduate Scholarships for this work. Simulations and analyses were performed on the Canadian Institute for Theoretical Astrophysics' Sunnyvale cluster and the University of Toronto's SciNet cluster. H.Y. acknowledges support from grants from the National Science Engineering Research Council of Canada and the Canada Research Chair program.

\appendix
\section{Analysis Pipeline Testing}
\label{app:analysis_testing}

Several aspects of the simulation analysis pipeline merit further testing. First, we would like to determine if the pipeline can recover known or measurable quantities such as the total mass/luminosity and half-light radii in single galaxies. This is accomplished by analysing a sample of spherical, pure Sersic profile plus dark matter halo galaxies generated with GalactICS. This allows us to simultaneously test whether GalactICS can generate equilibrium Sersic profile models (which is how the bulges of progenitor spirals are initialized) and whether the analysis pipeline can successfully recover input parameters at arbitrary resolutions. We analyse these models before simulating them in any way. In ~\appref{numerics}, we examine the results of simulating these simple models with PARTREE to test numerical convergence.

We can also use the group simulations themselves to test the analysis pipeline. Although we do not know the structural parameters of merger remnants a priori - indeed, they do not necessarily follow a single Sersic profile at all - the total luminosity is known in groups which have merged to a single remnant. Similarly, we can directly measure a half-light radius from mock images with no PSF or sky background in these cases and compare to observational estimates from the SDSS-equivalent mock images. This procedure allows us to determine whether single Sersic profile fits can simultaneously recover the total luminosity of a galaxy and its half-light radius.

In addition to Sersic fits from GALFIT, we fit de Vaucouleurs profiles with GALMORPH and measure non-parametric Petrosian radii to determine if these size measures can consistently recover the true half-light radius of a galaxy.

\subsection{Sersic plus Halo Models}

Our reference Sersic plus halo models consist of a single Sersic profile bulge and a dark halo with the same baryonic mass ratio as our fiducial M31 models. We produce $\mathrm{n_{s}}=2$ and $\mathrm{n_{s}}=4$ models to cover most of the range of typical elliptical surface brightness profiles. We create models with $R_{eff}$ of 2, 4, 8 and 16 kpc, again covering ranges of typical elliptical galaxies and massive spiral bulges. The 2-kpc model is slightly larger than the 1.5-kpc bulge in our fiducial M31 model. All models are in virial equilibrium and follow a size-luminosity relation $\log(R_{eff})=0.7\log(L_{r})$, with one model exactly on this relation and an extra model either over- or under-luminous for its size. Each model is imaged at mock redshifts of 0.01, 0.025 and 0.1. These models will be used in the future to test recovery of scaling relations. However, for now we are mainly interested in whether the pipeline can recover the known values of $\mathrm{n_{s}}$, $R_{eff}$ and $L$ for each model and whether the systematics depend on any of those parameters.

In addition to varying the galaxy luminosity as a function of size, each set of Sersic plus halo models is simulated at three resolutions. The lowest resolution has 15,000 star and 40,000 dark matter particles, identical to the lowest resolution model used in the simulations. The resolution increases by a factor of 8 each step such that the highest resolution model has 7,680,000 star and 20,480,000 dark particles, or at least a factor of two more than the total particle counts of the most massive group simulations. In principle these models should be rescaled versions of each other; however, the nominal SDSS PSF and signal-to-noise ratio set a physical scale for mock images, while our fixed softening length sets another physical scale for the simulation.

\subsection{Sersic Quantities}

For $\mathrm{n_{s}}=2$ models imaged at z=0.025, GALFIT Sersic fits show excellent agreement with expectations, even at low numerical resolution. With just 15,000 star particles, sizes are recovered to within $1\pm0.5\%$ for 2 kpc radius, although larger galaxies have underestimated sizes to the level of $3\pm1\%$ at $R_{eff}=16$ kpc. However, $\mathrm{n_{s}}$ is underestimated by $10\%$ for $R_{eff}=2$ kpc galaxies, which improves to $4\pm1\%$ at $R_{eff}=16$ kpc. Luminosities, in turn, are underestimated at fairly constant levels of $3\%$, with standard deviations increasing with size from $0.1$ to $1\%$. Similar trends are found at medium resolution but with smaller amplitudes - the largest errors on $\mathrm{n_{s}}$ are just $1.4\pm0.3\%$ at 16 kpc, while errors on $R_{eff}$ are at most $3.5\pm0.5\%$ at 2 kpc and shrink to half of that value at 2 kpc.

Errors on parameters are reduced by about a factor of two by imaging at nearby redshifts (z=0.01) and increase by about the same factor by imaging at z=0.1. These errors are not eliminated by increasing the image size (and shrinking the PSF relative to $R_{eff}$) but shrink dramatically at the highest numerical resolution, to well under $1\%$ in $\mathrm{n_{s}}$ and $L$ and about $1\%$ in $R_{eff}$. This suggests that these parameters are in principle completely recoverable with SDSS-equivalent imaging and good sky subtraction.

We have also fit $\mathrm{n_{s}}=4$ de Vaucouleurs profiles using GALMORPH, as in \citet{HydBer09a}. GALMORPH fits to the $\mathrm{n_{s}}=2$ models show expectedly poor results. Sizes are overestimated by factors from 1.6 (at 2kpc) to 2.4 (at 16kpc) while luminosities are overestimated by $30-60\%$. These results are not entirely unexpected -- $\mathrm{n_{s}}=4$ models have shallower outer profiles and hence more light at large radii compared to profiles with lower $\mathrm{n_{s}}=2$. However, they do demonstrate that other free parameters such as the effective radius and mean surface brightness cannot adjust to compensate for an incorrect profile choice, and so pure de Vaucouleurs profiles are not a good choice to fit ellipticals if their underlying surface brightness profiles are truly Sersic profile with $\mathrm{n_{s}}$ significantly lower than 4.

For $\mathrm{n_{s}}=4$ models, GALFIT free-$\mathrm{n_{s}}$ fits show curiously constant fractional errors on sizes, consistently underestimating $R_{eff}$ by $8-9\pm1\%$. Underestimates of $\mathrm{n_{s}}$ vary from a substantial $14\pm1\%$ at 2 kpc to $4\pm0.6\%$ at 16 kpc. Luminosity underestimates shrink from $5\pm0.6\%$ to $1.5\pm0.3\%$. These errors are not improved by imaging at lower redshift and so are unrelated to the relative size of the PSF. Instead, they are reduced substantially by increasing numerical resolution. At the highest resolution, size estimates shrink to $5-6\pm1\%$ at all sizes, while $\mathrm{n_{s}}$ underestimates now scale from $8-2\%$.

Since typical resolutions for central group galaxies are at the medium level or slightly higher, we expect that $\mathrm{n_{s}}$ is underestimated by $7\%$ on average for a pure de Vaucouleurs fit, with better performance at smaller $\mathrm{n_{s}}$. Size estimates are a minimum of $2\%$ lower for large galaxies and up to $10\%$ off for $R_{eff}=2$kpc.

By contrast, the GALMORPH fits with fixed $\mathrm{n_{s}}=4$ accurately recover sizes and luminosities to better than $2\%$ even at the smallest sizes and at medium resolution. We conclude that even in ideal situations, free-$\mathrm{n_{s}}$ fits will systematically underestimate sizes and luminosities at the 5\% level, whereas fixed-$\mathrm{n_{s}}$ fits only perform better if the exact value of $\mathrm{n_{s}}$ is known. In \appref{numerics} we detail how these results change after 2 Gyr of simulation with PARTREE using a fixed 100 pc softening length, as in the simulations.

\subsection{Group Simulation Results}

\begin{figure*}
\includegraphics[width=0.5\textwidth]{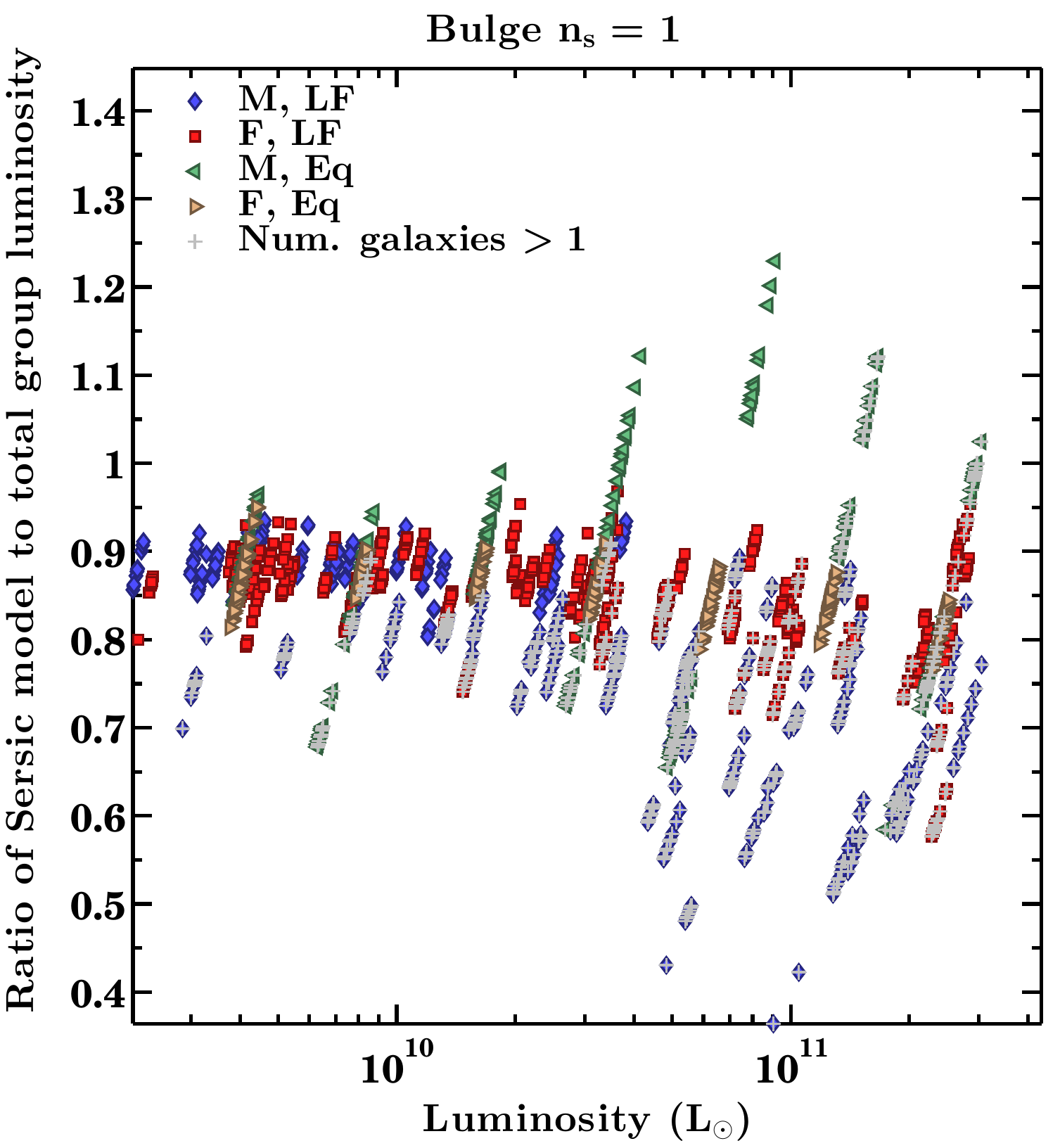}
\includegraphics[width=0.5\textwidth]{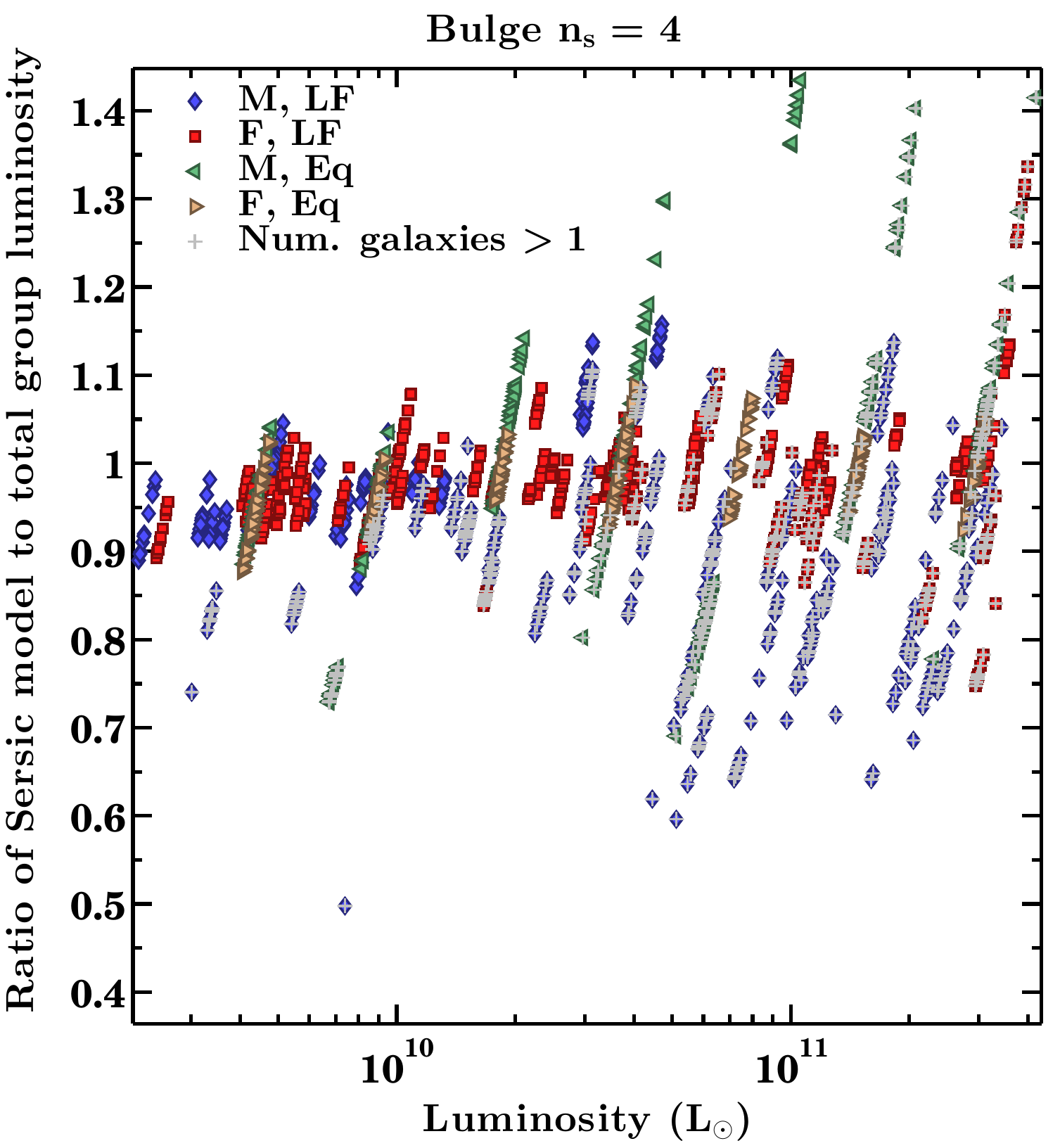}
\caption{Ratio of Sersic model luminosities to total group luminosity. Groups with multiple galaxies are highlighted, since the fraction contained in the satellites is not well-constrained. Mergers of equal-mass spirals ('Eq') tend to show the largest deviations from unity.
\label{fig:sersiclrat}}
\end{figure*}

As \figref{sersiclrat} shows, Sersic models generally do an acceptable job recovering central galaxy luminosities. For the $\mathrm{n_{s}}=1$ sample, model luminosities are typically 85-90\% of the total in groups with no satellites with relatively small scatter. The Sersic luminosities of $\mathrm{n_{s}}=4$ central galaxies appear to have little or no systematic deviations from the true luminosities, although the scatter appears somewhat larger than in the $\mathrm{n_{s}}=1$ case. The largest discrepancies are found for groups with many mergers, particularly equal mass mergers, in which case the models can overestimate the central galaxy's luminosity by at least 20-30\%, largely due to runaway growth of the effective radius and Sersic index. However, in most cases Sersic profiles appear to be appropriate fits to the galaxies. The underestimation of $\mathrm{n_{s}}=1$ merger luminosities appears to be a systematic effect. The underestimation of luminosities in groups with satellite galaxies is difficult to quantify, as the total luminosity in satellites is not easily separable from that of the central galaxy.

\begin{figure*}
\includegraphics[width=0.5\textwidth]{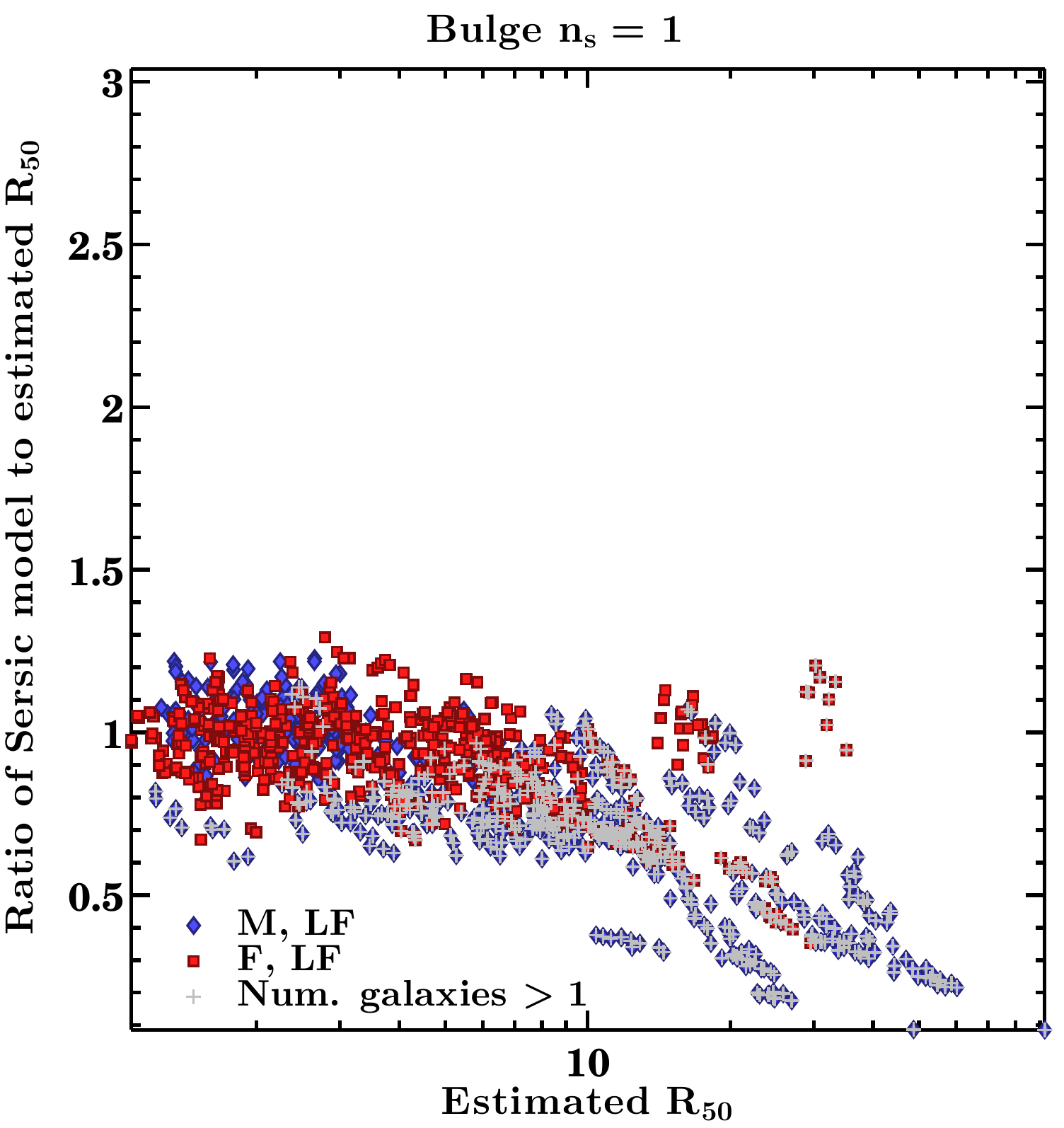}
\includegraphics[width=0.5\textwidth]{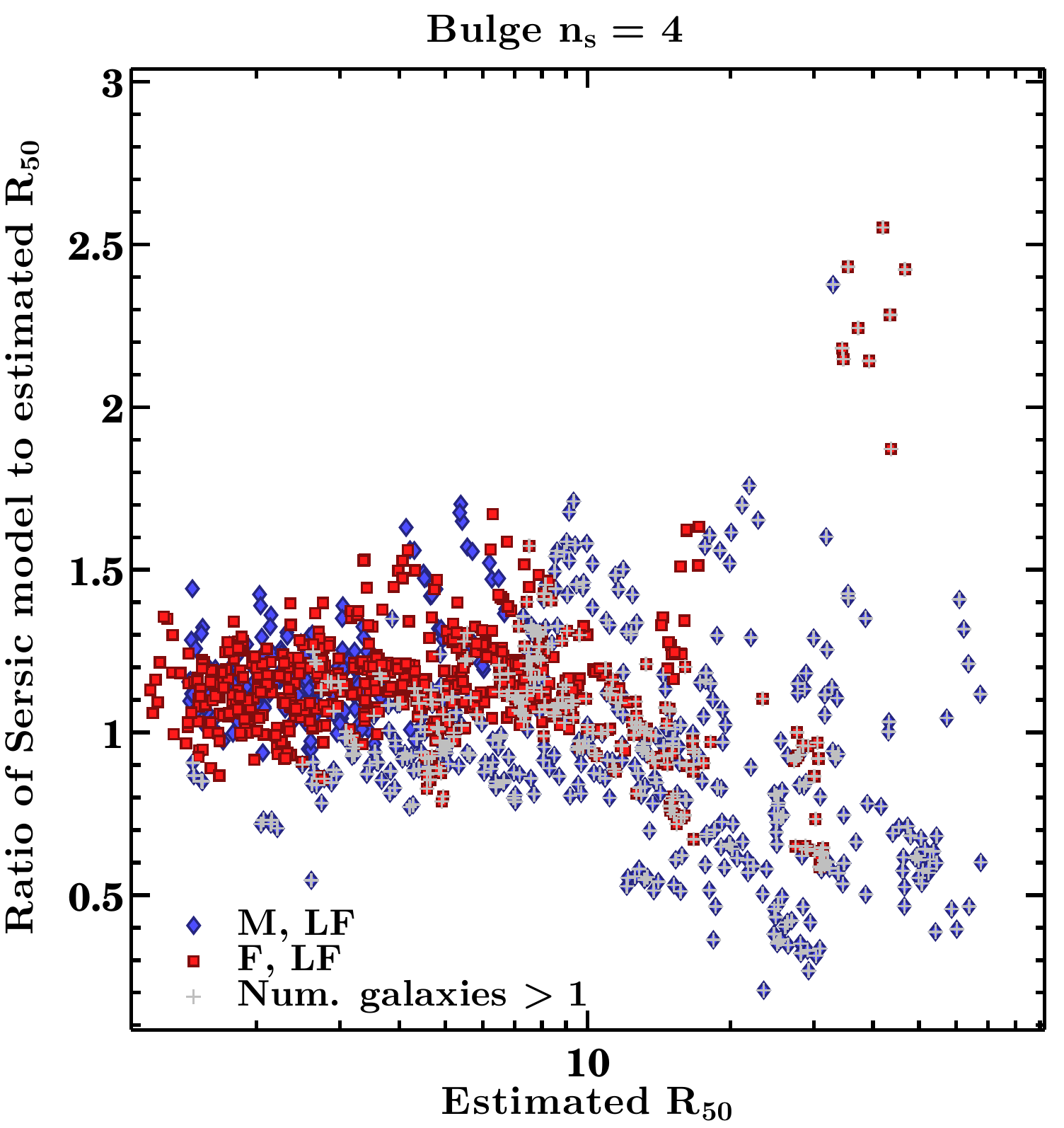}
\caption{Ratio of Sersic model effective radii to estimated half-light radii. Half-light radii are estimated by measuring the radius at which the enclosed light in a sky- and satellite-subtracted image equals half of the total group luminosity, and so are strictly larger than a true half-light radius if there are satellite galaxies. Regardless, Sersic half-light radii show considerably larger scatter relative to the 'true' half-light radius than do Sersic model luminosities to the total group luminosity (\figref{sersiclrat}).
\label{fig:sersicrrat}}
\end{figure*}

Testing whether half-light radii are recovered is also complicated by the presence of satellite galaxies. Nonetheless, we attempt to measure how closely $R_{eff}$ matches the 'true' half-light radius $R_{50}$ of the central galaxy in \figref{sersicrrat}. We estimate $R_{50}$ as the radius enclosing half of the group luminosity in a given sky- and satellite-subtracted image, using the same best-fit ellipse as the Sersic model. The ratio should be unity if there are no satellite galaxies in the group and less than unity if there are. As \figref{sersicrrat} shows, half-light radii are more difficult to measure than total galaxy luminosities - or rather, errors on half-light radii from Sersic fits are considerably larger than for luminosities, which likely contributes to the significant scatter in the Sersic size-luminosity relation compared to the Faber-Jackson relation. 

At first glance, the large scatter in the ratio of Sersic model to 'true' half-light radius might suggest that much of the error in the Sersic size-luminosity relation is due to systematics rather than any intrinsic scatter. However, the size-luminosity relation using total group luminosity and 'true' half-light radius still shows significant scatter (0.08 dex) even when limited to galaxies with no satellites. A much larger sample of higher-resolution simulations would be required to determine if this scatter is due to numerical effects or genuinely intrinsic.

\subsection{Petrosian Radii}

As in SDSS, the Petrosian radius $R_{P}$ is given by the radius at which the mean surface brightness in the ring bounded by $0.8 R_{P} < r < 1.25 R_{P}$ is 0.2 times the mean surface brightness within $R_{P}$. As a non-parametric size measure, it requires no fitting to measure, unlike the Sersic $R_{eff}$. Since the Sersic profile is an analytical solution, one can compute $R_{P}$ uniquely for any given $\mathrm{n_{s}}$. For $\mathrm{n_{s}}=$ 3 to 6, $R_{P}/R_{eff}$ ranges from 1.5 to 2. The Petrosian magnitude of a galaxy is often estimated as the flux contained within a radius of a factor $N_{P}$ larger than this Petrosian radius; SDSS uses $N_{P}=2$. Petrosian magnitudes effectively measure half-light radii within $3-4 R_{eff}$ rather than the nominal $8R_{eff}$ bounding box for the FITS images used to derive SDSS-equivalent magnitudes. We measure Petrosian radii using both circular apertures and elliptical apertures, using the best-fit ellipse from Sersic model fits in the latter case.

\begin{figure}
\includegraphics[width=0.5\textwidth]{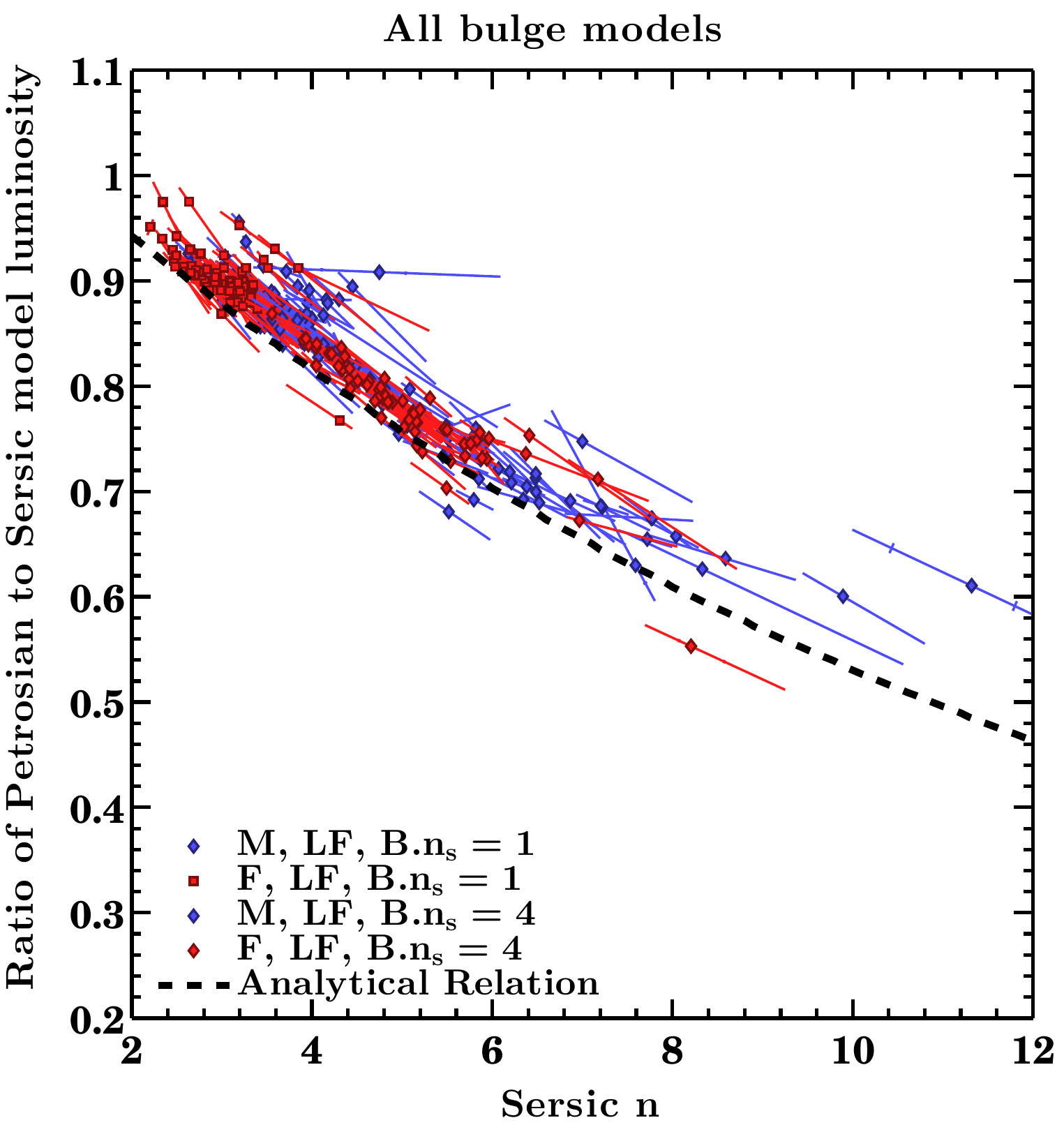}
\includegraphics[width=0.5\textwidth]{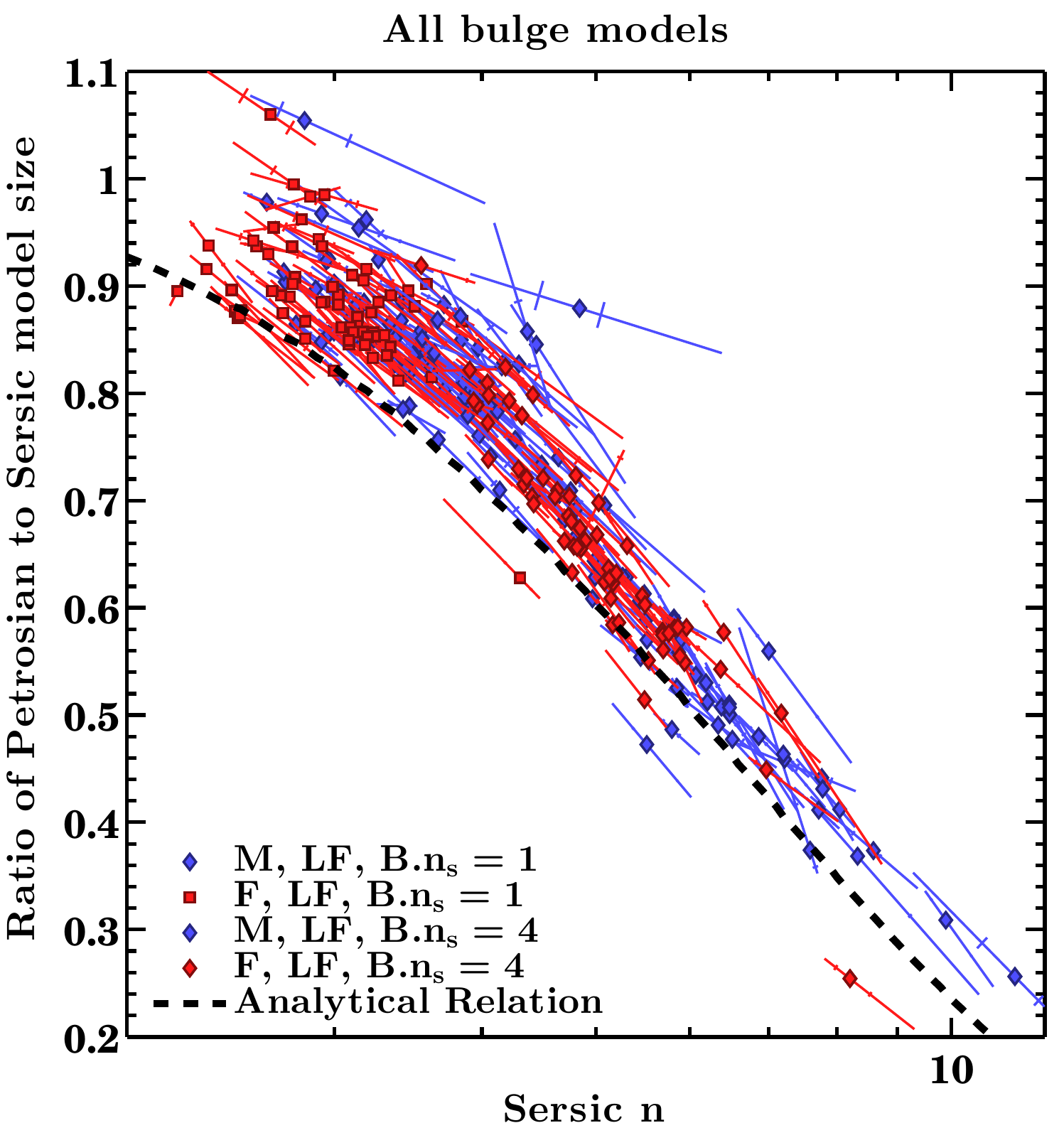}
\caption{Ratio of Petrosian model to Sersic model luminosities and sizes. Petrosian luminosities are derived from the elliptical Petrosian half-light radii measured within twice the Petrosian radius. Petrosian sizes and luminosities generally follow the analytical relation for a pure Sersic profile, underestimating sizes and luminosities by larger fractions for large Sersic indices. 
\label{fig:l2repratn}}
\end{figure}

Unfortunately, as shown in \figref{l2repratn}, Petrosian luminosities appear to underestimate the true galaxy luminosity by a similar amount to the analytical relation for purely circular profiles (see \citet{GraDri05} for a reference to various Sersic quantities). Sizes are also underestimated to a similar degree as predicted for a pure circular Sersic profile, which suggests that most galaxies do not deviate greatly from a pure Sersic profile. The slight excess could be due to a number of factors, including the Sersic models underestimating the true half-light radii and/or Sersic indices, radial variations in the ellipticity or shape of the isophotes, or deviations of the underlying profile from a pure Sersic model, all of which are plausible. In principle, one can correct for this 'missing' flux using fitting formulae valid for a wide range of Sersic or other profiles \citep{GraDriPet05}, but this seems unnecessary given that the Sersic fits appear sufficient and are available for all of the simulations and observational catalogs alike.

\section{Numerical Convergence}
\label{app:numerics}

We test the numerical convergence of the spherical Sersic plus halo models by simulating every galaxy for 2 Gyr at 3 different resolutions (differing in particle number by a factor of 8 in each step). We also test a subset of the group simulations at similar resolutions. All measurements are made using the same analysis pipeline as the results above; the images also have the same nominal redshift of $z=0.025$.

Convergence is generally quite good. With a 0.2 Myr timestep, total energy is conserved to better than one part in $10^{5}$. With the initial conditions re-centered to the barycenter, linear momentum remains small. The net angular momentum vector is the least well conserved quantity in Sersic plus halo models; each orthogonal component can vary by up to 5\% of the net rotation. However, the total angular momentum is usually dominated by a small number of dark matter halo particles at large distances from the galaxy center. Angular momentum conservation for baryons in isolated galaxies is considerably better, and deviations of 1 to 2\% are typical for groups where the bulk of the angular momentum is initially in galaxy orbits.

Having tested input parameter (Sersic index and effective radius) recovery with the analysis pipeline, we now turn to examining how these same parameters evolve in a 100 pc softened potential with a fixed, 0.2 Myr timestep, as in the group simulations. While idealized, these simulations are comparable to both the central ellipticals (which are slowly rotating and close to Sersic profiles, albeit somewhat flattened) and the bulges of the input spirals (which are smaller than the Sersic models and also slightly flattened by the presence of the disk) and will give estimates for how galaxy structure is affected by numerical resolution.

\subsection{Sersic plus Halo Model Convergence}

For a typical model ($R_{eff}$=8 kpc) at very high resolution (7.68 million star particle), convergence of all parameters is achieved at the 1 to 2\% level, with sizes, Sersic indices and dispersions shrinking slightly over 2 Gyr. Convergence is considerably worse for the $\mathrm{n_{s}=4}$ model and is strongly resolution dependent. A factor eight drop to high resolution (0.96 million star particles) approximately double errors in all parameters to 2-4\%. For medium resolution (120,000 star particle), $\mathrm{n_{s}=4}$ models, parameters can shrink by over 10\% - typical values being 5 to 15\% for $\mathrm{n_{s}}$ (4 to 3.4), 15\% for sizes (8 kpc to 6.8 kpc) and 5\% for dispersions. Thus, for larger ellipticals to be suitably resolved, a million or more stellar particles are required, especially if the profiles are as or more centrally concentrated than an $\mathrm{n_{s}=4}$ model. Less centrally concentrated models such as $\mathrm{n_{s}=2}$ are much less sensitive to numerical resolution and can be resolved by 100,000 stellar particles with at most 3 to 4\% level drops in sizes and Sersic index. Only 4 simulations in the sample have fewer than 720,000 stellar particles, so central remnants are largely unaffected by numerical relaxation after formation regardless of their central concentration.

Unfortunately, the results are not as encouraging for smaller models. For the smallest $R_{eff}$=2 kpc model at low (15,000 star particle) resolution, Sersic indices shrink up to 50\% (from 2 to 1.5, or 4 to 2.3). Sizes typically drop by less than 10\%, but dispersions also shrink up to 20\%. At high resolution, Sersic indices converge at the 5 to 15\% level (from 2 to 1.9 and 4 to 3.4). Sizes remain constant for $\mathrm{n_{s}=2}$ and drop at most 5\% for $\mathrm{n_{s}=4}$, with dispersions also shrinking by 3 to 5\%. Typical remnants are resolved at close to this high resolution, so the greatest effect would be on the Sersic indices of small, high $\mathrm{n_{s}}$ ellipticals. 

The greater concern with these results is the relaxation that occurs in the bulges of progenitor spirals. The effective radius of the M31 model is 1.5 kpc, but most galaxies are scaled to smaller sizes than this, with 0.5 to 1 kpc bulge $R_{eff}$. Moreover, in groups with larger numbers of galaxies, total particle counts are larger, but individual spirals can have as few as 60,000 stellar particles, of which only 20,000 are in the bulge. The bulge is partially stabilized (and flattened) by the disk, but the disk forms a core near the center of the galaxy, and so one might expect the behaviour of these compact, marginally resolved bulges to be similar to the Sersic plus halo models. We will now test this hypothesis with convergence studies of group mergers.

\subsection{Group Simulation Convergence}

We test numerical convergence in the groups by running a selected sample with a factor of eight higher and lower resolution and comparing parameters after the usual elapsed times (5.0, 7.7 and 10.3 Gyr). As all of the groups are resolved with an average of over a million stellar particles, numerical convergence is expected to be good once groups have merged. However, as detailed above, the least massive spirals in more massive groups are not as well resolved, so not all groups are expected to be converged at our standard resolution.

\subsubsection{Parameter Recovery}

\begin{figure*}
\includegraphics[width=0.49\textwidth]{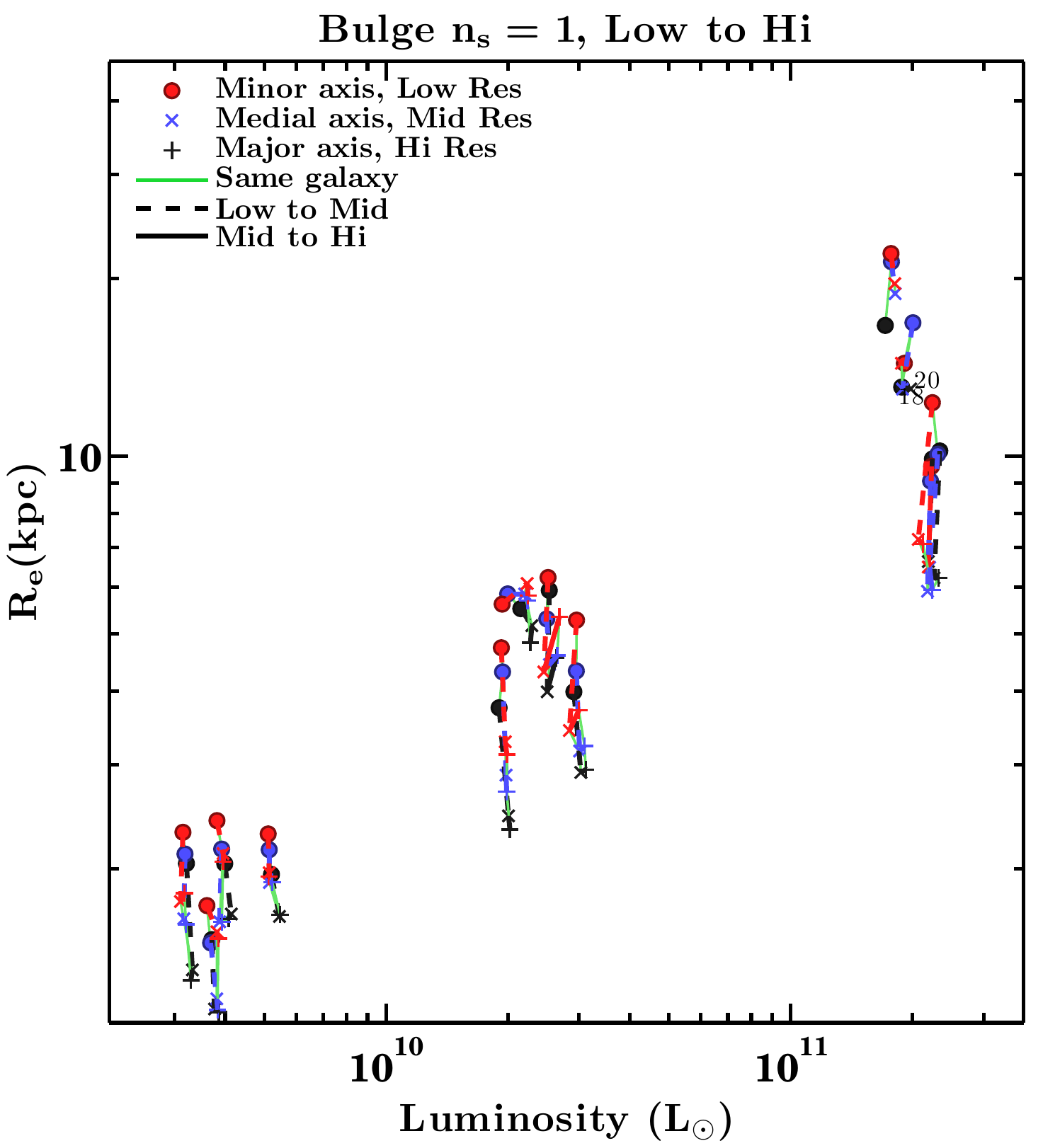}
\includegraphics[width=0.49\textwidth]{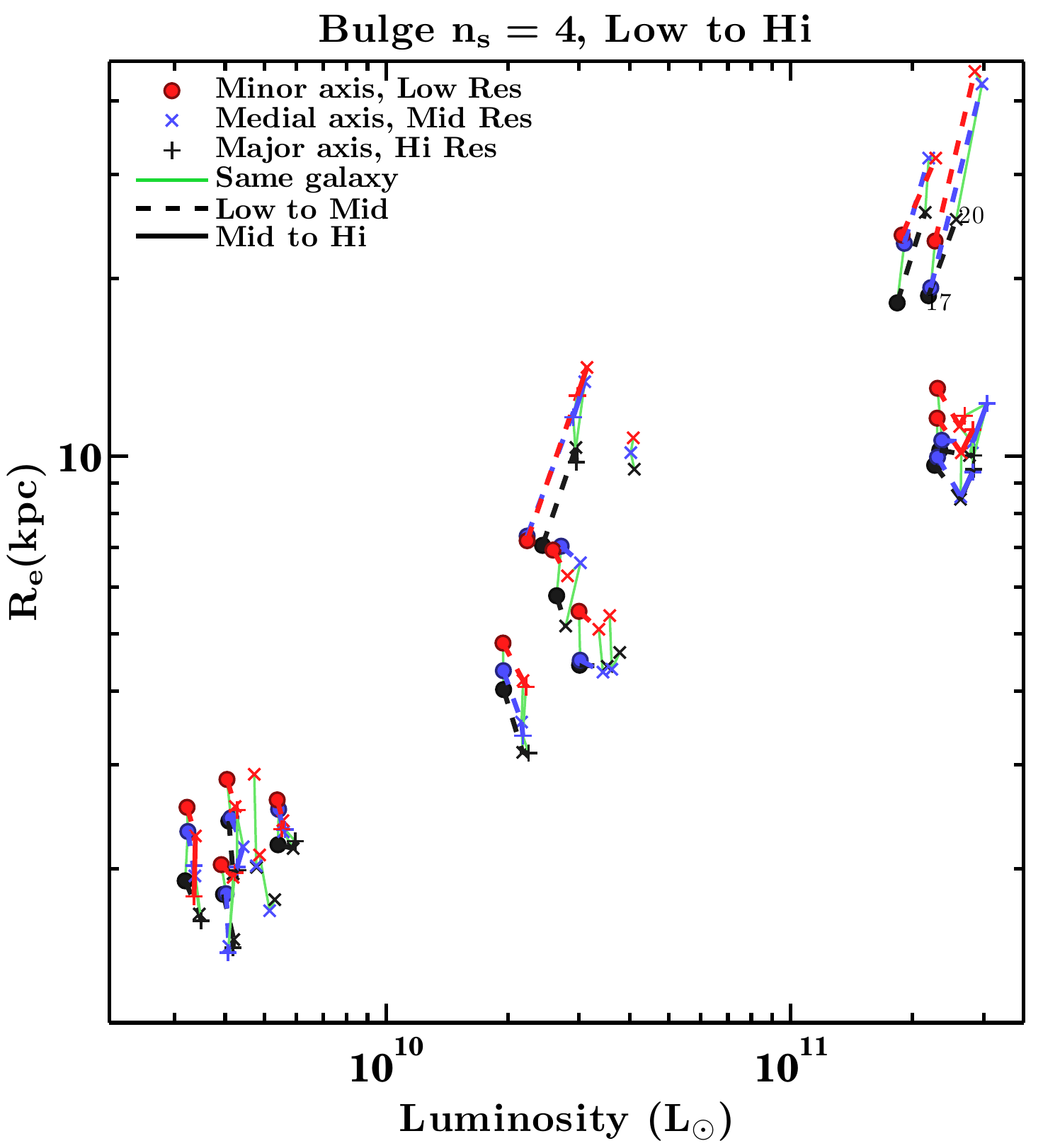}
\includegraphics[width=0.49\textwidth]{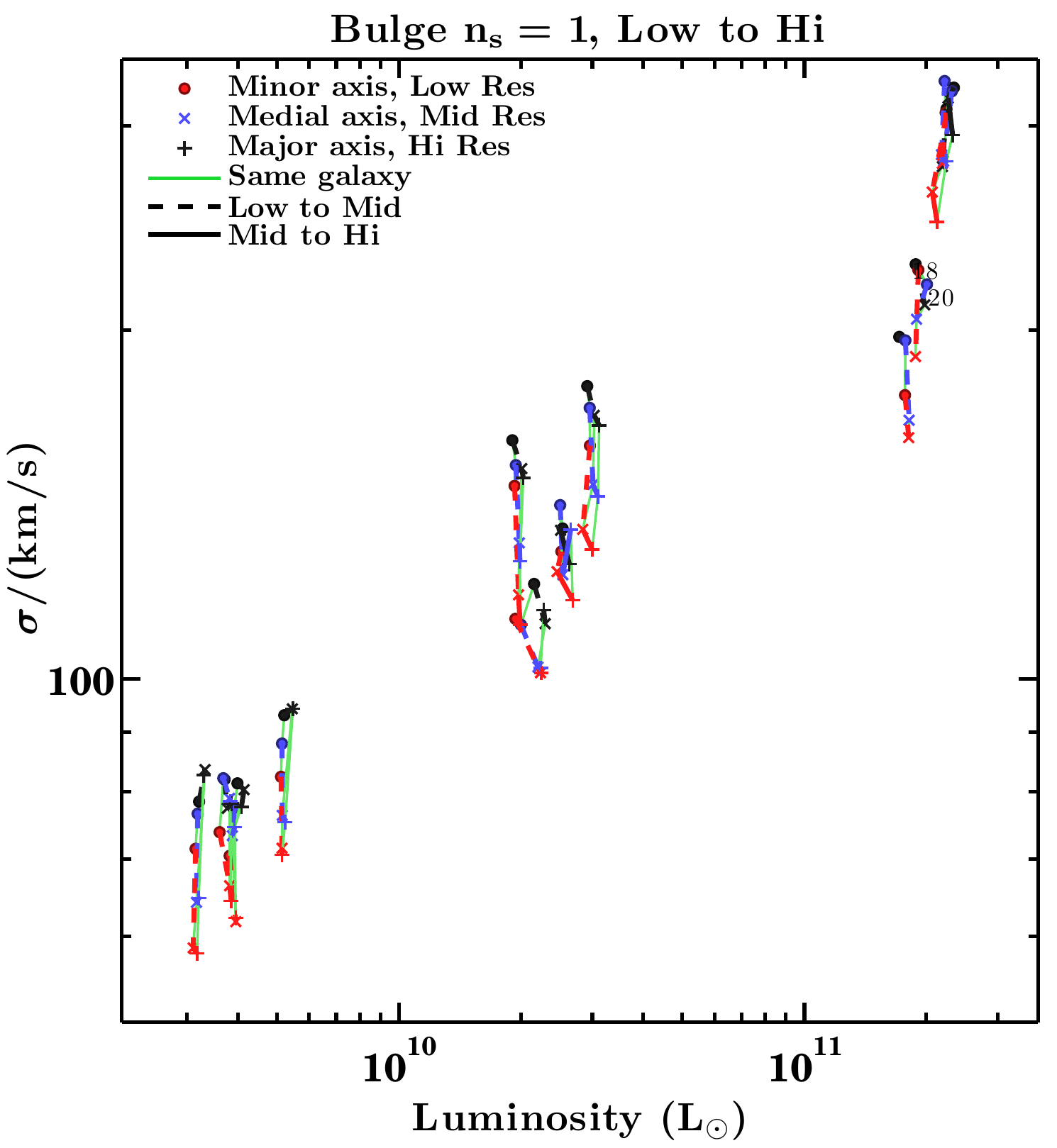}
\includegraphics[width=0.49\textwidth]{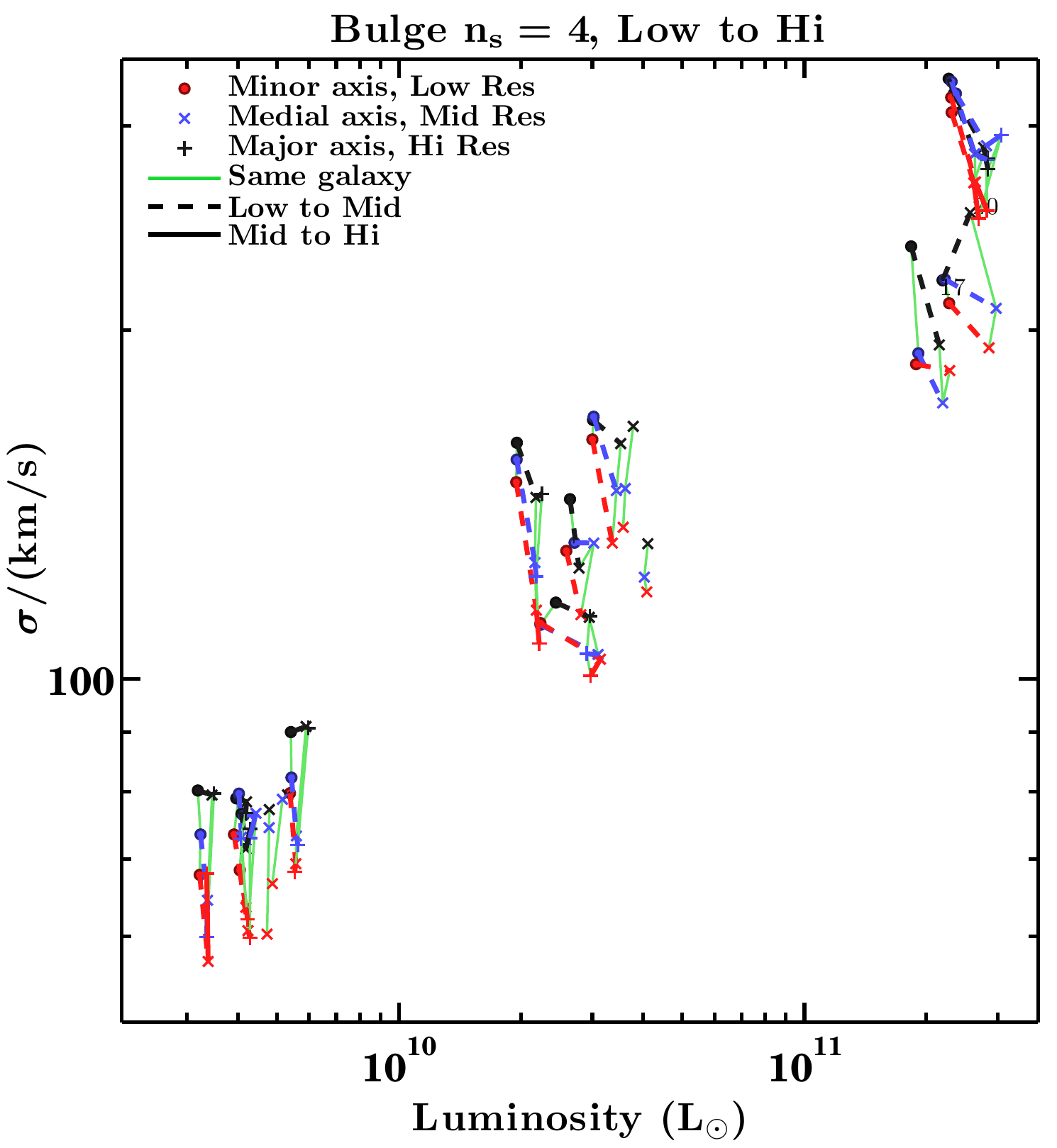}
\caption{Numerical convergence about size-luminosity and size-$\mathrm{\sigma}$ relations for principal axis projections of selected groups after 10.3 Gyr. At low resolutions, galaxies generally have larger sizes and dispersions than at higher resolution.
\label{fig:l_re_conv}}
\end{figure*}

\figref{l_re_conv} shows convergence for several identical groups on the size-luminosity and size-$\mathrm{\sigma}$ relations after 10.3 Gyr. Central remnant luminosities are fairly constant across all resolutions, but low resolutions can have slightly lower values. Sizes and dispersions are larger at low resolutions. Both trends continue from fiducial/medium to high resolution, although it is not as extreme - sizes are usually not more than 10 percent smaller between medium and high resolution.

\begin{figure*}
\includegraphics[width=0.49\textwidth]{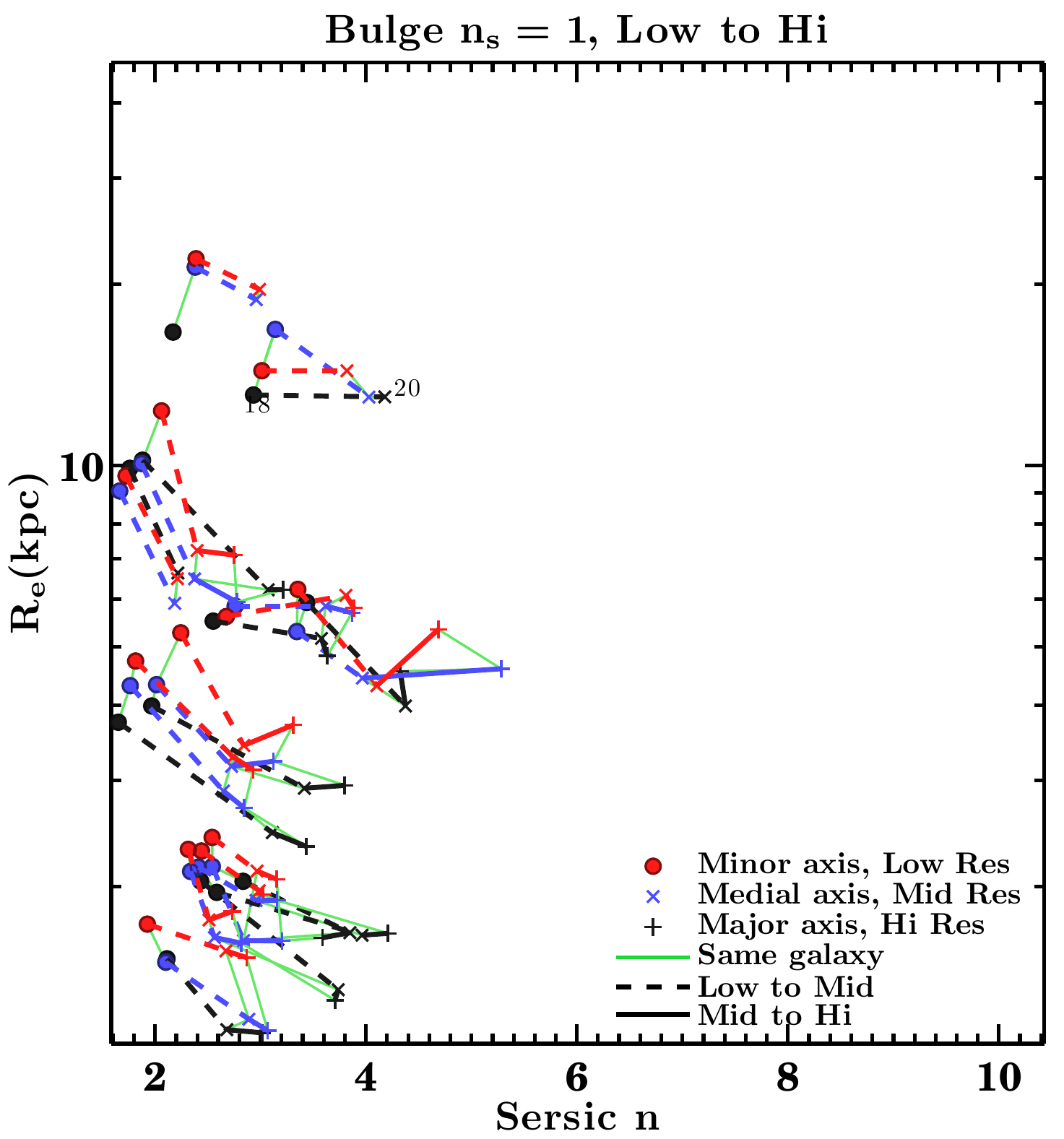}
\includegraphics[width=0.49\textwidth]{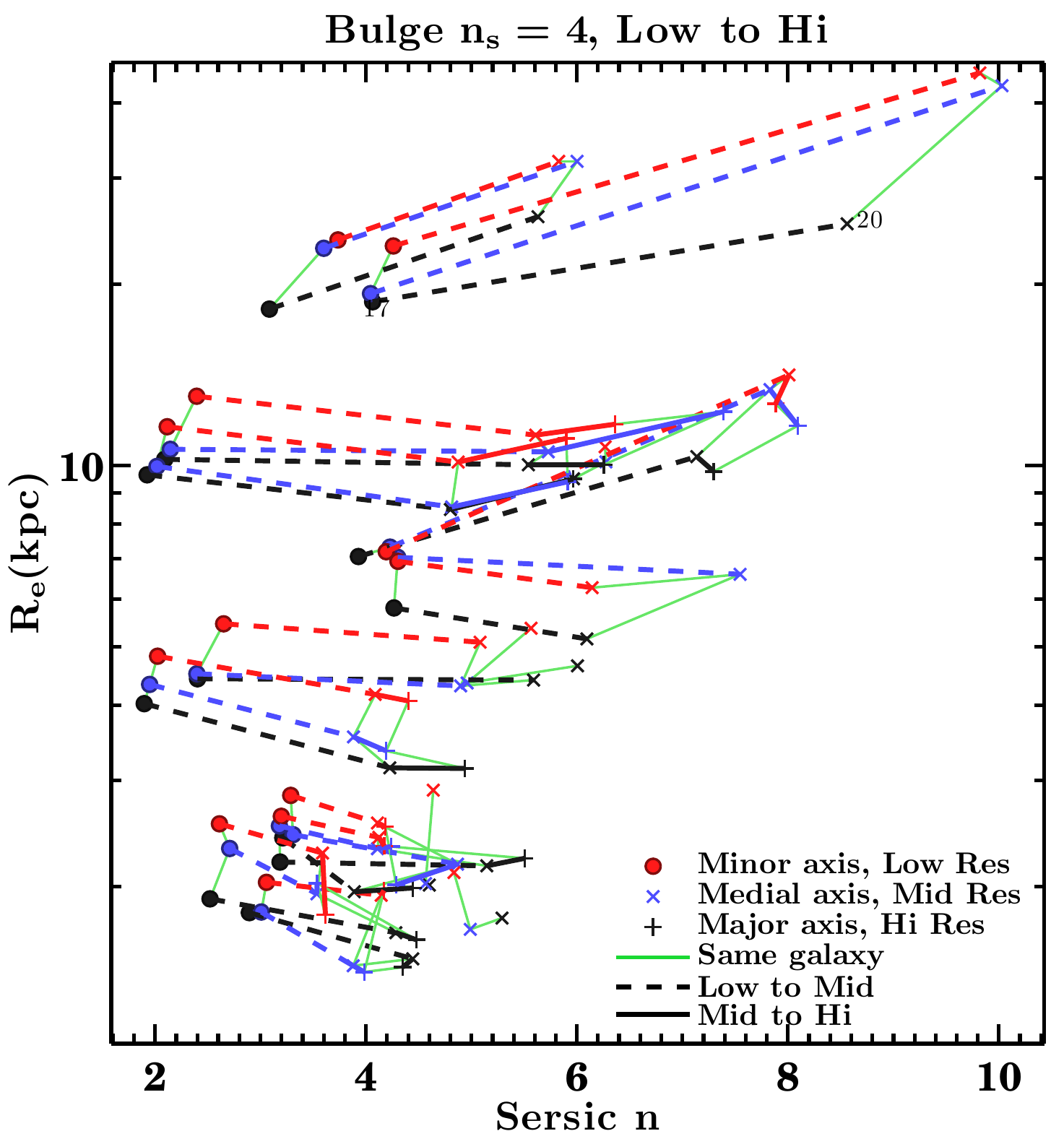}
\caption{Numerical convergence about the size-$\mathrm{n_{s}}$ relation for principal axis projections of selected groups. At low resolutions, galaxies have lower $\mathrm{n_{s}}$ by a factor of 1-2 than at medium or high resolution.
\label{fig:n_re_conv}}
\end{figure*}

Sersic indices are systematically lower at low resolution by a factor of 1 to 2 (\figref{n_re_conv}). The trend persists at high resolution, although $\mathrm{n_{s}}$ typically increases by a smaller factor of 0.2 to 0.3 between mid to high resolutions. Of the four parameters tested, then, luminosity appears to be the most robust, while the Sersic index is most sensitive to resolution effects. The effects on sizes are too small to fully reconcile the mismatch between sizes of faint simulated galaxies compared to observed ellipticals (\figref{sizelum}). Dispersions generally decrease with increasing resolution, and so numerical effects also cannot explain the lower intercept of the simulated Faber-Jackson relation compared to that of observed ellipticals (\figref{fj}).

In general, increasing resolution by a factor of eight produces similar trends in the group simulations as in isolated Sersic plus halo models - Sersic indices increase, while sizes and dispersions decrease. The effects are not very large going from our standard (medium) to high resolution but are considerable when stepping down to low resolution. We recommend that a minimum of a million stellar particles be used to adequately resolve spheroidal galaxies. While luminosities and masses remain converged at low resolution, sizes and dispersions are overestimated. Sersic indices are especially untrustworthy, being systematically offset lower by one or two from higher resolutions.

\section{Scaling Relations at Different Times}
\label{app:evolution}

The scaling relations presented in \subsecref{scalerel} are nominally for a zero-redshift galaxy population, assuming evolution from z=2. We can instead consider scaling relations at younger ages, assuming a fixed formation time for all groups. This is equivalent to assuming evolution from z=1 or z=0.5, since the only initial redshift-dependent parameter in the initial conditions is the group size. One might also consider combining groups from different snapshots into a single sample to simulate a sample with galaxies of different ages; however, this is best left to purely cosmological initial conditions with known merger trees and formation times.

\begin{table}
\caption{Sersic model size-luminosity relations at different times}
Simulations, Sersic model L and $r_{eff}$, Unweighted \\
\begin{tabular}{cccccc}
\hline 
B.$\mathrm{n_{s}}$ & Time & Sample & Slope & Intercept & R.M.S. \\
\hline
All & 5.0 & All & 0.51 $\pm$ 0.01 & -4.73 $\pm$ 0.06 & 0.11 \\ 
All & 7.7 & All & 0.53 $\pm$ 0.01 & -4.88 $\pm$ 0.05 & 0.11 \\ 
All & 10.3 & All & 0.58 $\pm$ 0.01 & -5.32 $\pm$ 0.06 & 0.12 \\    
\hline
All & 5.0 & Many & 0.57 $\pm$ 0.01 & -5.29 $\pm$ 0.10 & 0.11 \\ 
All & 7.7 & Many & 0.58 $\pm$ 0.01 & -5.30 $\pm$ 0.06 & 0.10 \\ 
All & 10.3 & Many & 0.62 $\pm$ 0.01 & -5.69 $\pm$ 0.06 & 0.10 \\ 
\hline
All & 5.0 & Few & 0.47 $\pm$ 0.01 & -4.36 $\pm$ 0.06 & 0.10 \\ 
All & 7.7 & Few & 0.50 $\pm$ 0.01 & -4.61 $\pm$ 0.06 & 0.10 \\ 
All & 10.3 & Few & 0.54 $\pm$ 0.01 & -4.96 $\pm$ 0.08 & 0.12 \\ 
\hline
\end{tabular}

\tablecomments{Sersic model size-luminosity relations of simulations after different times have elapsed (in Gyr) or, equivalently, assuming different formation redshifts (0.5, 1.0 and 2.0). Data are for ten equally-spaced, randomly oriented projections of each galaxy.}
\label{tab:l_reff_evol}
\end{table}

With these caveats in mind, we now present predictions for the evolution of the slope and scatter of selected scaling relations assuming a fixed formation time for all groups. The best-fit relations measured in \tabref{l_reff_evol} show slight evolution with time in the slopes (increasing) and intercepts (decreasing) and limited evolution in scatter. The steepening of the slope and lowering of the intercept would seem to suggest that brighter ellipticals grow off the relation at later times while fainter ellipticals grow slowly, if it all - in our case largely by construction, since the Few-merger sample does not have any late-time mergers. This interpretation is complicated by the fact that some of the largest groups do not have a relaxed, early-type central remnant formed in the earlier time steps and so are not included in the sample at earlier times but are included later on. Thus, as in most observational catalogs, not all of the descendants can necessarily be clearly identified with a previous early-type ancestor.

\begin{table}
\caption{Sersic model Faber-Jackson relations of simulations after different times}
Simulations, Sersic model L, Unweighted \\
\begin{tabular}{cccccc}
\hline 
B.$\mathrm{n_{s}}$ & Time & Sample & Slope & Intercept & R.M.S. \\
\hline
All & 5.0 & All & 0.31 $\pm$ 0.00 & -1.08 $\pm$ 0.02 & 0.05 \\ 
All & 7.7 & All & 0.30 $\pm$ 0.00 & -0.99 $\pm$ 0.02 & 0.04 \\ 
All & 10.3 & All & 0.28 $\pm$ 0.00 & -0.86 $\pm$ 0.02 & 0.04 \\ 
\hline
All & 5.0 & Many & 0.29 $\pm$ 0.00 & -0.89 $\pm$ 0.03 & 0.05 \\ 
All & 7.7 & Many & 0.28 $\pm$ 0.00 & -0.83 $\pm$ 0.02 & 0.04 \\ 
All & 10.3 & Many & 0.27 $\pm$ 0.00 & -0.72 $\pm$ 0.02 & 0.04 \\ 
\hline
All & 5.0 & Few & 0.32 $\pm$ 0.00 & -1.21 $\pm$ 0.02 & 0.04 \\ 
All & 7.7 & Few & 0.31 $\pm$ 0.00 & -1.09 $\pm$ 0.02 & 0.04 \\
All & 10.3 & Few & 0.30 $\pm$ 0.00 & -0.98 $\pm$ 0.03 & 0.04 \\ 
\hline
\end{tabular}

\tablecomments{Sersic model Faber-Jackson relations of simulations after different times have elapsed. Format as in \tabref{l_reff_evol}. The slopes generally flatten slightly while intercepts increase and scatter remains constant at 0.04 dex.}
\label{tab:fj_evol}
\end{table}

The best-fit Faber-Jackson relations measured in \tabref{l_reff_evol} also show slight evolution of the slope, but in the opposite sense (decreasing/flattening), with a corresponding increase in the intercept. However, the scatter remains largely unchanged at 0.04 dex.

\bibliography{taranu}

\begin{thebibliography}{81}
\expandafter\ifx\csname natexlab\endcsname\relax\def\natexlab#1{#1}\fi

\bibitem[{{Abazajian} {et~al.}(2004){Abazajian}, {Adelman-McCarthy},
  {Ag{\"u}eros}, {Allam}, {Anderson}, {Anderson}, {Annis}, {Bahcall}, {Baldry},
  {Bastian}, {Berlind}, {Bernardi}, {Blanton}, {Bochanski}, {Boroski},
  {Briggs}, {Brinkmann}, {Brunner}, {Budav{\'a}ri}, {Carey}, {Carliles},
  {Castander}, {Connolly}, {Csabai}, {Doi}, {Dong}, {Eisenstein}, {Evans},
  {Fan}, {Finkbeiner}, {Friedman}, {Frieman}, {Fukugita}, {Gal}, {Gillespie},
  {Glazebrook}, {Gray}, {Grebel}, {Gunn}, {Gurbani}, {Hall}, {Hamabe},
  {Harris}, {Harris}, {Harvanek}, {Heckman}, {Hendry}, {Hennessy}, {Hindsley},
  {Hogan}, {Hogg}, {Holmgren}, {Ichikawa}, {Ichikawa}, {Ivezi{\'c}}, {Jester},
  {Johnston}, {Jorgensen}, {Kent}, {Kleinman}, {Knapp}, {Kniazev}, {Kron},
  {Krzesinski}, {Kunszt}, {Kuropatkin}, {Lamb}, {Lampeitl}, {Lee}, {Leger},
  {Li}, {Lin}, {Loh}, {Long}, {Loveday}, {Lupton}, {Malik}, {Margon},
  {Matsubara}, {McGehee}, {McKay}, {Meiksin}, {Munn}, {Nakajima}, {Nash},
  {Neilsen}, {Newberg}, {Newman}, {Nichol}, {Nicinski}, {Nieto-Santisteban},
  {Nitta}, {Okamura}, {O'Mullane}, {Ostriker}, {Owen}, {Padmanabhan},
  {Peoples}, {Pier}, {Pope}, {Quinn}, {Richards}, {Richmond}, {Rix}, {Rockosi},
  {Schlegel}, {Schneider}, {Scranton}, {Sekiguchi}, {Seljak}, {Sergey},
  {Sesar}, {Sheldon}, {Shimasaku}, {Siegmund}, {Silvestri}, {Smith}, {Smol{\v
  c}i{\'c}}, {Snedden}, {Stebbins}, {Stoughton}, {Strauss}, {SubbaRao},
  {Szalay}, {Szapudi}, {Szkody}, {Szokoly}, {Tegmark}, {Teodoro}, {Thakar},
  {Tremonti}, {Tucker}, {Uomoto}, {Vanden Berk}, {Vandenberg}, {Vogeley},
  {Voges}, {Vogt}, {Walkowicz}, {Wang}, {Weinberg}, {West}, {White}, {Wilhite},
  {Xu}, {Yanny}, {Yasuda}, {Yip}, {Yocum}, {York}, {Zehavi}, {Zibetti}, \&
  {Zucker}}]{SDSSDR2}
{Abazajian}, K., {Adelman-McCarthy}, J.~K., {Ag{\"u}eros}, M.~A., {et~al.}
  2004, \aj, 128, 502

\bibitem[{{Abazajian} {et~al.}(2009){Abazajian}, {Adelman-McCarthy},
  {Ag{\"u}eros}, {Allam}, {Allende Prieto}, {An}, {Anderson}, {Anderson},
  {Annis}, {Bahcall}, \& et~al.}]{SDSSDR7}
{Abazajian}, K.~N., {Adelman-McCarthy}, J.~K., {Ag{\"u}eros}, M.~A., {et~al.}
  2009, \apjs, 182, 543

\bibitem[{{Aceves} \& {Vel{\'a}zquez}(2005)}]{AceVel05}
{Aceves}, H., \& {Vel{\'a}zquez}, H. 2005, \mnras, 360, L50

\bibitem[{{Aihara} {et~al.}(2011){Aihara}, {Allende Prieto}, {An}, {Anderson},
  {Aubourg}, {Balbinot}, {Beers}, {Berlind}, {Bickerton}, {Bizyaev}, {Blanton},
  {Bochanski}, {Bolton}, {Bovy}, {Brandt}, {Brinkmann}, {Brown}, {Brownstein},
  {Busca}, {Campbell}, {Carr}, {Chen}, {Chiappini}, {Comparat}, {Connolly},
  {Cortes}, {Croft}, {Cuesta}, {da Costa}, {Davenport}, {Dawson}, {Dhital},
  {Ealet}, {Ebelke}, {Edmondson}, {Eisenstein}, {Escoffier}, {Esposito},
  {Evans}, {Fan}, {Femen{\'{\i}}a Castell{\'a}}, {Font-Ribera}, {Frinchaboy},
  {Ge}, {Gillespie}, {Gilmore}, {Gonz{\'a}lez Hern{\'a}ndez}, {Gott}, {Gould},
  {Grebel}, {Gunn}, {Hamilton}, {Harding}, {Harris}, {Hawley}, {Hearty}, {Ho},
  {Hogg}, {Holtzman}, {Honscheid}, {Inada}, {Ivans}, {Jiang}, {Johnson},
  {Jordan}, {Jordan}, {Kazin}, {Kirkby}, {Klaene}, {Knapp}, {Kneib},
  {Kochanek}, {Koesterke}, {Kollmeier}, {Kron}, {Lampeitl}, {Lang}, {Le Goff},
  {Lee}, {Lin}, {Long}, {Loomis}, {Lucatello}, {Lundgren}, {Lupton}, {Ma},
  {MacDonald}, {Mahadevan}, {Maia}, {Makler}, {Malanushenko}, {Malanushenko},
  {Mandelbaum}, {Maraston}, {Margala}, {Masters}, {McBride}, {McGehee},
  {McGreer}, {M{\'e}nard}, {Miralda-Escud{\'e}}, {Morrison}, {Mullally},
  {Muna}, {Munn}, {Murayama}, {Myers}, {Naugle}, {Neto}, {Nguyen}, {Nichol},
  {O'Connell}, {Ogando}, {Olmstead}, {Oravetz}, {Padmanabhan},
  {Palanque-Delabrouille}, {Pan}, {Pandey}, {P{\^a}ris}, {Percival},
  {Petitjean}, {Pfaffenberger}, {Pforr}, {Phleps}, {Pichon}, {Pieri}, {Prada},
  {Price-Whelan}, {Raddick}, {Ramos}, {Reyl{\'e}}, {Rich}, {Richards}, {Rix},
  {Robin}, {Rocha-Pinto}, {Rockosi}, {Roe}, {Rollinde}, {Ross}, {Ross},
  {Rossetto}, {S{\'a}nchez}, {Sayres}, {Schlegel}, {Schlesinger}, {Schmidt},
  {Schneider}, {Sheldon}, {Shu}, {Simmerer}, {Simmons}, {Sivarani}, {Snedden},
  {Sobeck}, {Steinmetz}, {Strauss}, {Szalay}, {Tanaka}, {Thakar}, {Thomas},
  {Tinker}, {Tofflemire}, {Tojeiro}, {Tremonti}, {Vandenberg}, {Vargas
  Maga{\~n}a}, {Verde}, {Vogt}, {Wake}, {Wang}, {Weaver}, {Weinberg}, {White},
  {White}, {Yanny}, {Yasuda}, {Yeche}, \& {Zehavi}}]{SDSSDR8}
{Aihara}, H., {Allende Prieto}, C., {An}, D., {et~al.} 2011, \apjs, 193, 29

\bibitem[{{Barnes}(1985)}]{Bar85}
{Barnes}, J. 1985, \mnras, 215, 517

\bibitem[{{Barnes} \& {Hut}(1986)}]{BarHut86}
{Barnes}, J., \& {Hut}, P. 1986, \nat, 324, 446

\bibitem[{{Barnes}(1989)}]{Bar89}
{Barnes}, J.~E. 1989, \nat, 338, 123

\bibitem[{{Bekki}(2001)}]{Bek01}
{Bekki}, K. 2001, \apj, 546, 189

\bibitem[{{Blanton} {et~al.}(2005){Blanton}, {Schlegel}, {Strauss},
  {Brinkmann}, {Finkbeiner}, {Fukugita}, {Gunn}, {Hogg}, {Ivezi{\'c}}, {Knapp},
  {Lupton}, {Munn}, {Schneider}, {Tegmark}, \& {Zehavi}}]{BlaSchStr05}
{Blanton}, M.~R., {Schlegel}, D.~J., {Strauss}, M.~A., {et~al.} 2005, \aj, 129,
  2562

\bibitem[{{Bois} {et~al.}(2011){Bois}, {Emsellem}, {Bournaud}, {Alatalo},
  {Blitz}, {Bureau}, {Cappellari}, {Davies}, {Davis}, {de Zeeuw}, {Duc},
  {Khochfar}, {Krajnovi{\'c}}, {Kuntschner}, {Lablanche}, {McDermid},
  {Morganti}, {Naab}, {Oosterloo}, {Sarzi}, {Scott}, {Serra}, {Weijmans}, \&
  {Young}}]{BoiEmsBou11}
{Bois}, M., {Emsellem}, E., {Bournaud}, F., {et~al.} 2011, \mnras, 416, 1654

\bibitem[{{Bournaud} {et~al.}(2007){Bournaud}, {Jog}, \&
  {Combes}}]{BouJogCom07}
{Bournaud}, F., {Jog}, C.~J., \& {Combes}, F. 2007, \aap, 476, 1179

\bibitem[{{Boylan-Kolchin} {et~al.}(2006){Boylan-Kolchin}, {Ma}, \&
  {Quataert}}]{BoyMaQua06}
{Boylan-Kolchin}, M., {Ma}, C.-P., \& {Quataert}, E. 2006, \mnras, 369, 1081

\bibitem[{{Cappellari} {et~al.}(2011){Cappellari}, {Emsellem}, {Krajnovi{\'c}},
  {McDermid}, {Scott}, {Verdoes Kleijn}, {Young}, {Alatalo}, {Bacon}, {Blitz},
  {Bois}, {Bournaud}, {Bureau}, {Davies}, {Davis}, {de Zeeuw}, {Duc},
  {Khochfar}, {Kuntschner}, {Lablanche}, {Morganti}, {Naab}, {Oosterloo},
  {Sarzi}, {Serra}, \& {Weijmans}}]{CapEmsKra11}
{Cappellari}, M., {Emsellem}, E., {Krajnovi{\'c}}, D., {et~al.} 2011, \mnras,
  413, 813

\bibitem[{{Carlberg}(1986)}]{Car86}
{Carlberg}, R.~G. 1986, \apj, 310, 593

\bibitem[{{Carnevali} {et~al.}(1981){Carnevali}, {Cavaliere}, \&
  {Santangelo}}]{CarCavSan81}
{Carnevali}, P., {Cavaliere}, A., \& {Santangelo}, P. 1981, \apj, 249, 449

\bibitem[{{Ciotti} {et~al.}(2007){Ciotti}, {Lanzoni}, \&
  {Volonteri}}]{CioLanVol07}
{Ciotti}, L., {Lanzoni}, B., \& {Volonteri}, M. 2007, \apj, 658, 65

\bibitem[{{Courteau} {et~al.}(2007){Courteau}, {Dutton}, {van den Bosch},
  {MacArthur}, {Dekel}, {McIntosh}, \& {Dale}}]{CouDutvdB07}
{Courteau}, S., {Dutton}, A.~A., {van den Bosch}, F.~C., {et~al.} 2007, \apj,
  671, 203

\bibitem[{{Courteau} {et~al.}(2011){Courteau}, {Widrow}, {McDonald},
  {Guhathakurta}, {Gilbert}, {Zhu}, {Beaton}, \& {Majewski}}]{CouWidMcD11}
{Courteau}, S., {Widrow}, L.~M., {McDonald}, M., {et~al.} 2011, \apj, 739, 20

\bibitem[{{Cox} {et~al.}(2006){Cox}, {Dutta}, {Di Matteo}, {Hernquist},
  {Hopkins}, {Robertson}, \& {Springel}}]{CoxDutDiM06}
{Cox}, T.~J., {Dutta}, S.~N., {Di Matteo}, T., {et~al.} 2006, \apj, 650, 791

\bibitem[{{Croton} {et~al.}(2005){Croton}, {Farrar}, {Norberg}, {Colless},
  {Peacock}, {Baldry}, {Baugh}, {Bland-Hawthorn}, {Bridges}, {Cannon}, {Cole},
  {Collins}, {Couch}, {Dalton}, {De Propris}, {Driver}, {Efstathiou}, {Ellis},
  {Frenk}, {Glazebrook}, {Jackson}, {Lahav}, {Lewis}, {Lumsden}, {Maddox},
  {Madgwick}, {Peterson}, {Sutherland}, \& {Taylor}}]{CroFarNor05}
{Croton}, D.~J., {Farrar}, G.~R., {Norberg}, P., {et~al.} 2005, \mnras, 356,
  1155

\bibitem[{{de Vaucouleurs}(1959)}]{deV59}
{de Vaucouleurs}, G. 1959, Handbuch der Physik, 53, 275

\bibitem[{{de Zeeuw} {et~al.}(2002){de Zeeuw}, {Bureau}, {Emsellem}, {Bacon},
  {Carollo}, {Copin}, {Davies}, {Kuntschner}, {Miller}, {Monnet}, {Peletier},
  \& {Verolme}}]{deZBurEms02}
{de Zeeuw}, P.~T., {Bureau}, M., {Emsellem}, E., {et~al.} 2002, \mnras, 329,
  513

\bibitem[{{Dubinski}(1996)}]{Dub96}
{Dubinski}, J. 1996, New Astronomy, 1, 133

\bibitem[{{Dubinski}(1998)}]{Dub98}
---. 1998, \apj, 502, 141

\bibitem[{{Dutton} {et~al.}(2011){Dutton}, {Conroy}, {van den Bosch}, {Simard},
  {Mendel}, {Courteau}, {Dekel}, {More}, \& {Prada}}]{DutConvdB11}
{Dutton}, A.~A., {Conroy}, C., {van den Bosch}, F.~C., {et~al.} 2011, \mnras,
  416, 322

\bibitem[{{Emsellem} {et~al.}(2011){Emsellem}, {Cappellari}, {Krajnovi{\'c}},
  {Alatalo}, {Blitz}, {Bois}, {Bournaud}, {Bureau}, {Davies}, {Davis}, {de
  Zeeuw}, {Khochfar}, {Kuntschner}, {Lablanche}, {McDermid}, {Morganti},
  {Naab}, {Oosterloo}, {Sarzi}, {Scott}, {Serra}, {van de Ven}, {Weijmans}, \&
  {Young}}]{EmsCapKra11}
{Emsellem}, E., {Cappellari}, M., {Krajnovi{\'c}}, D., {et~al.} 2011, \mnras,
  414, 888

\bibitem[{{Faber} \& {Jackson}(1976)}]{FabJac76}
{Faber}, S.~M., \& {Jackson}, R.~E. 1976, \apj, 204, 668

\bibitem[{{Fakhouri} {et~al.}(2010){Fakhouri}, {Ma}, \&
  {Boylan-Kolchin}}]{FakMaBoy10}
{Fakhouri}, O., {Ma}, C.-P., \& {Boylan-Kolchin}, M. 2010, \mnras, 406, 2267

\bibitem[{{Feldmann} {et~al.}(2011){Feldmann}, {Carollo}, \&
  {Mayer}}]{FelCarMay11}
{Feldmann}, R., {Carollo}, C.~M., \& {Mayer}, L. 2011, \apj, 736, 88

\bibitem[{{Graham} \& {Driver}(2005)}]{GraDri05}
{Graham}, A.~W., \& {Driver}, S.~P. 2005, \pasa, 22, 118

\bibitem[{{Graham} {et~al.}(2005){Graham}, {Driver}, {Petrosian}, {Conselice},
  {Bershady}, {Crawford}, \& {Goto}}]{GraDriPet05}
{Graham}, A.~W., {Driver}, S.~P., {Petrosian}, V., {et~al.} 2005, \aj, 130,
  1535

\bibitem[{{Hernquist}(1992)}]{Her92}
{Hernquist}, L. 1992, \apj, 400, 460

\bibitem[{{Hernquist}(1993)}]{Her93}
---. 1993, \apj, 409, 548

\bibitem[{{Hernquist} {et~al.}(1993){Hernquist}, {Spergel}, \&
  {Heyl}}]{HerSpeHey93}
{Hernquist}, L., {Spergel}, D.~N., \& {Heyl}, J.~S. 1993, \apj, 416, 415

\bibitem[{{Hilz} {et~al.}(2013){Hilz}, {Naab}, \& {Ostriker}}]{HilNaaOst12}
{Hilz}, M., {Naab}, T., \& {Ostriker}, J.~P. 2013, \mnras, 429, 2924

\bibitem[{{Holmberg}(1941)}]{Hol41}
{Holmberg}, E. 1941, \apj, 94, 385

\bibitem[{{Hopkins} {et~al.}(2008){Hopkins}, {Cox}, \&
  {Hernquist}}]{HopCoxHer08}
{Hopkins}, P.~F., {Cox}, T.~J., \& {Hernquist}, L. 2008, \apj, 689, 17

\bibitem[{{Hopkins} {et~al.}(2009){Hopkins}, {Hernquist}, {Cox}, {Keres}, \&
  {Wuyts}}]{HopHerCox09}
{Hopkins}, P.~F., {Hernquist}, L., {Cox}, T.~J., {Keres}, D., \& {Wuyts}, S.
  2009, \apj, 691, 1424

\bibitem[{{Hopkins} {et~al.}(2010){Hopkins}, {Croton}, {Bundy}, {Khochfar},
  {van den Bosch}, {Somerville}, {Wetzel}, {Keres}, {Hernquist}, {Stewart},
  {Younger}, {Genel}, \& {Ma}}]{HopCroBun10}
{Hopkins}, P.~F., {Croton}, D., {Bundy}, K., {et~al.} 2010, \apj, 724, 915

\bibitem[{{Hyde} \& {Bernardi}(2009{\natexlab{a}})}]{HydBer09a}
{Hyde}, J.~B., \& {Bernardi}, M. 2009{\natexlab{a}}, \mnras, 394, 1978

\bibitem[{{Hyde} \& {Bernardi}(2009{\natexlab{b}})}]{HydBer09b}
---. 2009{\natexlab{b}}, \mnras, 396, 1171

\bibitem[{{Ishizawa} {et~al.}(1983){Ishizawa}, {Matsumoto}, {Tajima},
  {Kageyama}, \& {Sakai}}]{IshMatTaj83}
{Ishizawa}, T., {Matsumoto}, R., {Tajima}, T., {Kageyama}, H., \& {Sakai}, H.
  1983, \pasj, 35, 61

\bibitem[{{Khalatyan} {et~al.}(2008){Khalatyan}, {Cattaneo}, {Schramm},
  {Gottl{\"o}ber}, {Steinmetz}, \& {Wisotzki}}]{KhaCatSch08}
{Khalatyan}, A., {Cattaneo}, A., {Schramm}, M., {et~al.} 2008, \mnras, 387, 13

\bibitem[{{Kormendy}(1977)}]{Kor77}
{Kormendy}, J. 1977, \apj, 218, 333

\bibitem[{{Kormendy} {et~al.}(2009){Kormendy}, {Fisher}, {Cornell}, \&
  {Bender}}]{KorFisCor09}
{Kormendy}, J., {Fisher}, D.~B., {Cornell}, M.~E., \& {Bender}, R. 2009, \apjs,
  182, 216

\bibitem[{{Krajnovi{\'c}} {et~al.}(2012){Krajnovi{\'c}}, {Alatalo}, {Blitz},
  {Bois}, {Bournaud}, {Bureau}, {Cappellari}, {Davies}, {Davis}, {de Zeeuw},
  {Duc}, {Emsellem}, {Khochfar}, {Kuntschner}, {McDermid}, {Morganti}, {Naab},
  {Oosterloo}, {Sarzi}, {Scott}, {Serra}, {Weijmans}, \& {Young}}]{KraAlaBli12}
{Krajnovi{\'c}}, D., {Alatalo}, K., {Blitz}, L., {et~al.} 2012, \mnras, 278

\bibitem[{{McGee} {et~al.}(2009){McGee}, {Balogh}, {Bower}, {Font}, \&
  {McCarthy}}]{McGBalBow09}
{McGee}, S.~L., {Balogh}, M.~L., {Bower}, R.~G., {Font}, A.~S., \& {McCarthy},
  I.~G. 2009, \mnras, 400, 937

\bibitem[{{Moster} {et~al.}(2012){Moster}, {Macci{\`o}}, \&
  {Somerville}}]{MosMacSom12}
{Moster}, B.~P., {Macci{\`o}}, A.~V., \& {Somerville}, R.~S. 2012, ArXiv
  e-prints

\bibitem[{{Naab} {et~al.}(2009){Naab}, {Johansson}, \&
  {Ostriker}}]{NaaJohOst09}
{Naab}, T., {Johansson}, P.~H., \& {Ostriker}, J.~P. 2009, \apjl, 699, L178

\bibitem[{{Nair} {et~al.}(2011){Nair}, {van den Bergh}, \&
  {Abraham}}]{NaiAbr11}
{Nair}, P., {van den Bergh}, S., \& {Abraham}, R.~G. 2011, \apjl, 734, L31

\bibitem[{{Nair} \& {Abraham}(2010)}]{NaiAbr10}
{Nair}, P.~B., \& {Abraham}, R.~G. 2010, \apjs, 186, 427

\bibitem[{{Navarro} {et~al.}(1997){Navarro}, {Frenk}, \& {White}}]{NavFreWhi97}
{Navarro}, J.~F., {Frenk}, C.~S., \& {White}, S.~D.~M. 1997, \apj, 490, 493

\bibitem[{{Nipoti} {et~al.}(2003){Nipoti}, {Londrillo}, \&
  {Ciotti}}]{NipLonCio03}
{Nipoti}, C., {Londrillo}, P., \& {Ciotti}, L. 2003, \mnras, 342, 501

\bibitem[{{Nipoti} {et~al.}(2009{\natexlab{a}}){Nipoti}, {Treu}, {Auger}, \&
  {Bolton}}]{NipTreAug09}
{Nipoti}, C., {Treu}, T., {Auger}, M.~W., \& {Bolton}, A.~S.
  2009{\natexlab{a}}, \apjl, 706, L86

\bibitem[{{Nipoti} {et~al.}(2009{\natexlab{b}}){Nipoti}, {Treu}, \&
  {Bolton}}]{NipTreBol09}
{Nipoti}, C., {Treu}, T., \& {Bolton}, A.~S. 2009{\natexlab{b}}, \apj, 703,
  1531

\bibitem[{{Oser} {et~al.}(2012){Oser}, {Naab}, {Ostriker}, \&
  {Johansson}}]{OseNaaOst12}
{Oser}, L., {Naab}, T., {Ostriker}, J.~P., \& {Johansson}, P.~H. 2012, \apj,
  744, 63

\bibitem[{{Peng} {et~al.}(2002){Peng}, {Ho}, {Impey}, \& {Rix}}]{PenHoImp02}
{Peng}, C.~Y., {Ho}, L.~C., {Impey}, C.~D., \& {Rix}, H.-W. 2002, \aj, 124, 266

\bibitem[{{Peng} {et~al.}(2010){Peng}, {Ho}, {Impey}, \& {Rix}}]{PenHoImp10}
---. 2010, \aj, 139, 2097

\bibitem[{{Petrosian}(1976)}]{Pet76}
{Petrosian}, V. 1976, \apjl, 209, L1

\bibitem[{{Robertson} {et~al.}(2006){Robertson}, {Cox}, {Hernquist}, {Franx},
  {Hopkins}, {Martini}, \& {Springel}}]{RobCoxHer06}
{Robertson}, B., {Cox}, T.~J., {Hernquist}, L., {et~al.} 2006, \apj, 641, 21

\bibitem[{{Robotham} {et~al.}(2010){Robotham}, {Phillipps}, \& {de
  Propris}}]{RobPhideP10}
{Robotham}, A., {Phillipps}, S., \& {de Propris}, R. 2010, \mnras, 403, 1812

\bibitem[{{Schechter}(1976)}]{Sch76}
{Schechter}, P. 1976, \apj, 203, 297

\bibitem[{{Schmidt}(1968)}]{Sch68}
{Schmidt}, M. 1968, \apj, 151, 393

\bibitem[{{Shao} {et~al.}(2007){Shao}, {Xiao}, {Shen}, {Mo}, {Xia}, \&
  {Deng}}]{Sha07}
{Shao}, Z., {Xiao}, Q., {Shen}, S., {et~al.} 2007, \apj, 659, 1159

\bibitem[{{Simard} {et~al.}(2011){Simard}, {Mendel}, {Patton}, {Ellison}, \&
  {McConnachie}}]{SimMenPat11}
{Simard}, L., {Mendel}, J.~T., {Patton}, D.~R., {Ellison}, S.~L., \&
  {McConnachie}, A.~W. 2011, \apjs, 196, 11

\bibitem[{{Simard} {et~al.}(2002){Simard}, {Willmer}, {Vogt}, {Sarajedini},
  {Phillips}, {Weiner}, {Koo}, {Im}, {Illingworth}, \& {Faber}}]{SimWilVog02}
{Simard}, L., {Willmer}, C.~N.~A., {Vogt}, N.~P., {et~al.} 2002, \apjs, 142, 1

\bibitem[{{Simard} {et~al.}(2009){Simard}, {Clowe}, {Desai}, {Dalcanton}, {von
  der Linden}, {Poggianti}, {White}, {Arag{\'o}n-Salamanca}, {De Lucia},
  {Halliday}, {Jablonka}, {Milvang-Jensen}, {Saglia}, {Pell{\'o}}, {Rudnick},
  \& {Zaritsky}}]{SimCloDes09}
{Simard}, L., {Clowe}, D., {Desai}, V., {et~al.} 2009, \aap, 508, 1141

\bibitem[{{Stewart} {et~al.}(2008){Stewart}, {Bullock}, {Wechsler}, {Maller},
  \& {Zentner}}]{SteBulWec08}
{Stewart}, K.~R., {Bullock}, J.~S., {Wechsler}, R.~H., {Maller}, A.~H., \&
  {Zentner}, A.~R. 2008, \apj, 683, 597

\bibitem[{{Stoughton} {et~al.}(2002){Stoughton}, {Lupton}, {Bernardi},
  {Blanton}, {Burles}, {Castander}, {Connolly}, {Eisenstein}, {Frieman},
  {Hennessy}, {Hindsley}, {Ivezi{\'c}}, {Kent}, {Kunszt}, {Lee}, {Meiksin},
  {Munn}, {Newberg}, {Nichol}, {Nicinski}, {Pier}, {Richards}, {Richmond},
  {Schlegel}, {Smith}, {Strauss}, {SubbaRao}, {Szalay}, {Thakar}, {Tucker},
  {Vanden Berk}, {Yanny}, {Adelman}, {Anderson}, {Anderson}, {Annis},
  {Bahcall}, {Bakken}, {Bartelmann}, {Bastian}, {Bauer}, {Berman},
  {B{\"o}hringer}, {Boroski}, {Bracker}, {Briegel}, {Briggs}, {Brinkmann},
  {Brunner}, {Carey}, {Carr}, {Chen}, {Christian}, {Colestock}, {Crocker},
  {Csabai}, {Czarapata}, {Dalcanton}, {Davidsen}, {Davis}, {Dehnen},
  {Dodelson}, {Doi}, {Dombeck}, {Donahue}, {Ellman}, {Elms}, {Evans}, {Eyer},
  {Fan}, {Federwitz}, {Friedman}, {Fukugita}, {Gal}, {Gillespie}, {Glazebrook},
  {Gray}, {Grebel}, {Greenawalt}, {Greene}, {Gunn}, {de Haas}, {Haiman},
  {Haldeman}, {Hall}, {Hamabe}, {Hansen}, {Harris}, {Harris}, {Harvanek},
  {Hawley}, {Hayes}, {Heckman}, {Helmi}, {Henden}, {Hogan}, {Hogg}, {Holmgren},
  {Holtzman}, {Huang}, {Hull}, {Ichikawa}, {Ichikawa}, {Johnston}, {Kauffmann},
  {Kim}, {Kimball}, {Kinney}, {Klaene}, {Kleinman}, {Klypin}, {Knapp},
  {Korienek}, {Krolik}, {Kron}, {Krzesi{\'n}ski}, {Lamb}, {Leger},
  {Limmongkol}, {Lindenmeyer}, {Long}, {Loomis}, {Loveday}, {MacKinnon},
  {Mannery}, {Mantsch}, {Margon}, {McGehee}, {McKay}, {McLean}, {Menou},
  {Merelli}, {Mo}, {Monet}, {Nakamura}, {Narayanan}, {Nash}, {Neilsen},
  {Newman}, {Nitta}, {Odenkirchen}, {Okada}, {Okamura}, {Ostriker}, {Owen},
  {Pauls}, {Peoples}, {Peterson}, {Petravick}, {Pope}, {Pordes}, {Postman},
  {Prosapio}, {Quinn}, {Rechenmacher}, {Rivetta}, {Rix}, {Rockosi}, {Rosner},
  {Ruthmansdorfer}, {Sandford}, {Schneider}, {Scranton}, {Sekiguchi}, {Sergey},
  {Sheth}, {Shimasaku}, {Smee}, {Snedden}, {Stebbins}, {Stubbs}, {Szapudi},
  {Szkody}, {Szokoly}, {Tabachnik}, {Tsvetanov}, {Uomoto}, {Vogeley}, {Voges},
  {Waddell}, {Walterbos}, {Wang}, {Watanabe}, {Weinberg}, {White}, {White},
  {Wilhite}, {Wolfe}, {Yasuda}, {York}, {Zehavi}, \& {Zheng}}]{SDSSEDR}
{Stoughton}, C., {Lupton}, R.~H., {Bernardi}, M., {et~al.} 2002, \aj, 123, 485

\bibitem[{{Taranu} {et~al.}(2013){Taranu}, {Dubinski}, \& {Yee}}]{TarDubYee13b}
{Taranu}, D., {Dubinski}, J., \& {Yee}, H. 2013, {In Prep}

\bibitem[{{Toomre}(1977)}]{Too77}
{Toomre}, A. 1977, in Evolution of Galaxies and Stellar Populations, ed. B.~M.
  {Tinsley} \& R.~B.~G. {Larson}, D.~Campbell, 401

\bibitem[{{Toomre} \& {Toomre}(1972)}]{TooToo72}
{Toomre}, A., \& {Toomre}, J. 1972, \apj, 178, 623

\bibitem[{{Trujillo} {et~al.}(2011){Trujillo}, {Ferreras}, \& {de La
  Rosa}}]{TruFerdeL11}
{Trujillo}, I., {Ferreras}, I., \& {de La Rosa}, I.~G. 2011, \mnras, 415, 3903

\bibitem[{{Tully} \& {Fisher}(1977)}]{TulFis77}
{Tully}, R.~B., \& {Fisher}, J.~R. 1977, \aap, 54, 661

\bibitem[{{van der Kruit} \& {Freeman}(2011)}]{vdKFre11}
{van der Kruit}, P.~C., \& {Freeman}, K.~C. 2011, \araa, 49, 301

\bibitem[{{van Dokkum} {et~al.}(2010){van Dokkum}, {Whitaker}, {Brammer},
  {Franx}, {Kriek}, {Labb{\'e}}, {Marchesini}, {Quadri}, {Bezanson},
  {Illingworth}, {Muzzin}, {Rudnick}, {Tal}, \& {Wake}}]{vDoWhiBra10}
{van Dokkum}, P.~G., {Whitaker}, K.~E., {Brammer}, G., {et~al.} 2010, \apj,
  709, 1018

\bibitem[{{Watkins} {et~al.}(2010){Watkins}, {Evans}, \& {An}}]{Wat10}
{Watkins}, L.~L., {Evans}, N.~W., \& {An}, J.~H. 2010, \mnras, 406, 264

\bibitem[{{Weil} \& {Hernquist}(1996)}]{WeiHer96}
{Weil}, M.~L., \& {Hernquist}, L. 1996, \apj, 460, 101

\bibitem[{{Widrow} \& {Dubinski}(2005)}]{WidDub05}
{Widrow}, L.~M., \& {Dubinski}, J. 2005, \apj, 631, 838

\bibitem[{{Widrow} {et~al.}(2008){Widrow}, {Pym}, \& {Dubinski}}]{WidPymDub08}
{Widrow}, L.~M., {Pym}, B., \& {Dubinski}, J. 2008, \apj, 679, 1239

\bibitem[{{Yip} {et~al.}(2004){Yip}, {Connolly}, {Szalay}, {Budav{\'a}ri},
  {SubbaRao}, {Frieman}, {Nichol}, {Hopkins}, {York}, {Okamura}, {Brinkmann},
  {Csabai}, {Thakar}, {Fukugita}, \& {Ivezi{\'c}}}]{YipConSza04}
{Yip}, C.~W., {Connolly}, A.~J., {Szalay}, A.~S., {et~al.} 2004, \aj, 128, 585

\end{thebibliography}

\end{document}